\documentclass[11pt]{article}
\usepackage[dvipsnames]{xcolor}
\RequirePackage{doi}
\usepackage{hyperref}

\hypersetup{
   colorlinks,
   linktocpage,
    linkcolor=Maroon,
    filecolor=Maroon,      
    urlcolor=Maroon,
    citecolor=Maroon
}
\usepackage{slashed}
\usepackage[utf8]{inputenc}
\usepackage{amsmath, amssymb, amsthm, tabto, epigraph, mathtools, dsfont, verbatim, slashed, mathrsfs}
\usepackage{graphicx, wrapfig}
\usepackage{enumitem}
\usepackage{relsize}
\usepackage{subcaption}
\usepackage{csquotes}
\usepackage{bbm}
\usepackage{tabularx}
\usepackage{comment}
\usepackage{soul}
\usepackage{multirow}
\usepackage{adjustbox}

\usepackage[left=2.5cm,right=2.5cm,top=2.5cm,bottom=3cm]{geometry}
\usepackage{tikz}
\usetikzlibrary{arrows,decorations.markings,cd}
\tikzset{
->-/.style={decoration={markings, mark=at position .5 with {\arrow[scale=1.5]{stealth}}}, postaction={decorate}}
}
\usepackage{epsfig}
\usepackage{amssymb}
\usepackage{amsfonts}
\usepackage{amsbsy}
\usepackage{amsmath}
\usepackage{dsfont}
\usepackage{bbm}
\usepackage{upgreek}
\usepackage{amscd}
\usepackage[nosort]{cite}
\usepackage{graphicx}
\usepackage{mathrsfs}
\usepackage{amsmath,amsthm}
\usepackage{slashed}	
\usepackage{dsfont}
\usepackage[utf8]{inputenc}
\numberwithin{equation}{section}

\renewcommand{\thefootnote}{\fnsymbol{footnote}}

\makeatletter
\DeclareFontFamily{OMX}{MnSymbolE}{}
\DeclareSymbolFont{MnLargeSymbols}{OMX}{MnSymbolE}{m}{n}
\SetSymbolFont{MnLargeSymbols}{bold}{OMX}{MnSymbolE}{b}{n}
\DeclareFontShape{OMX}{MnSymbolE}{m}{n}{
    <-6>  MnSymbolE5
   <6-7>  MnSymbolE6
   <7-8>  MnSymbolE7
   <8-9>  MnSymbolE8
   <9-10> MnSymbolE9
  <10-12> MnSymbolE10
  <12->   MnSymbolE12
}{}
\DeclareFontShape{OMX}{MnSymbolE}{b}{n}{
    <-6>  MnSymbolE-Bold5
   <6-7>  MnSymbolE-Bold6
   <7-8>  MnSymbolE-Bold7
   <8-9>  MnSymbolE-Bold8
   <9-10> MnSymbolE-Bold9
  <10-12> MnSymbolE-Bold10
  <12->   MnSymbolE-Bold12
}{}

\let\llangle\@undefined
\let\rrangle\@undefined
\DeclareMathDelimiter{\llangle}{\mathopen}%
                     {MnLargeSymbols}{'164}{MnLargeSymbols}{'164}
\DeclareMathDelimiter{\rrangle}{\mathclose}%
                     {MnLargeSymbols}{'171}{MnLargeSymbols}{'171}
\makeatother

\DeclareFontShape{OT1}{cmr}{mx}{n}%
    {<->cmr10}{}
\newcommand{\mytitlefont}{\fontseries{mx}\selectfont}
\DeclareMathAlphabet{\titlemath}{OT1}{cmr}{mx}{n}

\def\ie{\begin{equation}\begin{aligned}}
\def\fe{\end{aligned}\end{equation}}

\allowdisplaybreaks

\begin{document}

\begin{titlepage}

\begin{center}

~\\[2cm]

{\fontsize{20pt}{0pt} \mytitlefont Non-Invertible Anyon Condensation\\   and Level-Rank Dualities}

~\\[0.1cm]

Clay C\'ordova\footnote{\href{clayc@uchicago.edu}{clayc@uchicago.edu}} and 
Diego Garc\'{i}a-Sep\'{u}lveda\footnote{\href{dgarciasepulveda@uchicago.edu}{dgarciasepulveda@uchicago.edu}} 
~\\[0.1cm]

{\it Kadanoff Center for Theoretical Physics \& Enrico Fermi Institute, University of Chicago}\\[4pt]

~\\[15pt]

\end{center}

\noindent We derive new dualities of topological quantum field theories in three spacetime dimensions that generalize the familiar level-rank dualities of Chern-Simons gauge theories. The key ingredient in these dualities is non-abelian anyon condensation, which is a gauging operation for topological lines with non-group-like i.e.\ non-invertible fusion rules.  We find that, generically, dualities involve such non-invertible anyon condensation and that this unifies a variety of exceptional phenomena in topological field theories and their associated boundary rational conformal field theories, including conformal embeddings, and Maverick cosets (those where standard algorithms for constructing a coset model fail.)  We illustrate our discussion in a variety of isolated examples as well as new infinite series of dualities involving non-abelian anyon condensation including: i) a new description of the parafermion theory as $(SU(N)_{2} \times Spin(N)_{-4})/\mathcal{A}_{N},$  ii) a new presentation of a series of points on the orbifold branch of $c=1$ conformal field theories as $(Spin(2N)_{2} \times Spin(N)_{-2} \times Spin(N)_{-2})/\mathcal{B}_{N}$, and iii) a new dual form of $SU(2)_{N}$ as $(USp(2N)_{1} \times SO(N)_{-4})/\mathcal{C}_{N}$ arising from conformal embeddings, where $\mathcal{A}_{N}, \mathcal{B}_{N},$ and $\mathcal{C}_{N}$ are appropriate collections of gauged non-invertible bosons.

\vfill

\begin{flushleft}
December 2023
\end{flushleft}

\end{titlepage}

\setcounter{tocdepth}{4}

\tableofcontents

\newpage

\section{Introduction}
\label{sec:intro}

In this paper we derive new dualities of topological quantum field theories (TQFTs) in three spacetime dimensions.  Our results generalize the celebrated level-rank dualities of Chern-Simons gauge theories, the most familiar of which are of the form:
\begin{equation}
\begin{tabular}{lll}
   $ SU(N)_{K}\leftrightarrow U(K)_{-N,-N}$, & $USp(N)_{K}\leftrightarrow USp(K)_{-N}$, & $SO(N)_{K}\leftrightarrow SO(K)_{-N}$.
    \end{tabular}
\end{equation}
For unitary groups these dualities have been explored in \cite{Nakanishi:1990hj,Naculich:1990pa, Mlawer:1990uv, Witten:1993xi, Douglas:1994ex, Naculich:2007nc, Hsin:2016blu}, the dualities for $SO$ and $USp$ were derived in \cite{Aharony:2016jvv}, while those for more general orthogonal groups were discussed in \cite{Cordova:2017vab}.  Additionally, there are also dualities involving exceptional groups derived in \cite{Cordova:2018qvg}. Beyond their intrinsic conceptual interest, these dualities are important e.g. in that they provide non-trivial evidence for proposals of phase diagrams of 3D gauge theories \cite{Hsin:2016blu, Gomis:2017ixy}, and establish the existence of families of time-reversal invariant TQFTs \cite{Cordova:2017kue}.

Typically, the starting point to prove these dualities is to establish an equivalence of the associated chiral algebras that appear on the edge of the TQFT equipped with suitable boundary conditions \cite{Witten:1988hf,Elitzur:1989nr}.  For connected, simple, and simply-connected gauge groups these are the familiar Kac-Moody current algebras (for an overview see e.g.\ \cite{DiFrancesco:1997nk}), while for other global forms of the gauge group, obtained by quotienting by central elements, they are instead extensions of the Kac-Moody algebras \cite{Moore:1989yh}. Often, the initial step in deriving these equivalences of chiral algebras is to find a larger chiral algebra that contains as a subalgebra the two chiral algebras of interest and then study how representations of the larger chiral algebra decompose under restriction to the subalgebras. Frequently, the larger algebra is taken to be a Kac-Moody algebra at level one, with the same central charge as that of the two subalgebras combined, in which case such embeddings fall under the so-called \emph{conformal embeddings}. In this technique, duality of chiral algebras is intimately related with conformal field theories (CFTs) that are derived from appropriate quotients of chiral algebras; namely, coset CFTs.  

A well-known example illustrates the general procedure.  Starting from the embedding:
\begin{equation}\label{unitaryembedintro}
SU(N)_{K}\times SU(K)_{N}\subset SU(NK)_{1},
\end{equation}
we learn that the chiral algebra $SU(N)_{K}$ can be presented as a coset:
\begin{equation}
    SU(N)_{K}\cong \frac{SU(NK)_{1}}{SU(K)_{N}}.
\end{equation}
Passing to the bulk TQFT one then obtains a duality of Chern-Simons theories:
\begin{equation} \label{UnitaryCSExample}
    SU(N)_{K}\cong \frac{SU(NK)_{1}\times SU(K)_{-N}}{\mathbb{Z}_{K}}.
\end{equation}
A particularly subtle point in the above is the appearance of the quotient by $\mathbb{Z}_{K},$ the common center of the gauge group. As we review in Section \ref{cosetinterfacebulkboundaryreview} this quotient necessarily appears so that the boundary CFT has a unique ground state without additional topological degrees of freedom.  We can also directly interpret this quotient as a gauging operation on the theory $SU(NK)_{1}\times SU(K)_{-N}$.  Specifically this theory has abelian anyons, i.e. lines with abelian fusion rules, which are bosonic and hence may condense.  In the language of higher symmetry \cite{Kapustin:2014gua,Gaiotto:2014kfa}, these are one-form global symmetries and condensing them is equivalent to gauging this one-form symmetry.  As an operation on the initial TQFT this condensation operation acts as a simple algorithm \cite{Moore:1989yh,Hsin:2018vcg}:
\begin{itemize}
    \item We remove all lines that braid non-trivially with the condensing abelian anyons.  Such removed lines are often said to be \emph{confined}.
    \item We identify any remaining lines that differ by fusion with the condensing abelian anyons.  This is the step of forming gauge orbits.
    \item If a remaining line $a$ is invariant under fusion with $s$ condensing abelian anyons, then in the resulting theory the line $a$ is split into $s$ distinct lines. 
\end{itemize}

This three-step gauging procedure allows us to treat many foundational examples of duality amongst TQFTs.  However, as originally noted in \cite{Dunbar:1992gh,Dunbar:1993hr}, there are certain cosets where this procedure of gauging by the common center to isolate a CFT with a unique vacuum fails.  Traditionally such cosets were often referred to as \emph{Maverick cosets} and in these cases, the construction of a standard CFT proceeds in an ad hoc manner.  Two infinite series of such cosets are known:
\begin{equation}\label{maverickintro}
    \frac{SU(k)_{2}}{Spin(k)_{4}}, \hspace{.1in}c=\frac{2(k-1)}{k+2}, \hspace{.2in}\text{and}\hspace{.2in} \frac{Spin(2N)_{2}}{Spin(N)_{2}\times Spin(N)_{2}},\hspace{.1in}c=1,
\end{equation}
as well as a finite list of exceptional cases summarized in Section \ref{maverickdualities} below.  One of the main results of this paper is to provide a uniform analysis of these cosets, and their implications for duality in 3D TQFTs.  

As we will exhibit, a key idea unifying these cosets is the appearance of \emph{non-abelian} bosonic anyons in the associated TQFTs, as first explored in \cite{Frohlich:2003hm,Bais:2008ni}. To obtain boundary CFTs with unique vacua and no additional topological degrees of freedom we must condense such non-abelian bosons. This procedure unifies the treatment of Maverick and more familiar cosets, and in fact is even crucial for a complete understanding of the conformal embeddings behind more familiar level-rank dualities as we discuss in Section \ref{conformalembeddings}. 

Non-abelian anyon condensation in 3D TQFTs can also be fruitfully described using the language of higher symmetry. Indeed, as already mentioned, abelian anyons are generators of one-form global symmetries i.e. they are line topological operators with abelian fusion rules.  Anyons with more general non-abelian fusion rules are therefore interpreted as non-invertible one-form symmetries.  Such generalized symmetries have recently been investigated particularly in spacetime dimension greater than two.  (See e.g. \cite{Cordova:2022ruw, Schafer-Nameki:2023jdn, Shao:2023gho} for recent reviews and lectures).  

Relatedly, on the boundary the bulk one-form symmetries restrict to ordinary, zero-form symmetries of the CFT.  For abelian anyons, these boundary zero-form symmetries are group-like, but for non-abelian anyons, the boundary zero-form symmetries are a general fusion category. In this context, early foundational work on non-invertible symmetries and gauging was done in \cite{Fuchs:2002cm,Fuchs:2003id, Fuchs:2004dz,Fuchs:2004xi,Frohlich:2004ef,Frohlich:2006ch}, and a rigorous mathematical treatment of condensing or gauging general symmetries was carried out in \cite{muger2001subfactors, MUGER2000151, kirillov2002q, davydov2013witt, Hung:2015hfa, kong2014anyon}. We briefly review this formalism in Appendix \ref{MathematicsSection}.  

From a more physical point of view, a treatment of non-invertible topological lines in 2D theories as symmetries was pioneered in \cite{Chang:2018iay, Thorngren:2019iar,Thorngren:2021yso} and further discussed in \cite{Huang:2021zvu, Lin:2022dhv}.  The relationship between the 3D bulk TQFT and 2D boundary CFT is a foundational example of the general paradigm of a bulk topological field theory controlling the generalized symmetry of the boundary theory \cite{Gaiotto:2014kfa, Gaiotto:2020iye, Freed:2022qnc, Kaidi:2022cpf, Freed:2022iao, Bhardwaj:2023ayw}. The general idea of gauging and condensing non-invertible symmetries has been explored in \cite{Bhardwaj:2017xup, Gaiotto:2019xmp, Kaidi:2021gbs, Roumpedakis:2022aik, Choi:2023xjw, Perez-Lona:2023djo} and the relationship between bulk and boundary non-invertible gauging has been utilized in \cite{Kaidi:2023maf, Zhang:2023wlu, Cordova:2023bja,Antinucci:2023ezl, Bhardwaj:2023idu, Bhardwaj:2023fca}.  Finally, recent discussions of gauging non-invertible symmetries in 2D CFTs, closely related to our analysis below, include in particular \cite{Choi:2023vgk, Diatlyk:2023fwf}.  Below we often make use of the language of generalized symmetries, interchangeably using non-abelian and non-invertible, as well as gauge and condense.

\subsection{An Invitational Example with Fibonacci Anyons}\label{sec:introG2}

As an illustrative example to show how non-abelian anyon condensation is in fact central to many dualities of TQFTs consider the exceptional  conformal embedding \cite{ PhysRevD.34.3092, PhysRevD.37.2231, boysal2010strange, kac2015conformal,Cordova:2018qvg}:
\begin{equation}\label{confembed}
    SU(2)_{1} \times SU(2)_{3} \hookrightarrow (G_{2})_{1}.
\end{equation}
In CFT, the existence of this embedding can be interpreted in two ways: 
\begin{itemize}
    \item The branching functions of the coset $\frac{(G_{2})_{1}}{SU(2)_{1}}$ give the characters of $SU(2)_{3}$.
    \item The branching functions of the coset $\frac{(G_{2})_{1}}{SU(2)_{3}}$ give the characters of $SU(2)_{1}.$
\end{itemize}
Translating the first statement to 3D TQFTs following \cite{Moore:1989yh} gives the duality
\begin{equation} \label{firstduality}
    SU(2)_{3} \cong (G_{2})_{1} \times SU(2)_{-1}.
\end{equation}
The simplicity of the TQFTs involved in this duality make it easy to verify explicitly.  For instance, $SU(2)_{k}$ has $k+1$ anyons obeying (truncated) fusion rules of $SU(2)$ representations. Meanwhile $(G_{2})_{1}$ is the theory of Fibonacci anyons, the simplest non-abelian TQFT consisting of two anyons $1$ and $\phi$ obeying:
\begin{equation}\label{fibfuse}
    \phi \times \phi=1 +\phi.
\end{equation}
Thus, for example both sides of the duality \eqref{firstduality} have four total anyons, and one may readily verify that their fusion rules and spins are identical.  

The previous --seemingly simple-- chain of ideas raises however an immediate puzzle. Suppose instead that we make use of the second implication of the embedding \eqref{confembed}, then proceeding blindly along the same steps would have lead to the proposed duality:
\begin{equation} \label{secondputativeduality}
    SU(2)_{1} 
   \overset{?}{\cong}    
  (G_{2})_{1} \times SU(2)_{-3}.
\end{equation}
We note that, as in \eqref{firstduality}, there is no common center of the gauge groups on the right-hand side. However, \eqref{secondputativeduality} is obviously false, for instance the number of lines (2 vs.\ 8) does not match. The resolution of this puzzle is that while $(G_{2})_{1} \times SU(2)_{-3}$ does not have any condensable abelian anyons, there are non-abelian bosonic anyons which may condense, and doing so leads to a correct, and novel, duality.  

To demonstrate this we begin with the correct duality \eqref{firstduality}. Reversing orientation (flipping the signs of all levels), and tensoring by $(G_{2})_{1}$, we obtain
\begin{equation}
    (G_{2})_{1} \times SU(2)_{-3} \cong (G_{2})_{1} \times (G_{2})_{-1} \times SU(2)_{1}. \label{tensoringg21andsu2minus3}
\end{equation}
We now use that $(G_{2})_{1} \times (G_{2})_{-1}$ is a Drinfeld double, i.e., it is of the form $G_{k} \times G_{-k}$.  In particular for any such theory it is known that one can gauge/condense all the anyons and obtain a trivial theory \cite{davydov2013witt, muger2001subfactors,Davydov:2011pp}. The novelty here is that such condensation is necessarily non-abelian. Concretely then, we write
\begin{equation}
    \mathcal{Z}(\mathbf{Fib})\equiv (G_{2})_{1} \times (G_{2})_{-1},
\end{equation} 
where the notation above indicates that $\mathcal{Z}(\mathbf{Fib})$ is the Drinfeld center of the Fibonacci anyons \eqref{fibfuse}. Condensing $\mathcal{Z}(\mathbf{Fib})$ in \eqref{tensoringg21andsu2minus3} then leads to a new duality:
\begin{equation} \label{su21intermsofnoninvertiblecondensation}
    SU(2)_{1} \cong \frac{(G_{2})_{1} \times SU(2)_{-3}}{\mathcal{Z}(\mathbf{Fib})}.
\end{equation}

In summary, $(G_{2})_{1} \times SU(2)_{-3}$ has non-abelian bosonic anyons, or in a different language, it has a non-anomalous non-invertible one-form symmetry, and gauging it, one finds an equivalence with the $SU(2)_{1}$ Chern-Simons gauge theory.
These non-abelian condensable bosons thus play the role of the common center of the gauge group in more familiar examples. Compare for instance with the more familiar duality \eqref{UnitaryCSExample}. In particular, this non-invertible one-form symmetry must be gauged to obtain the duality expected from the existence of the conformal embedding \eqref{confembed}. 

We study this example of non-abelian anyon condensation in more detail in Section \ref{IntroductionExample} below, and review the condensation of $\mathcal{Z}(\mathbf{Fib})$ to the trivial theory in Appendix \ref{G21DrinfeldToVacuum}.

\subsection{Summary of Selected Results}\label{sec:summaryofresults}

Having illustrated the ubiquitous nature of non-abelian anyon condensation let us summarize several key results derived using this formalism below.  

Note that the central charges of the first infinite family of Maverick cosets in \eqref{maverickintro}  match those of the parafermion CFTs \cite{Fateev:1985mm,DiFrancesco:1997nk}, so it is natural to suggest that this infinite Maverick family reproduces the parafermions. The parafermions also have two standard coset descriptions given by the $SU(2)_{k}/U(1)_{2k}$, or $(SU(k)_{1} \times SU(k)_{1})/SU(k)_{2}$ cosets \cite{DiFrancesco:1997nk}. We therefore conjecture the infinite series of dualities:
\begin{equation} 
    \frac{SU(k)_{2} \times Spin(k)_{-4}}{\mathcal{A}_{k}} \cong \frac{SU(2)_{k} \times U(1)_{-2k}}{\mathbb{Z}_{2}} \cong \frac{SU(k)_{1} \times SU(k)_{1} \times SU(k)_{-2}}{\mathbb{Z}_{k}},
\end{equation}
for some suitable collection of condensable non-abelian anyons $\mathcal{A}_{k}$ on the left-hand side (see eqs. \eqref{algebra1} and \eqref{algebra2} for a specific proposal for this collection of condensable anyons). Below in Section \ref{TSPMfromLevelRank} we verify this result explicitly for the first non-trivial case $k=3$.  In this case the parafermion theory in question coincides with the three-state Potts model and the algebra of non-abelian anyons is generated by:
\begin{equation}
    \mathcal{A}_{3} = (\mathbf{1},0) + (\mathbf{1},8) + (\mathbf{8},4),
\end{equation}
where we label $SU(3)$ representations by their dimension and $Spin(3)\cong SU(2)$ representations by their Dynkin index (i.e.\ their dimension is the label plus one.)  In particular, the anyon $(\mathbf{8},4)$ is non-abelian with fusion rule:
\begin{equation}
    (\mathbf{8},4)\times (\mathbf{8},4)=  \sum_{i=0}^{4}(\mathbf{1},2i) + (\mathbf{8},2i)\rightarrow (\mathbf{1},0)+  (\mathbf{1},8) + (\mathbf{8},4),
\end{equation}
where the first equation denotes the fusion in the full TQFT, and the arrow indicates its projection back to $\mathcal{A}_{3}.$

We also analyze the second infinite sequence of Maverick cosets in \eqref{maverickintro}.  Since all of these have $c=1$ they must correspond to rational points in the moduli space of $c=1$ CFTs.  We conjecture that these cosets correspond to the orbifold points of $U(1)_{2N}$ modulo its $\mathbb{Z}_{2}$ reflection symmetry, which we denote as $U(1)^{\mathrm{Orb}}_{2N}$.  Lifting to TQFTs leads to the proposal
\begin{equation}\label{orbintroresult}
    U(1)^{\mathrm{Orb}}_{2N} \cong \frac{Spin(2N)_{2} \times Spin(N)_{-2} \times Spin(N)_{-2} } {\mathcal{B}_{N}}.
\end{equation}
For a suitable collection of condensable non-abelian anyons $\mathcal{B}_{N}$ (see eqs. \eqref{algebra3} and \eqref{algebra4} for a specific proposal for this collection of condensable anyons). In particular we verify this for the first non-trivial case $N=3$ where:
\begin{equation}
    \mathcal{B}_{3}=(\mathbf{1},0,0) + (\mathbf{1},4,4) + (\mathbf{20}',0,4) + (\mathbf{20}',4,0) + (\mathbf{15},2,2).
\end{equation}
The fusion of the first four anyons above is abelian, but the last one is non-abelian with:
\begin{equation}
    (\mathbf{15},2,2)\times (\mathbf{15},2,2)=  \sum_{i,j=0}^{2}\left[(\mathbf{1},2i,2j) + (\mathbf{15},2i,2j)+(\mathbf{20}',2i,2j)\right]\rightarrow \mathcal{B}_{3},
\end{equation}
where again the first equality is the fusion in the full TQFT and the arrow indicates the restriction to $\mathcal{B}_{3}.$  We note that the Chern-Simons theory with the orbifold action is equivalent to changing the gauge group from $U(1)$ to $O(2)$. Level-rank dualities involving $O(N)$ were studied in \cite{Cordova:2017vab} and we have the equivalence $U(1)^{\mathrm{Orb}}_{2N} \cong O(2)^{0}_{2N,0}$ where the additional subscript and superscript on $O(2)$ indicate other possible levels.  Because of this equivalence \eqref{orbintroresult} can also be cast as a duality of orthogonal type Chern-Simons theories:
\begin{equation}
    O(2)^{0}_{2N,0}\cong \frac{Spin(2N)_{2} \times Spin(N)_{-2} \times Spin(N)_{-2} } {\mathcal{B}_{N}}. 
\end{equation}

Beyond analyzing these families of Maverick cosets we can also study many of the isolated examples in Section \ref{maverickdualities} and derive a variety of dualities.  All of these examples have $c<1$ and thus correspond to some (possibly non-diagonal) minimal model. For instance, the simplest of these leads to the Ising TQFT:
\begin{equation}
    \mathrm{Ising \ TQFT} \cong \frac{SU(4)_{1} \times SU(2)_{-10}}{\mathcal{A}},
\end{equation}
where 
\begin{equation}
    \mathcal{A} = (\mathbf{1}, 0) + (\mathbf{6},10) + (\mathbf{1},6) + (\mathbf{6},4),
\end{equation}
and both $(\mathbf{1},6)$ and $(\mathbf{6},4)$ have non-abelian fusion.

Armed with our improved understanding of non-abelian anyon condensation, we also revisit the conformal embeddings generalizing the example of Section \ref{sec:introG2} to obtain other level-rank dualities involving non-abelian anyon condensation.  For instance revisiting the embedding \eqref{unitaryembedintro} of unitary groups allows us to derive:
\begin{equation}
    SU(Nk)_{1} \cong \frac{SU(N)_{k} \times SU(k)_{N}}{\mathcal{A}_{N,k}},
\end{equation}
where $\mathcal{A}_{N,k}$ is a suitable collection of non-abelian anyons.\footnote{This collection corresponds to the anyons appearing in the branching rule of the identity anyon of $SU(Nk)_{1}$ in terms of the anyons of $SU(N)_{k} \times SU(k)_{N}$. The branching rules for the conformal embedding $SU(N)_{k} \times SU(k)_{N} \hookrightarrow SU(Nk)_{1}$ have been studied e.g. in \cite{Nakanishi:1990hj, Altschuler:1989nm, Walton:1988bs}.}  For instance for $N=3$ and $k=2$ this is the non-abelian anyon:
\begin{equation}
    \mathcal{A}_{3,2}=(\mathbf{1},0)+ (\mathbf{8},2).
\end{equation}

Similarly, a less explored example arises from the conformal embedding
\begin{equation} \label{additionalembeddingintro}
    SO(N)_{4} \times SU(2)_{N} \hookrightarrow USp(2N)_{1}.
\end{equation}
This implies the level-rank duality:
\begin{equation}
    SU(2)_{N} \cong \frac{USp(2N)_{1} \times SO(N)_{-4}}{\mathcal{A}_{N}},
\end{equation}
where for instance in the case $N=3$:
\begin{equation}
    \mathcal{A} = (\mathbf{1}, 0) + (\mathbf{14},4_{1}), 
\end{equation}
with non-abelian fusion:
\begin{equation}
    (\mathbf{14},4_{1}) \times (\mathbf{14},4_{1}) = (\mathbf{1},0) + (\mathbf{1},4_{1}) + (\mathbf{14}, 0) + (\mathbf{14},4_{1})\rightarrow (\mathbf{1},0) + (\mathbf{14},4_{1}).
\end{equation}

\setcounter{footnote}{0}

Many of the previous examples can be understood from the ``principle of coset inversion,'' first derived mathematically in \cite{Frohlich:2003hm}. We explain this in Section \ref{GeneralNonAbelianAnyonCondesationPicture} from a physics point of view and summarize it in mathematical form in Appendix \ref{MathematicsSection}. The principle may be summarized as follows: one may rearrange the numerator and denominator of any coset, provided we also allow for the possibility of non-abelian anyon condensation. This principle holds abstractly, independently of the description of the CFT in terms of WZW models.  For instance we can even revisit the well-known TQFT associated to the unitary minimal models
\begin{equation}    
M(k+3,k+2) \cong \frac{SU(2)_{k} \times SU(2)_{1} \times SU(2)_{-k-1}}{\mathbb{Z}_{2}},
\end{equation}
where $M(k+3,k+2)$ denotes the $k$-th minimal model TQFT, with $k=1$ the Ising model. Allowing for non-abelian anyon condensation implies that in general there is also an equivalence:
\begin{equation} 
    SU(2)_{k} \times SU(2)_{1} \cong \frac{M(k+3,k+2) \times SU(2)_{k+1}}{\mathcal{A}_{k}},
\end{equation}
which we explicitly check for $k=2$ in Section \ref{cosetinversionformulachecksection} (for $k=1$ this expression may be checked by the three-step gauging rule). 

In general, if two different TQFTs differ by anyon condensation, as in the examples above, it implies that a topological interface can be established between them (see below). As discussed e.g. in \cite{Kaidi:2021gbs}, this result indicates that, from a mathematical perspective, both TQFTs are Witt equivalent \cite{davydov2013witt}. Consequently, we find that the physical concept of level-rank duality corresponds to the mathematical concept of Witt equivalences of modular tensor categories.

Finally, more examples of duality involving non-abelian anyon condensation may be found from cosets of holomorphic CFTs at $c=24$ \cite{Schellekens:1992db} and the classification of RCFTs with a small number of primaries and a small central charge studied in \cite{Mukhi:2022bte, Rayhaun:2023pgc}.\renewcommand{\thefootnote}{\arabic{footnote}}\footnote{We thank B. Rayhaun for bringing this to our attention.} We hope to explore these examples in more detail in the future.

\section{Cosets, Interfaces, and Bulk-Boundary Correspondence} \label{cosetinterfacebulkboundaryreview}

In this section we review the relationship between coset CFTs and associated boundary conditions in topological Chern-Simons theories.  We pay particular attention to the interplay between gapless and gapped degrees of freedom at the boundary, whose understanding is important for determining dualities of the bulk topological theories.

Let us first recall the coset construction purely in the context of 2D CFT. Our starting point is a $G_{k}$ WZW theory based on a group $G$ and an integer level $k$. We take the group $G$ to be compact and simply-connected. Let $H$ be now a subgroup of $G$ such that the Lie algebra of $H$ embeds into the Lie algebra of $G$ with embedding index $\ell$. Then, there is an associated embedding of affine Lie algebras:
\begin{equation}
    H_{\tilde{k}} \subseteq G_{k}~, \hspace{.2in}\tilde{k}=\ell k~.
\end{equation}
 Coset CFTs are constructed by expanding a chiral algebra with characters $\chi_{\Lambda}(q)$ in terms of the characters of a smaller algebra with characters $\chi_{\lambda}(q)$:
\begin{equation} \label{branchingrules}
    \chi_{\Lambda}(q) = \sum_{\lambda} b_{\Lambda}^{\lambda}(q) \chi_{\lambda}(q), \quad q=e^{2 \pi i \tau},
\end{equation}
with $\tau$ the modular parameter. In our context, we assume the bigger chiral algebra is that of the $G_{k}$ WZW theory, and the smaller chiral algebra is that of the $H_{\tilde{k}}$ WZW theory, with $H_{\tilde{k}}$ embedded in $G_{k}$. The quantities $b_{\Lambda}^{\lambda}(q)$ are called \emph{branching functions}. The point of the previous expansion is to notice that, since the characters $\chi_{\Lambda}, \chi_{\lambda}$ are modular covariant, the branching functions inherit some form of modular covariance and can be thus thought of as the characters of a new CFT with torus partition function
\begin{equation}
    Z_{\mathrm{Coset \ CFT}} \, (T^{2}) = \sum_{\Lambda,\lambda} |b_{\Lambda}^{\lambda}(q)|^{2}. \label{cosetpartitionfunction}
\end{equation}
This is the so-called GKO coset construction \cite{Goddard:1984vk, Goddard:1986ee}, where for simplicity we have restricted ourselves to diagonal theories.

A subtlety in the construction above is the generic appearance of multiple copies of the vacuum and of many copies of the same chiral or Virasoro primary in the partition function. Relatedly, not all branching functions are non-zero, so the naive modular covariance of the branching functions in general requires further analysis.\footnote{\label{footnoteidentificationcurrent} Historically, these confusions led to the search for methods to remove such a degeneracy, such as the so-called ``identification current method'' or ``fixed point resolutions'' that made the final result into a CFT with a single vacuum, a non-degenerate modular S-matrix, etc. The result of such a procedure is what in some older literature is known as ``the'' coset CFT (See for instance \cite{DiFrancesco:1997nk, Fuchs:1995tq, Gepner:1989jq,Schellekens:1989uf}).} In modern terms, we recognize the vacuum degeneracy as the presence of a topological sector coupled to gapless, CFT degrees of freedom. In some circumstances it is desirable to remove this degeneracy by, roughly speaking, ``gauging away'' this topological sector resulting in a CFT with a unique vacuum. In general, we should therefore differentiate between two possible notions of cosets:
\begin{itemize}
    \item A coset CFT with degenerate vacua, where topological sectors are retained.
    \item A coset CFT with a unique vacuum state, where topological sectors have been removed.
\end{itemize}
The partition function \eqref{cosetpartitionfunction} corresponds to the first notion above. The distinction between these possibilities is particularly important in the special case where the topological sector is all there is, as occurs for example in the case of conformal embeddings discussed below.  

The phenomenon just described has been previously noticed and interpreted in terms of projection into universes and/or vacua (See \cite{Delmastro:2021otj, Komargodski:2020mxz}), and here we provide a further interpretation in the context of TQFTs with gapped and gapless boundaries, and lines ending or not perpendicularly at a topological junction.

Turning now to the relationship between 3D TQFTs and 2D CFTs, a natural question to ask is what are the TQFTs associated to these two possible notions of cosets. As shown in \cite{Moore:1989yh} the TQFT that reproduces the coset CFT with a single vacuum at the boundary is given --often, but crucially not always--
by the product Chern-Simons theory:
\begin{equation}\label{cscosteis}
    \frac{G_{k} \times H_{-\tilde{k}}}{Z}~,
\end{equation}
where $Z$ is the common center of groups $G$ and $H$. As in \cite{Moore:1989yh}, $Z$ is, for the time being, some abelian discrete group. Shortly, this assumption will be lifted.  With an eye towards future generalizations let us deduce why \eqref{cscosteis} is correct.  In particular, we would like to understand the difference between the boundary conditions in the Chern-Simons theories which differ by whether we gauge or not the common center. As we illustrate, this difference can be usefully phrased in terms of topological interfaces.

\begin{figure}[t] \hspace{-0.5cm}
\begin{subfigure}{.45\textwidth}
  \includegraphics[width=1.2\linewidth]{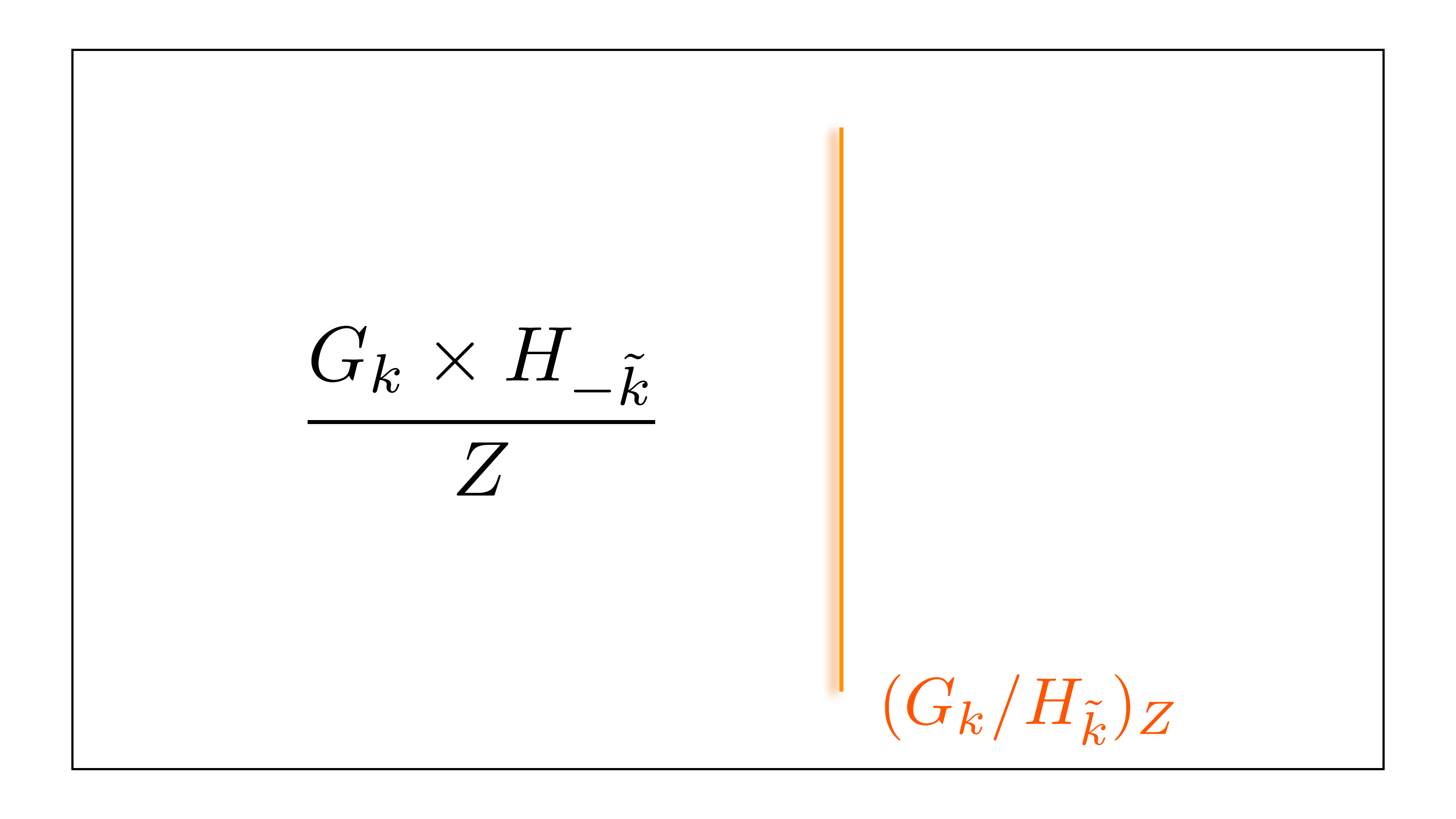}
\end{subfigure}%
\hspace{0.95cm}
\begin{subfigure}{.45\textwidth}
  \includegraphics[width=1.2\linewidth]{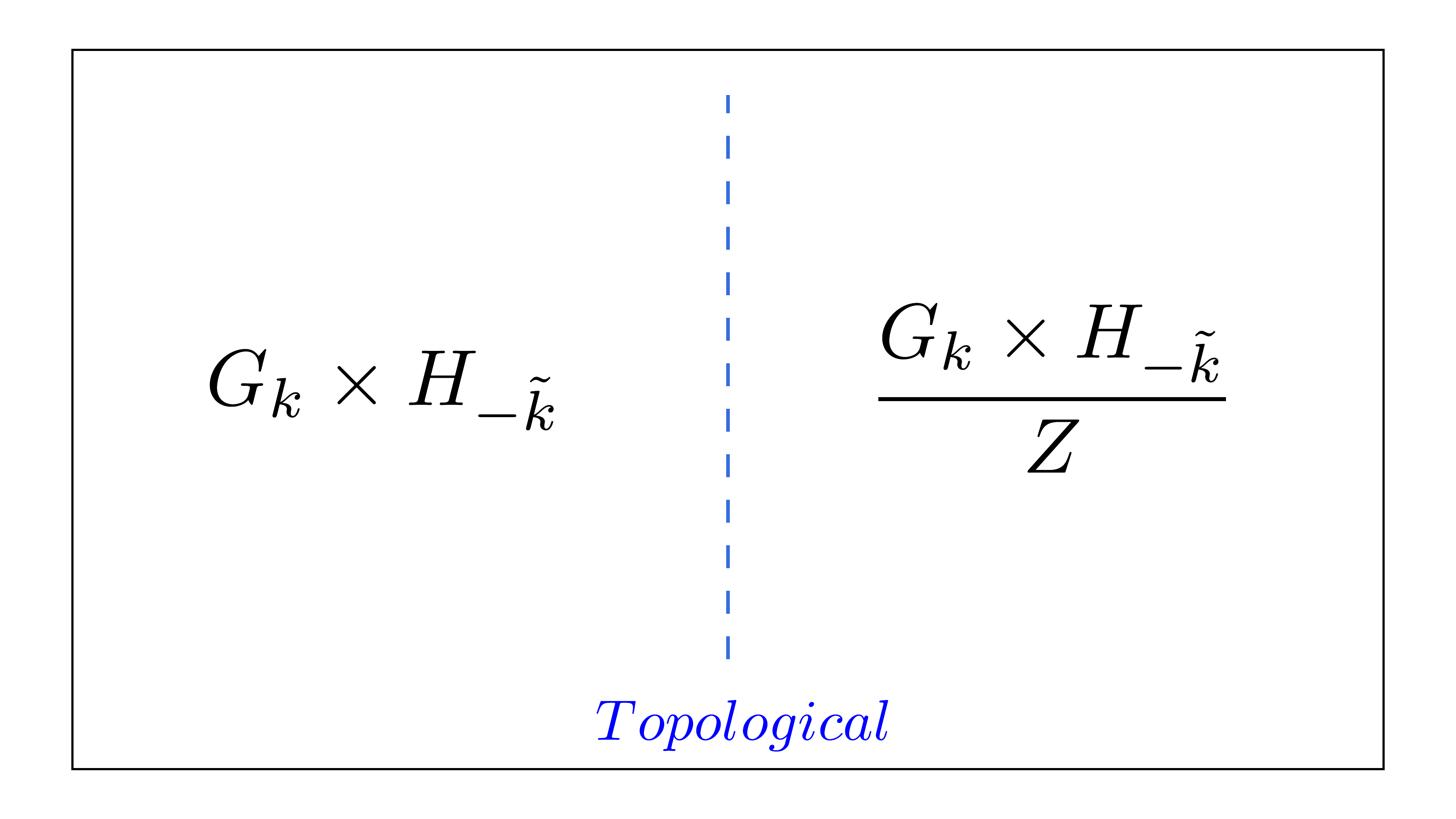}
\end{subfigure}
\caption{On the left: the theory $(G_{k} \times H_{-\tilde{k}}) / Z$, where the common center $Z$ of $G_{k}$ and $H_{-\tilde{k}}$ has been gauged, in the presence of the CFT coset boundary condition which is denoted $(G_{k}/H_{\tilde{k}})_{Z}$. On the right: a topological interface connecting the product $G_{k} \times H_{-\tilde{k}}$ without the common center gauged and $(G_{k} \times H_{-\tilde{k}})/Z$.}
\label{fig:test} 
\end{figure}

Consider first the case were we gauge the common center in the bulk as in \eqref{cscosteis}. This situation is depicted in the left in Figure \ref{fig:test},  where we denote the boundary condition as $(G_{k}/H_{\tilde{k}})_{Z}$ with a subindex to emphasize that the corresponding bulk has the center one-form symmetry gauged.\footnote{When discussing non-chiral 2D theories we should really picture the 3D theory with a left and right boundary component. In the following we illustrate one boundary component and we assume the same conditions on the left and right.  This means that we consider diagonal theories.} As mentioned above, this is the case where we have a theory with single vacuum on the boundary. Standard examples involving an abelian $\mathbb{Z}_{2}$ gauging that one could keep in mind are the minimal models $SU(2)_{k} \times SU(2)_{1} \times SU(2)_{-k-1} /\mathbb{Z}_{2}$, or the parafermions $SU(2)_{k} \times U(1)_{-2k}/\mathbb{Z}_{2}$. 

\begin{figure}[t]
        \centering
        \includegraphics[scale=0.14]{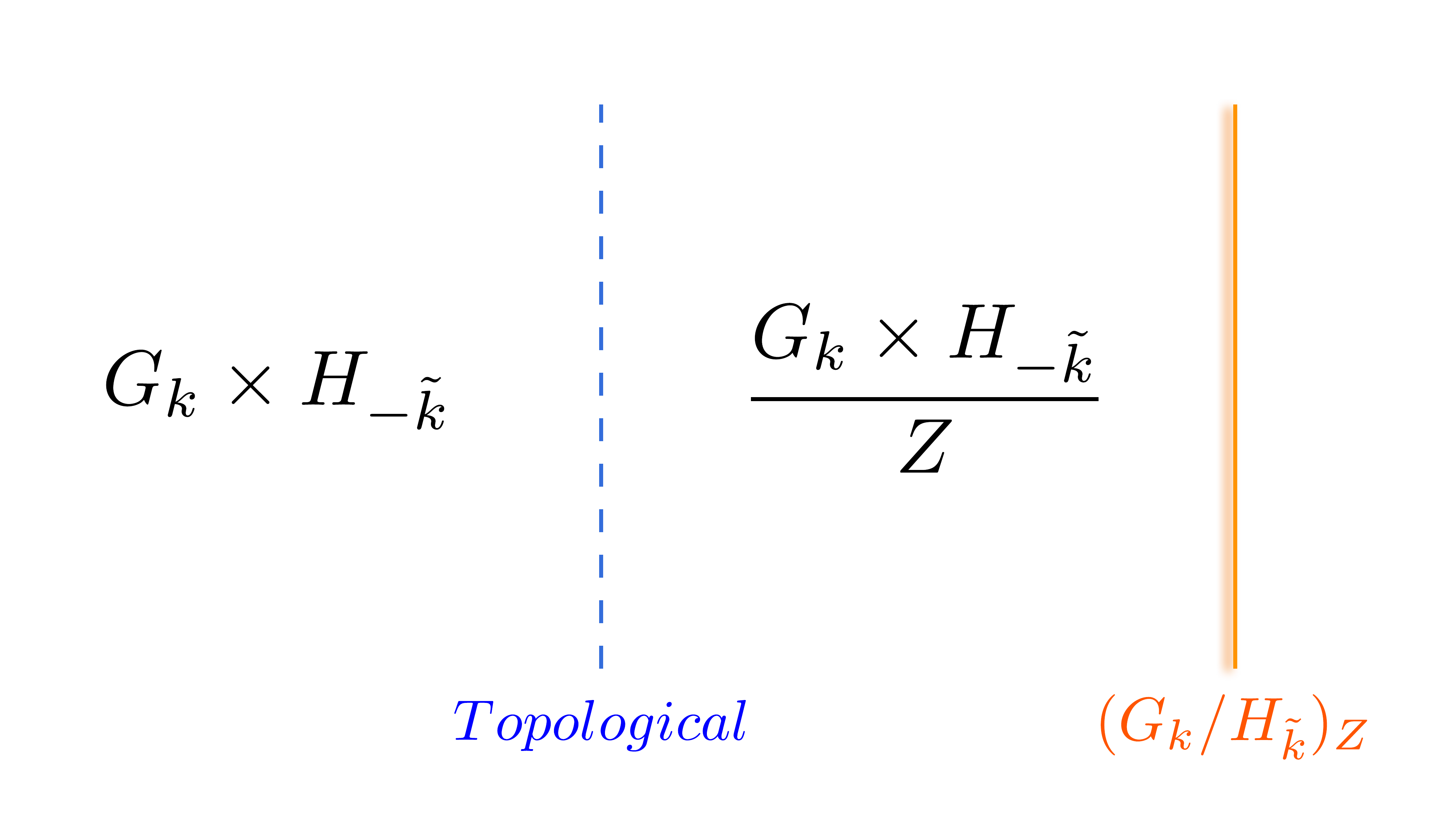} 
        \caption{Coset boundary condition with single vacuum $(G_{k}/H_{\tilde{k}})_{Z}$ in the presence of the topological interface joining $G_{k} \times H_{-\tilde{k}}$ and $(G_{k} \times H_{-\tilde{k}}) / Z$.} \label{Interfaceandsinglevacuumcondition}
\end{figure}

Separately, we note that in the Chern-Simons theory $G_{k} \times H_{-\tilde{k}},$ the common center indicates the presence of abelian anyons which define a gaugable one-form symmetry \cite{Moore:1989yh,Hsin:2018vcg}.   We can thus consider the topological interface generated by gauging the common center one-form symmetry on the right half of space, as depicted on the right in Figure \ref{fig:test}. Placing this topological interface together with the coset boundary $(G_{k}/H_{\tilde{k}})_{Z}$ of Figure \ref{fig:test}, we obtain the construction depicted in Figure \ref{Interfaceandsinglevacuumcondition}.

\begin{figure}[t]
        \centering
        \includegraphics[scale=0.14]{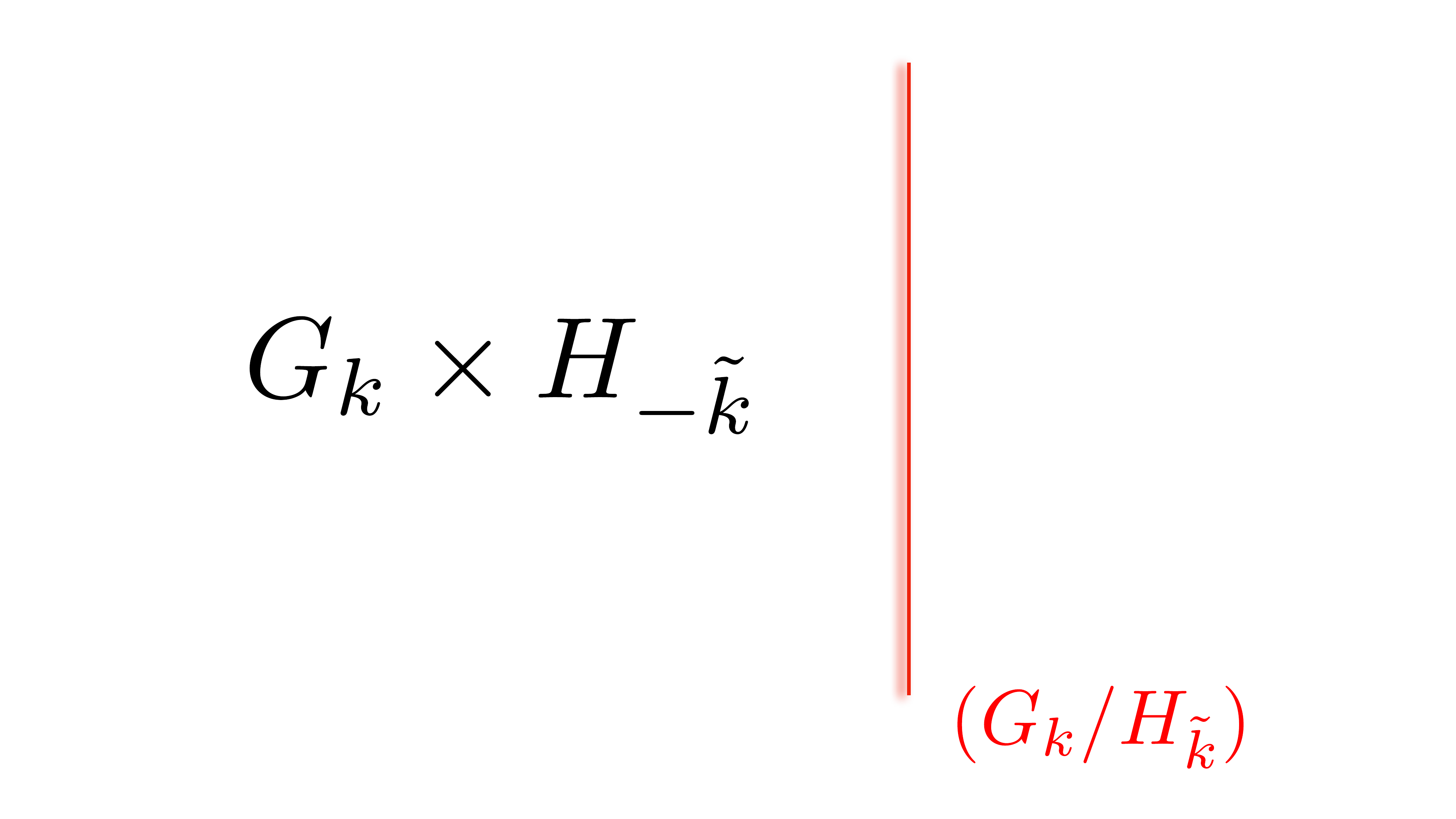} 
        \caption{Pushing the topological interface towards the boundary and fusing it with the coset boundary condition $(G_{k}/H_{\tilde{k}})_{Z}$ generates a gapless boundary for the $G_{k} \times H_{-\tilde{k}}$ Chern-Simons theory (with no common center symmetry gauged). By the nature of the construction, the result is generically a gapless boundary intertwined with a topological sector (i.e., there are now many vacua/topological local operators at the boundary) that is denoted as $(G_{k}/H_{\tilde{k}})$ (with no $Z$ subindex). The gapless boundary condition so generated corresponds to (the chiral version of) the associated gauged WZW model.} \label{interfaceboundarymerge}
\end{figure}

Since the interface is topological, we can move it towards the boundary to generate a new boundary condition for the theory without the common center one-form symmetry gauged. The latter boundary condition therefore differs from $(G_{k}/H_{\tilde{k}})_{Z}$ by some topological action, and so we call the new boundary condition $(G_{k}/H_{\tilde{k}})$ without a subindex to emphasize that the corresponding bulk has no one-form symmetry gauged. The result of all previous manipulations is depicted in Figure \ref{interfaceboundarymerge}.

In summary, we see that the key difference between the coset boundary condition when the bulk is given by $G_{k} \times H_{-\tilde{k}}$ and $(G_{k} \times H_{-\tilde{k}})/Z$ is determined by some topological degrees of freedom, here represented by the topological interface separating the product with and without the common center gauged. Notice the similarity with our comment above regarding the distinction between two notions of coset CFTs: one with and one without topological degrees of freedom removed.

Now that we understand how the boundary conditions $(G_{k} / H_{\tilde{k}})$ and $(G_{k} / H_{\tilde{k}})_{Z}$ are distinguished from each other we study how to setup the boundary condition $(G_{k}/H_{\tilde{k}})$ in $G_{k} \times H_{-\tilde{k}}$ recalling the well-known steps derived in \cite{Elitzur:1989nr,Moore:1989yh}.  The key point is that requiring the variation of the action to vanish implies the bulk equations of motion, but also the vanishing of a boundary term
\begin{equation} \label{deltaScontribution}
    \frac{k}{4\pi}\int_{\partial X} \mathrm{Tr}'_{G}(\delta A A) - \frac{\tilde{k}}{4\pi}\int_{\partial X} \mathrm{Tr}'_{H}(\delta B B),
\end{equation}
where $A$ and $B$ are gauge fields based on the Lie algebras of $G$ and $H$ respectively, $\mathrm{Tr}'_{G}$ and $\mathrm{Tr}'_{H}$ are the respective representation-independent traces, and  $\partial X$ is the boundary of our bulk 3D spacetime $X$. Imposing $A_{0} = 0$ and $B_{0} = 0 $ at the boundary gives the canonical chiral WZW boundary \cite{Elitzur:1989nr}. However, since we have taken $H$ to embed in $G$ there is another boundary condition where we ask for the gauge field $A$ projected onto the Lie algebra of $H$ to equal the gauge field $B$, which also makes \eqref{deltaScontribution} vanish. The action then reduces, following the steps of \cite{Elitzur:1989nr,Moore:1989yh} to an expression involving WZW actions:
\begin{equation}
    i k S_{WZW}(U) - i \tilde{k} S_{WZW}(V) + ik \int \mathrm{Tr}'_{G} \, \lambda (\partial_{\phi}U \, U^{-1} - \partial_{\phi} V \, V^{-1}),
\end{equation}
where $U$ and $V$ are the Maurer-Cartan fields that arise when we integrate the time components of $A$ and $B$ respectively in the bulk, and $\lambda$ is a Lagrange multiplier. As first described in \cite{Moore:1989yh}, changing variables and using the Polyakov-Wiegmann formula  gives the path integral of (the chiral version of) the gauged WZW model.\footnote{It is sometimes useful to express the trace in the algebra of $H$ in the path integral in terms of that of $G$ by noting that the embedding relates the normalization of the traces $\mathrm{Tr}'_{H} = \mathrm{Tr}'_{G}/\ell$ with $\ell$ the embedding index.} In summary, we see that what we have called above the $(G_{k}/H_{\tilde{k}})$ boundary condition corresponds to (the chiral version of) the gauged WZW action.

The path integral, clearly, takes into account contributions resulting from topological sectors. This can be illustrated for instance by taking $H_{\tilde{k}} = G_{k}$ so that we obtain the well-known $G_{k}/G_{k}$ topological coset field theory \cite{SPIEGELGLAS199221,Spiegelglas:1992jg,Witten:1991mm,Blau:1993tv} on the boundary, whose partition function over a Riemann surface $\Sigma$ has been evaluated (see \cite{Witten:1991mm}) to be $Z_{G_{k}/G_{k}} = \mathrm{dim}(\mathcal{V})$, with $\mathrm{dim}(\mathcal{V})$ the number of conformal blocks of the corresponding $G_{k}$ WZW model. On the torus, this evaluates to the number of chiral primaries of the underlying $G_{k}$ WZW model. We can also view this $G_{k}/G_{k}$ boundary condition as an instance of the universal gapped boundary for a bulk theory of the form $\mathcal{C} \times \overline{\mathcal{C}}$, with $\mathcal{C} = G_{k}$.\footnote{An easy way to see that the previous gapped boundary always exists is to notice that upon unfolding the interface is simply the identity interface between $\mathcal{C}$ and $\mathcal{C}$.}

Gauging the center one-form symmetry $Z$ in the bulk implies that the gauge group on the boundary is no longer $H$, but a non-simply-connected version based on the same Lie algebra. The corresponding summation over bundles/insertion of anyons generating $Z$ descends into a corresponding summation on the gauged WZW model on the boundary. The corresponding model based on a non-simply-connected gauge group has a single vacuum and the so-called ``identification current method'' has been applied to remove multiple copies of the same chiral primary, as studied in \cite{hori1994global} (see also \cite{Sharpe:2021srf} for recent related discussions). This corresponds to the boundary condition that we have called  $(G_{k}/H_{\tilde{k}})_{Z}$ above.

\begin{figure}[t]
        \centering
        \includegraphics[scale=0.15]{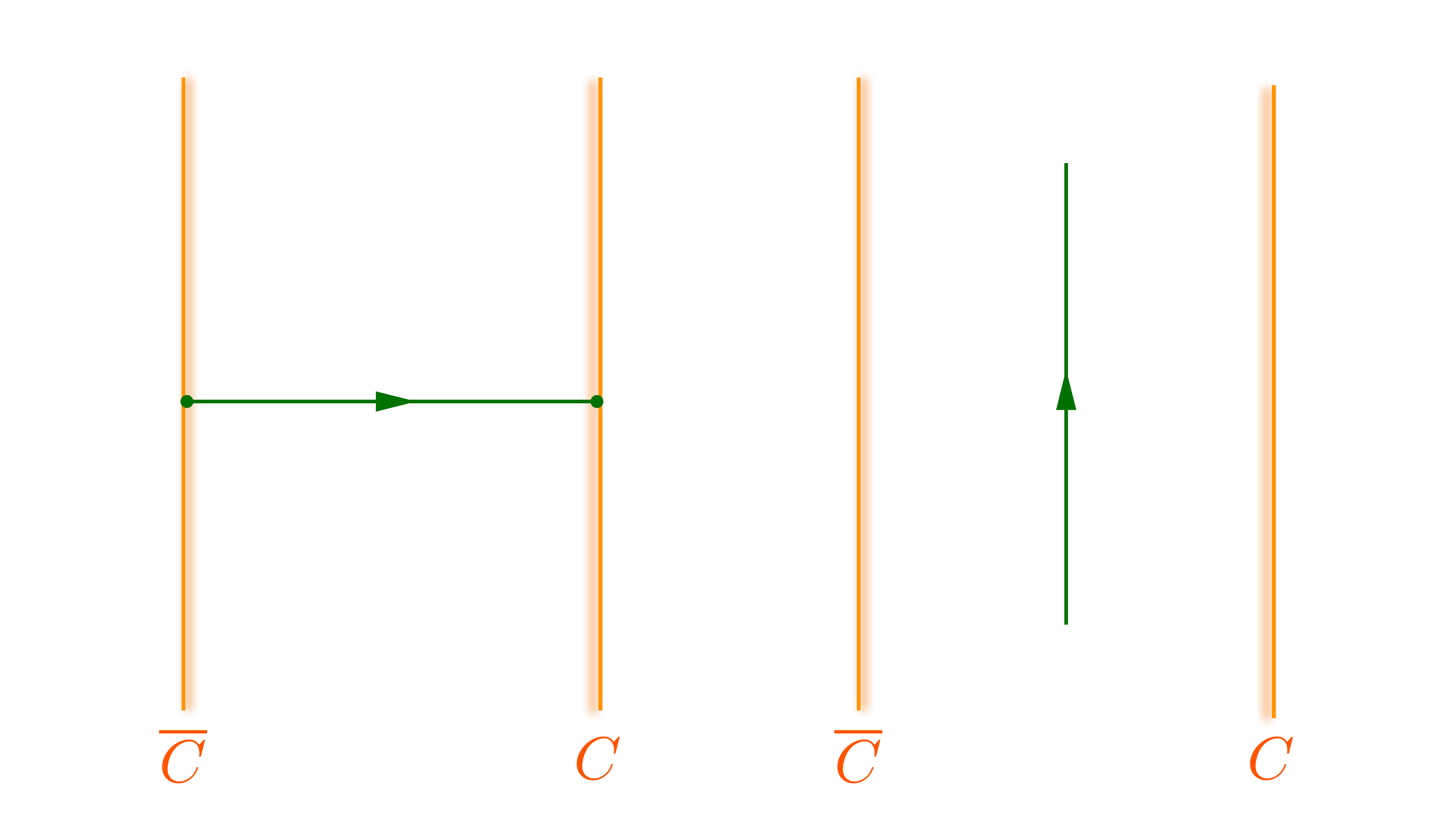} 
        \caption{In the canonical CFT boundary condition (in orange), all lines of the bulk CS theory end perpendicularly at the boundary on a (non-topological) local operator of the WZW theory based on the same group and level. Relatedly, pushing a line to the boundary in parallel to it gives rise to a Verlinde line of the corresponding WZW theory.} \label{CanonicalCFTBC}
\end{figure}

\begin{figure}[t]
        \centering
        \includegraphics[scale=0.15]{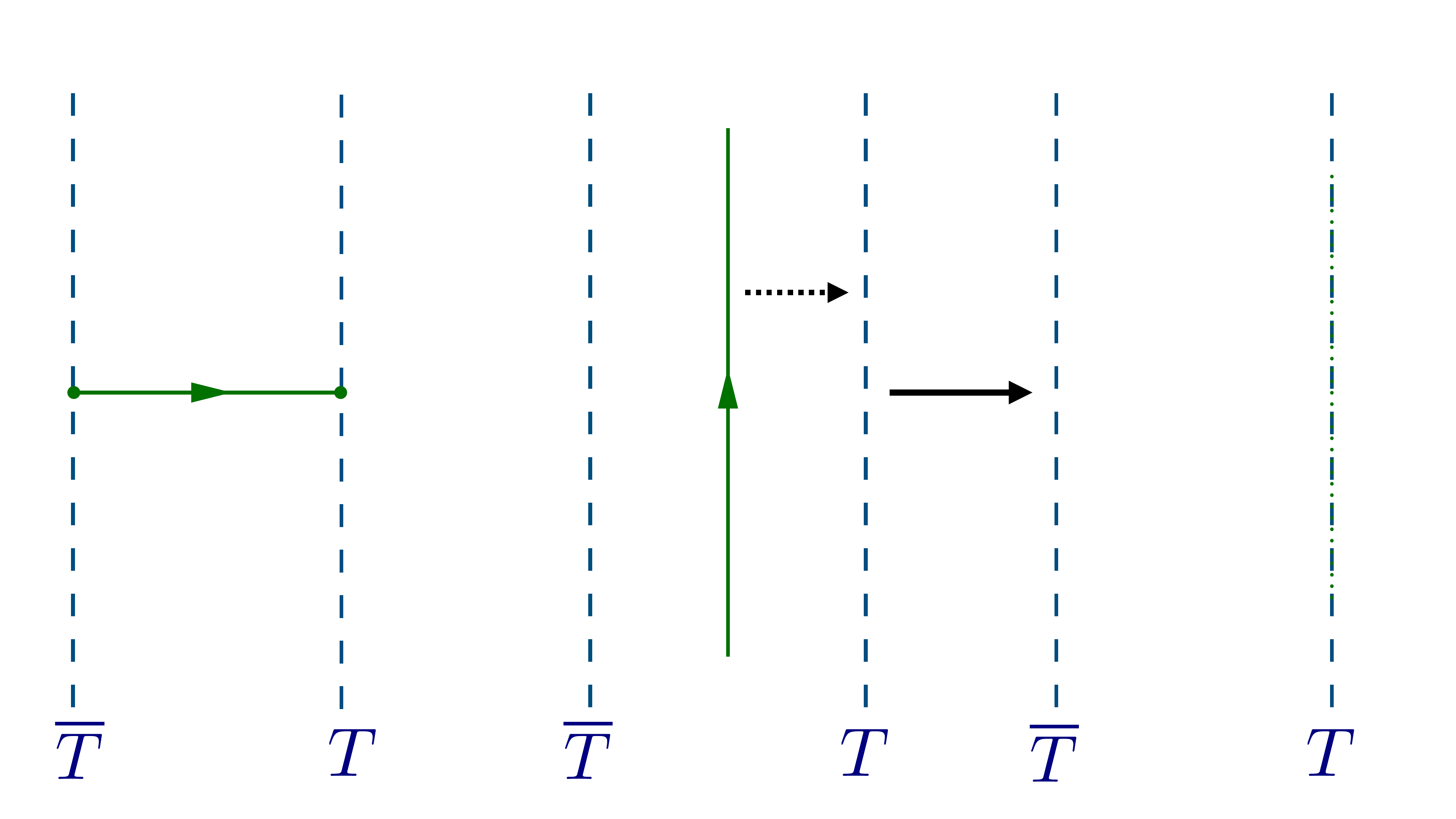} 
        \caption{In a topological boundary condition not all lines of the bulk theory end perpendicularly to the boundary. Only those generating a Lagrangian algebra can end (see Appendix \ref{FrobeniusAlgebras} for a precise mathematical definition), resulting in a topological local operator at the junction. When $Z$ is abelian as it is assumed in this section, the Lagrangian algebra is moreover a Lagrangian subgroup. Under parallel fusion, those lines in the Lagrangian subgroup become invisible at the boundary.} \label{TopBC}
\end{figure}

Let us now interpret the prior story in terms of lines ending at the boundary. To do this consider first two extreme scenarios: that of the canonical (chiral) CFT boundary with no topological sectors other than the identity, illustrated in Figure \ref{CanonicalCFTBC}, and that of a purely topological boundary, illustrated in Figure \ref{TopBC}.

In the CFT boundary case all lines can end perpendicularly at the boundary on a non-topological junction, which generates the local operators of the RCFT \cite{Witten:1988hf}. Under parallel fusion with the boundary, the bulk lines becomes the Verlinde lines \cite{Verlinde:1988sn,Petkova:2000ip,Drukker:2010jp,Gaiotto:2014lma} of the RCFT \cite{Komargodski:2020mxz}. In particular, this point of view is one way in which we can explain why the local operators and Verlinde lines of an RCFT follow the same fusion ring; namely, that of the bulk MTC \cite{Moore:1988qv}, and why furthermore Verlinde lines in an RCFT generate a modular tensor category (MTC) instead of merely a fusion category. Importantly, because the bulk lines form a MTC, and all lines can end perpendicularly at the boundary with this choice of boundary condition, it trivially follows that the set of lines that can end perpendicularly at the boundary have non-trivial mutual braiding. That is, the braiding matrix projected to those lines that can end perpendicularly at the boundary is non-degenerate (i.e., the projected braiding matrix has maximal rank).

We can compare the discussion above with the case of a purely topological boundary. Since we are assuming $Z$ to be abelian, only a square root of the total number of simple lines can end perpendicularly at the boundary \cite{Levin:2013gaa,Barkeshli_2013,Kapustin:2010hk,Fuchs:2012dt}, generating a topological junction in the process, and under parallel fusion with the boundary such lines becomes invisible\footnote{Technically, one summarizes this statement saying that such lines participate in a Lagrangian subgroup.} (See Figure \ref{TopBC}). In particular, this set of lines has a size strictly smaller than that of the simple objects of the full bulk MTC, and in which all lines braid trivially with each other. That is, the braiding matrix obtained from \eqref{braiding} projected to those lines that can end perpendicularly at the boundary is maximally degenerate (i.e., the projected braiding matrix has rank 1).

We now come back to the situation of the gauged WZW boundary condition $(G_{k}/H_{\tilde{k}})$. The general situation is complicated because of the variety of ways that the topological sector can intertwine with the CFT sector. However, we can get an idea of the general situation by considering the example $SU(2)_{1}\times SU(2)_{1}\times SU(2)_{-2}$ which is (up to the $\mathbb{Z}_{2}$ common center) the standard coset description of the Ising model. The branching rules of the characters are:
\begin{align}
    & \chi^{SU(2)_{1}}_{0} \chi^{SU(2)_{1}}_{0} = \chi^{I}_{0} \, \chi^{SU(2)_{2}}_{0} + \chi^{I}_{v} \, \chi^{SU(2)_{2}}_{2}, \\[0.3cm]
    & \chi^{SU(2)_{1}}_{0} \chi^{SU(2)_{1}}_{1} = \chi^{I}_{\sigma} \, \chi^{SU(2)_{2}}_{1}, \\[0.3cm]
    & \chi^{SU(2)_{1}}_{1} \chi^{SU(2)_{1}}_{0} = \chi^{I}_{\sigma} \, \chi^{SU(2)_{2}}_{1}  \\[0.3cm]
    & \chi^{SU(2)_{1}}_{1} \chi^{SU(2)_{1}}_{1} = \chi^{I}_{0} \, \chi^{SU(2)_{2}}_{2} + \chi^{I}_{v} \, \chi^{SU(2)_{2}}_{0}.
\end{align}
Comparing with \eqref{cosetpartitionfunction}, we obtain a total of six operators in the CFT, so not all lines of the bulk end at a non-trivial local operator of the boundary theory. Furthermore, every operator of the CFT sector appears twice, with a non-trivial topological operator appearing in the branching space associated to the $\mathbb{Z}_{2}$ generator in the bulk: the line $(1,1,2)$ in $SU(2)_{1} \times SU(2)_{1} \times SU(2)_{-2}$. The doubling of the spectrum can be then thought of as fusing this topological operator with the different CFT sectors.

We can also obtain the same conclusion by observing\footnote{This is a consequence of the fact that $\mathrm{Ising} \cong (SU(2)_{1} \times SU(2)_{1} \times SU(2)_{-2})/\mathbb{Z}_{2}$. Ungauging the $\mathbb{Z}_{2}$ by gauging the quantum zero-form symmetry allows us to write $SU(2)_{1} \times SU(2)_{1} \times SU(2)_{-2}$ in terms of $\mathrm{Ising}$ and a twisted $\mathbb{Z}_{2}$ gauge theory. Comparing the spectrum of the lines on both sides we can convince ourselves that such twisted $\mathbb{Z}_{2}$ gauge theory is $U(1)_{2} \times U(1)_{-2}$.}
\begin{equation}
    SU(2)_{1} \times SU(2)_{1} \times SU(2)_{-2} \cong \mathrm{Ising} \times U(1)_{2} \times U(1)_{-2}~.
\end{equation}
Then the gauged WZW boundary condition corresponds to setting the CFT boundary condition on the $\mathrm{Ising}$ factor, but the topological boundary condition on the $U(1)_{2} \times U(1)_{-2}$ factor, which explains both why not all lines end perpendicularly at the boundary, and why the spectrum of Ising is doubled. Thus we see that the $(G_{k}/H_{\tilde{k}})$ boundary on the one hand is similar to the CFT boundary condition in that the boundary is non-topological and there exists a subset of the bulk lines ending perpendicularly at the boundary such that their mutual braiding is non-trivial and the junctions at the end of the lines are non-topological. On the other hand, there also exists lines that generate topological junctions which have trivial mutual braiding. The gauged WZW boundary condition $(G_{k}/H_{\tilde{k}})$ is like the topological boundary condition in that the braiding of those lines ending perpendicularly is degenerate, but unlike the purely topological boundary, not maximally so. That is, the braiding matrix projected to those lines that can end perpendicularly at the boundary is degenerate, but with neither minimal nor maximal rank.

In hindsight, this explains why historically the naive proposal \eqref{cosetpartitionfunction} led to a degenerate CFT with multiple vacua -- since there are multiple topological junctions from the bulk-- and degenerate modular $S$-matrix --since the braiding of the bulk lines ending at the boundary is degenerate--. Upon gauging $Z$ in the bulk to obtain $(G_{k} \times H_{-\tilde{k}})/Z$ what we are doing is making those lines generating topological junctions transparent and identified with each other, so in the boundary the corresponding degeneracy is removed, and we find a standard 2D RCFT: that which the methods in the older literature provided via the ``field identification'' and ``fixed point resolution'' methods.

\subsection*{Application to Level-Rank Dualities}

A well-known and important application of coset CFTs is to level-rank dualities of 3d TQFTs \cite{Nakanishi:1990hj,Naculich:1990pa, Mlawer:1990uv, Witten:1993xi, Douglas:1994ex, Naculich:2007nc, Hsin:2016blu,Aharony:2016jvv,Cordova:2017vab,Cordova:2018qvg}.  Consider for instance the conformal embedding
\begin{equation}\label{firstconfembed}
    SU(N)_{k} \times SU(k)_{N} \hookrightarrow SU(Nk)_{1}.
\end{equation}
 At the level of the branching rules \eqref{branchingrules}, the embedding \eqref{firstconfembed} means that the characters of $SU(Nk)_{1}$ can be expanded in terms of products of those of $SU(N)_{k}$ and $SU(k)_{N}$. We can then construct the coset CFT with single vacuum $(SU(Nk)_{1}/SU(N)_{k})_{Z}$, and by the previous conformal embedding the result must be equivalent to the $SU(k)_{N}$ WZW theory. We have then the equivalence of chiral algebras
\begin{equation}
    SU(k)_{N} \longleftrightarrow \bigg( \frac{SU(Nk)_{1}}{SU(N)_{k}} \bigg)_{Z} \, ,
\end{equation}
Expressing now both sides in terms of their bulk TQFTs we obtain the duality
\begin{equation} \label{standardunitaryembedding}
    SU(k)_{N} \cong \frac{SU(NK)_{1} \times SU(N)_{-k}}{\mathbb{Z}_{N}}.
\end{equation}

More generally, the logic of the above example is simply that if we find two different descriptions of the same chiral algebra either in terms of a WZW theory on one side, and a coset description on the other, or by relating two different coset descriptions on either sides, we can use the relationship between bulk and boundary to find dual descriptions of the same bulk TQFT. 

In particular, we must take care to ensure that the vacuum degeneracy on both sides matches.  Often we implicitly assume that all such degeneracy is removed, as in \eqref{standardunitaryembedding} by a suitable gauging.  In familiar examples this is achieved by gauging a one-form symmetry in the bulk (the common center group), but in the examples to follow the required gauging will be more subtle.

\section{Dualities via Non-Invertible Anyon Condensation}  \label{noninvertibleextensionsubsection}

In the previous section, we have outlined how to construct (bosonic) dualities of 3D TQFTs by making clever use of the boundary theory and a proper understanding of the coset construction of 2D RCFTs. However, most of our discussion relied on the standard assumption \cite{Moore:1989yh} that gauging the common center $Z$ of $G_{k} \times H_{-\tilde{k}}$  is sufficient to find a CFT with a single vacuum. However, already in \cite{Moore:1989yh} it was observed that there are exceptions to the idea that gauging by an abelian common center $Z$ in $G_{k} \times H_{-\tilde{k}}$ is sufficient to remove all boundary topological degrees of freedom and obtain a standard boundary CFT with a unique vacuum, as in the case of the conformal embeddings.

In the context of the GKO coset construction, it is also known that there are exceptions to the methods mentioned in footnote \ref{footnoteidentificationcurrent} to construct CFTs with single vacuum from the ``naive'' partition function \eqref{cosetpartitionfunction}. In these exceptions there are additional selection rules (i.e.\, vanishing branching functions) and field identifications that are not a consequence of group-theoretical selection rules and thus cannot be submitted to the ``field identification method'' \cite{Gepner:1989jq} mentioned above to find a CFT with single vacuum and non-degenerate $S$-matrix. Importantly, again the conformal embeddings appear as part of such exceptions (see \cite{DiFrancesco:1997nk, Pedrini:1999iy}). We stress that if the conformal embedding has two factors in the denominator we mean the full quotient appears as an exception, and not a standard coset based on just one of the factors. Furthermore, it is known that a handful of gapless cosets also share this feature. In the literature, these cosets have been referred to as ``Maverick cosets,'' and the known examples have been constructed long ago in \cite{Pedrini:1999iy, Dunbar:1992gh, Dunbar:1993hr, Fuchs:1995tq}. In some sense, Maverick cosets are then a hybrid in between the standard cosets and the conformal embeddings. To the extent of the authors' knowledge, there is no known classification of Maverick cosets. 

The previous observations suggest that in order to understand these exceptions we must relax the assumption that the bulk is given by $G_{k} \times H_{-\tilde{k}}$ gauged by an abelian one-form symmetry $Z$. Indeed, in the past few years it has been understood that the notion of gauging can be extended to the case where the symmetries are not necessarily group-like \cite{Bais:2008ni,  Eliens:2013epa, Hung:2015hfa, neupert2016boson, Bhardwaj:2017xup, Gaiotto:2019xmp, Yu:2021zmu, Huang:2021zvu, Kaidi:2021gbs,Roumpedakis:2022aik, Choi:2023xjw, Choi:2023vgk, Diatlyk:2023fwf, Perez-Lona:2023djo} , and it is natural to explore the consequences of such generalized gauging in the present context. In the following discussion we will not need a rigorous presentation of non-abelian anyon condensation, and content ourselves with a physical presentation. In Appendix \ref{MathematicsSection} we present a summary of the rigorous statements that we use, and outline them in a physics nomenclature in the next subsection. More extensive treatments of non-abelian anyon condensation from a physics perspective may be found in \cite{Bais:2008ni,kong2014anyon,Eliens:2013epa,neupert2016boson}. See \cite{Burnell_2018} for a review.

To begin, it is instructive to study the example of the exceptional conformal embedding discussed briefly in Section \ref{sec:introG2}, which shows the inevitability of non-abelian anyon condensation in the general construction of coset CFTs, and in particular the conformal embeddings. The example is given by the embedding
 \begin{equation}
    SU(2)_{1} \times SU(2)_{3} \hookrightarrow (G_{2})_{1}.
\end{equation}
In 2D CFT terms, this conformal embedding is translated to the following branching rules:
\begin{align} \label{examplebranchingrules1}
    & \chi^{(G_{2})_{1}}_{\mathbf{1}} = \chi^{SU(2)_{1}}_{0} \, \chi^{SU(2)_{3}}_{0} + \chi^{SU(2)_{1}}_{1} \, \chi^{SU(2)_{3}}_{3}, \\[0.3cm]
    & \chi^{(G_{2})_{1}}_{\mathbf{7}} = \chi^{SU(2)_{1}}_{0} \, \chi^{SU(2)_{3}}_{2} + \chi^{SU(2)_{1}}_{1} \, \chi^{SU(2)_{3}}_{1},  \label{examplebranchingrules2}
\end{align}
where we have labeled the integrable representations of $(G_{2})_{1}$ by the dimensionality of the representation $\mathbf{R}$, and those of $SU(2)_{k}$ as standard by an integer $i=0,\ldots,k$. The notation is analogous whenever we consider negative levels in TQFT.

These branching rules can be regarded in many forms. In the simplest scenario, we can consider the coset $(G_{2})_{1}/SU(2)_{1}$, and the branching rules above show that the branching functions are the characters of $SU(2)_{3}$. Specifically, using the torus partition function for coset CFTs \eqref{cosetpartitionfunction}, we find
\begin{equation} \label{firstcoset}
    Z_{T^{2}}\bigg[ \frac{(G_{2})_{1}}{SU(2)_{1}} \bigg] = \big| \chi^{SU(2)_{3}}_{0} \big|^{2} + \big| \chi^{SU(2)_{3}}_{1} \big|^{2} + \big| \chi^{SU(2)_{3}}_{2} \big|^{2} + \big| \chi^{SU(2)_{3}}_{3} \big|^{2} = Z_{T^{2}}[SU(2)_{3}].
\end{equation}
The result is already an ordinary 2D CFT, with no vacuum degeneracy and a non-degenerate modular $S$-matrix. Clearly, this means that there are no topological sectors to consider, and the coset theory is simply
\begin{equation}
    (G_{2})_{1}/SU(2)_{1} = SU(2)_{3}. \label{firstcosettheory}
\end{equation}

More interestingly, we can consider the coset $(G_{2})_{1}/SU(2)_{3}$, and then the branching rules \eqref{examplebranchingrules1}-\eqref{examplebranchingrules2} tell us to regard the characters of the $SU(2)_{1}$ theory as the branching functions in the corresponding coset decomposition. More precisely, using \eqref{cosetpartitionfunction} again:
\begin{equation} \label{secondcoset}
    Z_{T^{2}}\bigg[ \frac{(G_{2})_{1}}{SU(2)_{3}} \bigg] = 2 \, \big|\chi^{SU(2)_{1}}_{0} \big|^{2} + 2 \, \big| \chi^{SU(2)_{1}}_{1} \big|^{2} = 2 Z_{T^{2}}[SU(2)_{1}],
\end{equation}
from which it is clear that now we do obtain a topological sector. However, notice that the group $G_{2}$ has trivial center, so we cannot possibly interpret the previous degeneracy in terms of some $\mathbb{Z}_{2}$ common center as in more standard description of cosets. Even more concretely, if we translate the previous observation to the associated 3D TQFTs, it is straightforward to check that $(G_{2})_{1} \times SU(2)_{-3}$ simply does not have an abelian anyon in its spectrum.\footnote{See Tables \ref{g2lv1table} and \ref{su2lv3table} later in the paper to check this statement.} 

The resolution of this puzzle is that while $(G_{2})_{1} \times SU(2)_{-3}$  has no abelian boson in its spectrum, it does have a non-abelian one consisting of the product of the line $\mathbf{7} \in (G_{2})_{1}$ and the line $2 \in SU(2)_{-3}$. Moreover, this is the exact combination giving the additional topological local operator in the branching rules \eqref{examplebranchingrules2}. Therefore, if this boson were to end on the boundary, it would explain the degeneracy above, and noticing that both $\mathbf{7} \in (G_{2})_{1}$ and $2 \in SU(2)_{3}$ follow Fibonacci fusion rules, provides us with a natural guess for the topological sector present on top of the CFT sector:
\begin{equation}
    (G_{2})_{1}/SU(2)_{3}  = \mathbf{Fib}^{\mathrm{TQFT}} \times SU(2)_{1}^{\mathrm{CFT}} = \frac{(G_{2})_{1}}{(G_{2})_{1}} \times SU(2)_{1}^{\mathrm{CFT}}. \label{secondcosettheory}
\end{equation}
Shortly, we will give a more rigorous derivation of this topological sector using level-rank duality and studying choices of boundary conditions. In passing, if we follow this proposal, it is clear that we can label the states in the coset $(G_{2})_{1}/SU(2)_{3}$ as $(0,i)$ and $(\phi,i)$, with fusion rules
\begin{align}
    &(0,i) \times (0,j) = (0,i+j), \\[0.3cm]
    &(0,i) \times (\phi,j) = (\phi,i+j), \\[0.3cm]
    &(\phi,i) \times (\phi,j) = (0,i+j) + (\phi,i+j),
\end{align}
where $i,j=0,1$ labels the integrable representations in the $SU(2)_{1}$ CFT factor.

Gauging by the non-abelian boson in the bulk would then remove the topological sector on the boundary, leaving us only with the non-degenerate CFT sector. This is clearly what we would have naively expected from the branching rules; that is, the $SU(2)_{1}$ WZW theory. Concretely:
\begin{equation}
    SU(2)_{1} \cong \frac{(G_{2})_{1} \times SU(2)_{-3}}{\mathcal{Z}(\mathbf{Fib})}.
\end{equation}
Below in Section \ref{IntroductionExample} we will verify this statement by a direct computation using non-abelian anyon condensation. Therefore, this is a situation where the coset RCFT lives at the boundary of a bulk TQFT whose only non-trivial boson is non-abelian. Correspondingly, to find the RCFT with single vacuum on the boundary (in this example $SU(2)_{1}$) we have to identify fields in the branching rules that are not related by some abelian action as in e.g., the case of the Ising model at the end of Section \ref{cosetinterfacebulkboundaryreview}. In other words, to find the RCFT with single vacuum in the boundary we would have to gauge/condense the non-abelian boson in the bulk.

Generically then, gauging non-abelian anyons in the bulk leads to some RCFT in the boundary, which may however also appear as a boundary condition in a different-looking bulk TQFT. Matching these different descriptions to one another is one way to guess dualities of TQFTs, which we may then verify directly as a statement solely in the context of 3D TQFTs.

Finally, one can also consider the full coset $\frac{(G_{2})_{1}}{SU(2)_{1} \times SU(2)_{3}}$, which is topological since the central charge vanishes. Indeed, computing the coset torus partition function \eqref{cosetpartitionfunction}, one obtains:
\begin{equation} \label{thirdcoset}
    Z_{T^{2}}\bigg[\frac{(G_{2})_{1}}{SU(2)_{1} \times SU(2)_{3}}\bigg] = 4.
\end{equation}
 Using similar steps as above the concrete proposal for the 2D TQFT is:
\begin{equation}
\frac{(G_{2})_{1}}{SU(2)_{1} \times SU(2)_{3}} = \mathbf{Fib}^{\mathrm{TQFT}} \otimes \mathbb{Z}_{2}^{\mathrm{TQFT}}= \frac{(G_{2})_{1}}{(G_{2})_{1}} \otimes \frac{SU(2)_{1}}{SU(2)_{1}} \label{thirdcosettheory}
\end{equation}

The previous discussion can also be derived by examining choices of boundary conditions implied by the TQFT duality:
\begin{equation} \label{Levelrank1}
    (G_{2})_{1} \times SU(2)_{-1} \cong SU(2)_{3},
\end{equation}
where it is straightforward to see that the spectrum of lines match on both sides. This shows that we can set the canonical $SU(2)_{3}$ RCFT boundary condition in the product $(G_{2})_{1} \times SU(2)_{-1}$, which gives the coset result \eqref{firstcoset}. On the other hand, we can take the orientation reversal of the previous duality and tensor the resulting expression by $(G_{2})_{1}$ which allows us to write
\begin{equation} \label{levelrank2}
    (G_{2})_{1} \times SU(2)_{-3} \cong (G_{2})_{1} \times (G_{2})_{-1} \times SU(2)_{1}.
\end{equation}
Then, by this duality, in the bulk theory $(G_{2})_{1} \times SU(2)_{-3}$ we are allowed to take as a boundary condition a canonical $SU(2)_{1}$ RCFT boundary condition for the $SU(2)_{1}$ factor, but the purely topological boundary condition for the factor $(G_{2})_{1} \times (G_{2})_{-1}$ given by the diagonal Lagrangian algebra (see Appendix \ref{FrobeniusAlgebras} for a precise mathematical definition). Recall this type of topological boundary is given by the $G_{k}/G_{k}$ 2D TQFT, which has as many topological local operators as the $G_{k}$ WZW theory has chiral primaries, with the same fusion rules as the $G_{k}$ WZW theory. The possibility of choosing this ``combined CFT-Topological'' boundary condition explains both the degeneracy of two in the torus partition function \eqref{secondcoset} as well as the precise Fibonacci TQFT factor in \eqref{secondcosettheory}, since $(G_{2})_{1}$ has Fibonacci fusion rules.

Finally, we can tensor \eqref{levelrank2} by a $SU(2)_{-1}$ factor to find 
\begin{equation} 
    (G_{2})_{1} \times SU(2)_{-1} \times SU(2)_{-3} \cong (G_{2})_{1} \times (G_{2})_{-1} \times SU(2)_{1} \times SU(2)_{-1},
\end{equation}
which allows us to set a purely topological boundary condition for $(G_{2})_{1} \times SU(2)_{-1} \times SU(2)_{-3}$ by choosing the purely topological boundary in both the $(G_{2})_{1} \times (G_{2})_{-1}$ and $SU(2)_{1} \times SU(2)_{-1}$ factors given by the corresponding diagonal Lagrangian algebras. Similarly, the duality explains the resulting 2D TQFT \eqref{thirdcosettheory} in the coset $(G_{2})_{1}/(SU(2)_{1} \times SU(2)_{3})$, and the corresponding torus partition function \eqref{thirdcoset}.

Now that we have convinced ourselves of the a priori generic appearance of non-abelian anyon condensation in the context of level-rank dualities, we will explore more examples where this phenomenon occurs. One such instance will be in the case of the conformal embeddings, as hinted above. In \cite{Bais:2008ni} it was suggested that the failure of the ``field identification method'' in the case of the Maverick cosets could be explained in terms of non-abelian anyon condensation, and we use this observation to propose dualities by first inspecting the resulting CFT, checking directly via non-abelian anyon condensation as a statement in the bulk TQFT, and then comparing different coset descriptions with the same TQFT data. Finally, a proper physical interpretation of certain mathematical results outlined in the next subsection will make manifest that non-abelian anyon condensation already appears even in standard examples of the coset construction, such as in the well-known coset description of the minimal models. We summarize the precise mathematical statements in Appendix \ref{MathematicsSection}.

Below in this section, we will first discuss in generality the main setup that describes the interplay between dualities and non-abelian anyon condensation, and then we consider the concrete case of Maverick cosets since the underlying CFT (with single vacuum) is non-trivial and it is most straightforward to repeat the arguments pertaining to the case of gauging by an abelian one-form symmetry based on the boundary CFT. The case of the conformal embeddings is slightly more complicated, so we discuss them later in Section \ref{conformalembeddings}. Since they have vanishing central charge, it seems like applying analogous arguments would lead to a rather trivial duality with the empty theory. This is indeed correct, but certain mathematical arguments will allow us to find non-trivial dualities nevertheless.

\subsection{The General Picture} \label{GeneralNonAbelianAnyonCondesationPicture}

To proceed further, we need to formalize the statements about the coset construction that we have used to convince ourselves of the general necessity of non-abelian condensation. The mathematical framework that we use that addresses such general scenario, and that is behind the results discussed in this subsection, corresponds to the formalism of local modules of special symmetric commutative Frobenius algebras described in \cite{Frohlich:2003hm}. To avoid a heavy mathematical digression, we summarize the definitions and results in this language in appendix \ref{MathematicsSection}, and in this subsection we approach such results from a physical point of view.

The main point is to unpack the results of \cite{Frohlich:2003hm} in the language of non-invertible one-form symmetry gauging. According to this result, starting from the decomposition of an affine Lie algebra, or more generally, a chiral algebra described by a MTC $\mathcal{M}$ in terms of a smaller one described by a MTC $\mathcal{M}'$ (as in \eqref{branchingrules}) we can write
\begin{equation} \label{MTCEmbedding}
    \mathcal{M} \cong (\mathcal{C} \times  \mathcal{M}')/\mathcal{A},
\end{equation}
where $\mathcal{C}$ is the MTC describing the coset theory and the quotient by $\mathcal{A}$ stands for some one-form gauging that is not necessarily abelian, and under rather general conditions over $\mathcal{A}$ (see Appendix \ref{MathematicsSection} for the more precise statement) we can ``solve'' for $\mathcal{C}$ and obtain the coset MTC in terms of $\mathcal{M}$ and $\mathcal{M}'$ as:
\begin{equation} \label{IsolatingMTCcoset}
    \mathcal{C} \cong (\mathcal{M} \times \overline{\mathcal{M}'})/\mathcal{B},
\end{equation}
for some new one-form gauging $\mathcal{B}$. The latter is what in the work of Moore and Seiberg \cite{Moore:1989yh} was identified as the common center of the affine Lie algebras. However, notice that from the current point of view we have no reason to believe that this is generally the case. Indeed, the conformal embeddings and Maverick cosets are explicit counterexamples to the common center rule, but they are still comfortably described by \eqref{IsolatingMTCcoset} when we allow for non-invertible anyon condensation.

The previous is the situation found more often, but when the general conditions over $\mathcal{A}$ are not strictly fulfilled a mild variation of the previous theorem still holds. Namely, there exist (generically non-invertible) one-form symmetries $\mathcal{T}_{1}$ and $\mathcal{T}_{2}$ such that:
\begin{equation} 
    \mathcal{C}/\mathcal{T}_{1} \cong (\mathcal{M} \times \overline{\mathcal{M}'})/\mathcal{T}_{2}.
\end{equation}
Essentially, what the conditions over $\mathcal{A}$ do is to ensure that the chiral algebra described by $\mathcal{C}$ in \eqref{MTCEmbedding} already appears ``extended,'' as otherwise the latter form of the theorem will perform the extension in any case.

We may consider our previous $SU(2)_{1} \times SU(2)_{3} \hookrightarrow (G_{2})_{1}$ example from this point of view. Taking $\mathcal{M} = (G_{2})_{1}$, we can take $\mathcal{M}'$ to be $SU(2)_{1}$ or $SU(2)_{3}$, with $\mathcal{C}$ the remaining factor. When $\mathcal{M}' = SU(2)_{1}$, $\mathcal{B}$ is trivial, and \eqref{IsolatingMTCcoset} reproduces \eqref{firstduality}. When $\mathcal{M}' = SU(2)_{3}$, however, the previous formalism allows $\mathcal{B}$ to be non-trivial and non-invertible. As a result, \eqref{IsolatingMTCcoset} readily reproduces \eqref{su21intermsofnoninvertiblecondensation}, which we previously reproduced by cleverly manipulating the factors. Equations \eqref{MTCEmbedding} and \eqref{IsolatingMTCcoset} represent the abstract form of such manipulations, and hold for general MTCs.

\subsubsection*{An Interesting Corollary}

Now, we notice the following rather interesting corollary of Eqs. \eqref{MTCEmbedding} and \eqref{IsolatingMTCcoset}. That is, from the CFT perspective, \eqref{MTCEmbedding} may be interpreted as the existence of some branching rules \eqref{branchingrules} for $\mathcal{M}$ in terms of the $\mathcal{M}'$ data. But on the TQFT perspective by itself, it means that not only we can obtain $\mathcal{C}$ in terms of $\mathcal{M}$ and $\mathcal{M}'$ via \eqref{IsolatingMTCcoset} --which is the standard form of the coset construction-- but that we can write $\mathcal{M}$ in terms of the coset $\mathcal{C}$ and $\mathcal{M}'$ after some one-form symmetry gauging. The gauging is generically by a non-invertible symmetry, even if the gauging by $\mathcal{B}$ in \eqref{IsolatingMTCcoset} is abelian and given by the common center. Because of this inversion property between \eqref{MTCEmbedding} and \eqref{IsolatingMTCcoset}, we refer to such expressions as the ``coset inversion formulas'' or ``coset inversion theorem'' in the following.

To illustrate this coset inversion theorem, we may consider the standard coset description for minimal models $(SU(2)_{k} \times SU(2)_{1} \times SU(2)_{-k-1})/\mathbb{Z}_{2}$. By the formulas above, we expect that $SU(2)_{k} \times SU(2)_{1}$ can be written in terms of the $k$-th minimal model as
\begin{equation} \label{MinimalModelCheck}
    SU(2)_{k} \times SU(2)_{1} \cong \frac{M(k+3,k+2) \times SU(2)_{k+1}}{\mathcal{A}_{k}},
\end{equation}
for some generically non-invertible gauging $\mathcal{A}_{k}$, where $M(k+3,k+2)$ stands for the $k$-th minimal model with $k=1$ the Ising model. As a check, it is readily shown that the combination of the $(1,3)$ primary\footnote{As usual, we have denoted primaries in minimal models as pairs $(r,s)$ in the Kac table \cite{DiFrancesco:1997nk}.} in the $M(k+3,k+2)$ minimal model and the line in the adjoint representation in $SU(2)_{k+1}$ are such that $h^{M(k+3,k+2)}_{1,3} + h^{SU(2)_{k+1}}_{2} = k+1/(k+3) + 2/(k+3) = 1$, so the product line for any $k$ is a boson that is generically non-abelian. For $k=1$ the gauging on the right-hand side by this combination is a standard abelian $\mathbb{Z}_{2}$ gauging. However, already at $k=2$ it can be seen that the gauging generically involves a non-invertible boson. It is amusing to check that this is indeed the case --so that \eqref{MinimalModelCheck} holds-- which we verify in Section \ref{cosetinversionformulachecksection} below by a direct computation on non-abelian anyon condensation.

\subsection{Maverick Cosets and Dualities} \label{maverickdualities}

In this section we provide some explicit proposal of dualities involving non-abelian anyon condensation based on the many observations that we have made previously in this work. The case of the conformal embeddings is treated separately in Section \ref{conformalembeddings}. 

In the following we will need the explicit expression for the Maverick cosets. The list of Maverick cosets known to date \cite{ Pedrini:1999iy, Dunbar:1992gh, Dunbar:1993hr, Fuchs:1995tq} and their central charges is:
\begin{align}
    & \hspace{1.0cm} \frac{SU(k)_{2}}{Spin(k)_{4}}, && \hspace{-1.5cm} 
 c = \frac{2(k-1)}{k+2}, \label{firstmaverickfamily} \\[0.4cm]
    & \frac{Spin(2N)_{2}}{Spin(N)_{2} \times Spin(N)_{2}},   && \hspace{-1.5cm} c = 1, \label{secondmaverickfamily} \\[0.4cm]
    & \hspace{1.0cm}  \frac{(E_{6})_{2}}{USp(16)_{2}}, && \hspace{-1.5cm} c = 6/7, \label{thirdmaverick} \\[0.4cm] 
    & \hspace{1.2cm}  \frac{(E_{7})_{2}}{SU(8)_{2}}, && \hspace{-1.5cm} c = 7/10,  \label{fourthMaverick} \\[0.4cm] 
    & \hspace{1.0cm}  \frac{(E_{8})_{2}}{Spin(16)_{2}}, && \hspace{-1.5cm} \label{fifthMaverick} c = 1/2, \\[0.4cm]
    & \hspace{0.6cm}  \frac{(E_{8})_{2}}{SU(2)_{2} \times (E_{7})_{2}}, && \hspace{-1.5cm} c = 7/10,  \label{sixthMaverick} \\[0.4cm]
    & \hspace{0.25cm}  \frac{(E_{7})_{2}}{SU(2)_{2} \times Spin(12)_{2}}, && \hspace{-1.5cm} c = 8/10,  \\[0.4cm]
    & \hspace{1.3cm}  \frac{SU(4)_{1}}{SU(2)_{10}}, && \hspace{-1.5cm} c = 1/2,
     \\[0.4cm]
    & \hspace{1.3cm}  \frac{Spin(7)_{1}}{SU(2)_{28}}, && \hspace{-1.5cm} c = 7/10.\label{lastmaverick}
\end{align}
It is straightforward to check that for the Maverick cosets $G_{k} / H_{\tilde{k}}$ above there is indeed a non-abelian boson in the spectrum of the associated $G_{k} \times H_{-\tilde{k}}$ TQFT.

\subsubsection*{First Family of Maverick Dualities}

As a warm-up, let us start considering the simplest example of a Maverick coset corresponding to the $k=3$ case in the first infinite family of Maverick cosets \eqref{firstcosettheory}:
\begin{equation}\label{oneof3}
    SU(3)_{2}/SU(2)_{8},
\end{equation}
where we have used the exceptional isomorphism of chiral algebras $Spin(3)_{k} = SU(2)_{2k}$ \cite{Aharony:2016jvv}. The central charge of such coset is $c=4/5$ meaning it could in principle be the Tetracritical Ising Model or the three-state Potts model (TSPM). Fortunately, already in \cite{Dunbar:1992gh,Dunbar:1993hr} from an analysis of the branching rules it was noticed that the result actually corresponds to the TSPM, which is known to allow for other coset descriptions. For instance, $SU(2)_{3}/U(1)_{6}$, or $(SU(3)_{1} \times SU(3)_{1})/SU(3)_{2}$ are coset description that reproduce the TSPM \cite{DiFrancesco:1997nk}.

Translating to TQFT then by the rules of \cite{Moore:1989yh}, we observe that both of the standard cosets $SU(2)_{3}/U(1)_{6}$, or $(SU(3)_{1} \times SU(3)_{1})/SU(3)_{2}$ translate to the TQFTs 
\begin{equation}\label{2and3of3}
    \frac{SU(2)_{3} \times U(1)_{-6}}{\mathbb{Z}_{2}}, \quad \mathrm{or} \quad \frac{SU(3)_{1} \times SU(3)_{1} \times SU(3)_{-2}}{\mathbb{Z}_{3}}
\end{equation}
respectively, both of which are readily seen to match the spectrum of the TSPM after applying the three-step gauging procedure \cite{Moore:1989yh}. However, translating the Maverick coset version of the TSPM to TQFTs is not so straightforward. Obviously, $SU(3)_{2} \times SU(2)_{-8}$ has too many lines to identify it by itself with the TSPM, so we must gauge some lines away in order to make them coincide. However, $SU(2)$ and $SU(3)$ have a trivial common center, so it does not seem possible to use the standard gauging procedure of \cite{Moore:1989yh} to do so. Notice that $SU(3)_{2} \times SU(2)_{-8}$ does have a non-trivial abelian boson in its spectrum; namely $(\mathbf{1},8)$, \footnote{Here and below we have used our notation of denoting a line in $SU(3)_{2}$ by the dimension of its representation in bold letters, and a line in $SU(2)_{k}$ by the standard integer $i=0,1,\ldots k$.} but gauging it gives $SU(3)_{2} \times SO(3)_{-4}$, which is also does not match the spectrum of the TSPM.

The solution to this conundrum, of course, is that $SU(3)_{2} \times SU(2)_{-8}$ has yet another non-trivial boson in its spectrum: $(\mathbf{8},4)$, and it is a non-abelian one! Clearly, we can now attempt to condense it in order to reproduce the spectrum of the TSPM. It goes without saying, the non-abelian anyon condensation calculation indeed reproduces the spectrum of the TSPM. The prior calculation was done in \cite{Bais:2008ni} as a test example of the formalism of non-abelian anyon condensation. In Appendix \ref{ExampleforsimplestMaverick} we provide an alternative argument based on exceptional conformal embeddings that simplifies the calculation, on top of relating quite directly this Maverick coset with conformal embeddings and the standard arguments for level-rank duality.

In summary, \eqref{oneof3} and \eqref{2and3of3} are three different coset descriptions for the same underlying 3D TQFT (i.e., we have found three different Chern-Simons-like descriptions of the same underlying MTC data), two of which are rather standard and one which involves non-abelian anyon condensation. Since all of these descriptions describe the same underlying theory, we can now relate all the previous Chern-Simons-like descriptions of the TSPM to one another, obtaining the dualities:
\begin{equation} \label{TSPMDuality}
    \frac{SU(3)_{2} \times SU(2)_{-8}}{\mathcal{A}_{3}} \cong \frac{SU(2)_{3} \times U(1)_{-6}}{\mathbb{Z}_{2}} \cong \frac{SU(3)_{1} \times SU(3)_{1} \times SU(3)_{-2}}{\mathbb{Z}_{3}},
\end{equation}
where the first of these involves non-abelian anyon condensation by the algebra object $\mathcal{A}_{3} = (\mathbf{1},0) + (\mathbf{1},8) + (\mathbf{8},4)$. Of course, starting with these expressions we can now attempt to ``make the dualities proliferate'' by gauging zero-form or one-form symmetries on either side, turning-on background fields possibly with discrete torsion, coupling to spin structure or combining these dualities with previously known ones, etc. We do not attempt this here, and we set it aside for future work.

This example can be generalized to the complete first infinite family of Maverick cosets \eqref{firstmaverickfamily} noticing that the central charges match those of the parafermion CFTs \cite{Fateev:1985mm,DiFrancesco:1997nk}, so it is natural to suggest that this infinite Maverick family reproduces the parafermions. Indeed, the three-state Potts model corresponds to the $\mathbb{Z}_{3}$ parafermion, so we could have foreseen the previous results by merely matching the central charges to that of the parafermions.

The parafermions also have two standard coset descriptions given by the $SU(2)_{k}/U(1)_{2k}$, or $(SU(k)_{1} \times SU(k)_{1})/SU(k)_{2}$ cosets \cite{DiFrancesco:1997nk}, which generalize the cosets of the TSPM above. Using the same arguments (although now with lack of a general calculation valid for any $k$), we expect the infinite family of dualities
\begin{equation} \label{ParafermionMaverickDuality}
    \boxed{\frac{SU(k)_{2} \times Spin(k)_{-4}}{\mathcal{A}_{k}} \cong \frac{SU(2)_{k} \times U(1)_{-2k}}{\mathbb{Z}_{2}} \cong \frac{SU(k)_{1} \times SU(k)_{1} \times SU(k)_{-2}}{\mathbb{Z}_{k}}},
\end{equation}
for some suitable algebra object $\mathcal{A}_{k}$ on the left-hand side. For $k=3$ this reproduces the result for the TSPM above. We can actually guess a general form for $\mathcal{A}_{k}$ by matching quantum dimensions for the first few values of $k$ in between the left-hand side OF \eqref{ParafermionMaverickDuality} and the parafermion TQFT (see Eqn. \eqref{quantumdimensionconstraint}). For odd $k = 2n+1$, $n \geq 1$, we obtain
\begin{equation} \label{algebra1}
    \mathcal{A}_{2n+1} = (\mathbf{1}, \overline{\mathbf{1}}) + (\mathbf{1}, \overline{4\mathbf{w}_{1}}) + \sum_{i=1}^{n-1} (\mathbf{w}_{i}+\mathbf{w}_{2n+1-i}, \overline{2\mathbf{w}_{i}}) + (\mathbf{w}_{n} + \mathbf{w}_{n+1}, \overline{4\mathbf{w}_{\sigma}}),
\end{equation}
while for $k$ even $k = 2n$, $n \geq 1$, we obtain
\begin{align} \label{algebra2}
    \mathcal{A}_{2n} = (\mathbf{1}, \overline{\mathbf{1}}) + (\mathbf{1}, \overline{4\mathbf{w}_{1}}) & + (2\mathbf{w}_{n}, \overline{4\mathbf{w}_{s}}) + (2\mathbf{w}_{n}, \overline{4\mathbf{w}_{c}}) \nonumber \\[0.1cm] & + \sum_{i=1}^{n-2}(\mathbf{w}_{i} + \mathbf{w}_{2n-i}, \overline{2\mathbf{w}_{i}}) + (\mathbf{w}_{n-1} + \mathbf{w}_{n + 1}, \overline{2\mathbf{w}_{s} + 2\mathbf{w}_{c}}).
\end{align}
Here we have denoted the lines in the Chern-Simons theories in terms of the fundamental weights $\mathbf{w}_{i}$ of the corresponding Lie algebras. For spin groups $\mathbf{w}_{\sigma}$ stands for the spinorial weight of $Spin(2n+1)$ and $\mathbf{w}_{s}$ and $\mathbf{w}_{c}$ stand for the spinor and cospinor weights of $Spin(2n)$. A bar has also been used here to denote the lines belonging to Chern-Simons theories with negative level.

\subsubsection*{Second Family of Maverick Dualities}

We study now the second infinite family of Maverick cosets \eqref{secondmaverickfamily}. Since the value of the central charge is one there are now two natural possibilities for such infinite family: either an infinite family corresponding to the free boson branch of $c=1$ RCFTs, or an infinite family corresponding to the orbifold branch of $c=1$ RCFTs. 

To decide for one of these families, we will have to perform an explicit calculation in the simplest case of the second infinite family of Maverick cosets \eqref{secondmaverickfamily}, corresponding to $N=3$. For the sake of presentation, we do not perform this calculation here but show this computation in Section \ref{orbifoldexample} below. The calculation shows that for $N=3$, the result of the non-abelian anyon condensation is actually the orbifold of $U(1)_{6}$, which we denote $U(1)^{\mathrm{Orb}}_{6}$. Based on this result, we conjecture that after some condensation of non-abelian bosons, the first infinite family of Maverick cosets leads to the orbifold branch of $c=1$ RCFTs. This suggests the following dualities between theories at the orbifold branch and Chern-Simons-like theories based on Spin groups:
\begin{equation} \label{secondduality}
    \boxed{U(1)_{2N}^{\mathrm{Orb}} \cong \frac{Spin(2N)_{2} \times Spin(N)_{-2} \times Spin(N)_{-2}}{\mathcal{B}_{N}}}
\end{equation}
for some suitable algebras $\mathcal{B}_{N}$. The case we will verify explicitly below corresponds to $N=3$, and it is straightforward to check that in the cases $N=1$ and $N=2$ the formula still holds.

A quick argument in support of the previous proposal that does not rely on going through the whole computation in Section \ref{orbifoldexample} is given as follows. We can use the exceptional isomorphisms of chiral algebras
\begin{align}
    &Spin(3)_{k} \cong SU(2)_{2k} \, , \\[0.3cm]
    &Spin(4)_{k} \cong SU(2)_{k} \times SU(2)_{k} \, ,  \\[0.3cm]
    &Spin(6)_{k} \cong SU(4)_{k}, \, 
\end{align}
to massage the expression for the Maverick coset \eqref{secondmaverickfamily} at $N=3$ in the following way:
\begin{align}
    Spin(6)_{2} \times Spin(3)_{-2} \times Spin(3)_{-2} &\cong SU(4)_{2} \times SU(2)_{-4} \times SU(2)_{-4} \\[0.3cm] &\cong SU(4)_{2} \times Spin(4)_{-4}.
\end{align}
That is, using the exceptional isomorphisms of chiral algebras we have written the Maverick coset \eqref{secondmaverickfamily} at $N=3$ as the Maverick coset in the first infinite Maverick family \eqref{firstmaverickfamily} at $k=4$. This is the next example in such family after the TSPM at $k=3$ studied above.

Let us assume now that the first infinite family of Maverick cosets \eqref{firstmaverickfamily} is given by the parafermion CFTs, as pointed out above. Then, the $\mathbb{Z}_{4}$ parafermion can actually be identified with the $U(1)_{6}^{\mathrm{Orb}}$ orbifold CFT \cite{DiFrancesco:1997nk}, and we have the explicit result
\begin{align}
     U(1)_{6}^{\mathrm{Orb}} \cong \frac{SU(4)_{2} \times Spin(4)_{-4}}{\mathcal{A}_{4}} \cong \frac{Spin(6)_{2} \times Spin(3)_{-2} \times Spin(3)_{-2}}{\mathcal{B}_{3}}, \label{orbifoldu6duality}
\end{align}
for some appropriate algebra objects $\mathcal{A}_{4} \cong \mathcal{B}_{3}$. In Section \ref{orbifoldexample} below we will verify this statement by a direct computation on non-abelian anyon condensation. Naturally, since this specific case turns out to be a parafermion, we can actually further identify \eqref{orbifoldu6duality} with the other Chern-Simons-like expressions on Eqn. \eqref{ParafermionMaverickDuality} for $k=4$. 

As above, we can guess a general form for $\mathcal{B}_{N}$ by matching quantum dimensions for the first few values of $N$ in between the right-hand side of \eqref{secondduality} and the orbifold branch TQFT (see Eqn. \eqref{quantumdimensionconstraint}). For odd $N = 2n+1$, $n \geq 1$, we obtain
\begin{align} \label{algebra3}
    \mathcal{B}_{2n+1} = (\mathbf{1}, \overline{\mathbf{1}}, \overline{\mathbf{1}}) + (\mathbf{1}, \overline{2\mathbf{w}_{1}}, \overline{2\mathbf{w}_{1}}) & + (2\mathbf{w}_{1}, \overline{2\mathbf{w}_{1}}, \overline{\mathbf{1}}) + (2\mathbf{w}_{1}, \overline{\mathbf{1}}, \overline{2\mathbf{w}_{1}}) \nonumber \\[0.1cm] & + (\mathbf{w}_{s} + \mathbf{w}_{c}, \overline{2\mathbf{w}_{\sigma}}, \overline{2\mathbf{w}_{\sigma}}) + \sum_{i=1}^{n-1} (\mathbf{w}_{2i}, \overline{\mathbf{w}_{i}}, \overline{\mathbf{w}_{i}}),
\end{align}
while for $N$ even $N = 2n$, $n \geq 1$, we obtain
\begin{align} \label{algebra4}
    \mathcal{B}_{2n} = & (\mathbf{1}, \overline{\mathbf{1}}, \overline{\mathbf{1}}) + (2\mathbf{w}_{s}, \overline{2\mathbf{w}_{s}}, \overline{2\mathbf{w}_{s}})  + (\mathbf{1}, \overline{2\mathbf{w}_{1}}, \overline{2\mathbf{w}_{1}}) + (2\mathbf{w}_{s}, \overline{2\mathbf{w}_{c}}, \overline{2\mathbf{w}_{c}}) \nonumber \\[0.2cm] & + (2\mathbf{w}_{1}, \overline{2\mathbf{w}_{1}}, \overline{\mathbf{1}}) + (2\mathbf{w}_{c}, \overline{2\mathbf{w}_{c}}, \overline{2\mathbf{w}_{s}}) + (2\mathbf{w}_{1}, \overline{\mathbf{1}}, \overline{2\mathbf{w}_{1}}) + (2\mathbf{w}_{c}, \overline{2\mathbf{w}_{s}}, \overline{2\mathbf{w}_{c}}) \nonumber \\[0.1cm] & + (\mathbf{w}_{2(n-1)}, \overline{\mathbf{w}_{s} + \mathbf{w}_{c}}, \overline{\mathbf{w}_{s} + \mathbf{w}_{c}}) + \sum_{i=1}^{n-2} (\mathbf{w}_{2i}, \overline{\mathbf{w}_{i}}, \overline{\mathbf{w}_{i}}).
\end{align}
As in the previous infinite family, we have denoted the lines in the Chern-Simons theories in terms of the fundamental weights $\mathbf{w}_{i}$ of the corresponding Lie algebras. For spin groups $\mathbf{w}_{\sigma}$ stands for the spinorial weight of $Spin(2n+1)$ and $\mathbf{w}_{s}$ and $\mathbf{w}_{c}$ stand for the spinor and cospinor weights of $Spin(2n)$. A bar has also been used here to denote the lines belonging to Chern-Simons theories with negative level.

\subsubsection*{Isolated Maverick Dualities}

We now study a few of the isolated Maverick cosets \eqref{thirdmaverick}-\eqref{lastmaverick}, all of which have $c<1$ and thus correspond to some (possibly non-diagonal) minimal model. The simplest of these is an interesting example leading to the Ising TQFT:
\begin{equation}
    \mathrm{Ising \ TQFT} \cong \frac{SU(4)_{1} \times SU(2)_{-10}}{\mathcal{A}},
\end{equation}
for some appropriate algebra object $\mathcal{A}$. We verify this example by a direct computation in appendix \ref{Ising Maverick Duality}. The Maverick coset \eqref{fifthMaverick} similarly leads to the Ising model and can also be checked by an explicit calculation. 

Minimal models, and in particular the Ising model, have many (standard) coset descriptions. Using these many descriptions, and equating them to the Maverick result above we obtain many instances of dualities involving a gauging by a non-invertible symmetry. Explicitly, for instance:
\begin{align}
    & \frac{SU(4)_{1} \times SU(2)_{-10}}{\mathcal{A}} \cong \frac{(E_{8})_{2} \times Spin(16)_{-2}}{\mathcal{A}'} \cong \frac{SU(2)_{1} \times SU(2)_{1} \times SU(2)_{-2}}{\mathbb{Z}_{2}} \nonumber \\[0.4cm]
    \cong & \frac{SU(2)_{2} \times U(1)_{-4}}{\mathbb{Z}_{2}} \cong (E_{8})_{1} \times (E_{8})_{1} \times (E_{8})_{-2} \cong \frac{USp(4)_{1} \times SU(2)_{-1} \times SU(2)_{-1}}{\mathbb{Z}_{2}},
\end{align}
for some appropriate algebra objects $\mathcal{A}$ and $\mathcal{A}'$.

We can play a similar strategy with the Tricritical Ising TQFT, which also has many (standard) coset descriptions as well as three Maverick descriptions, \eqref{fourthMaverick}, \eqref{sixthMaverick}, and \eqref{lastmaverick}. Following the same steps, we obtain:
\begin{align}
    & \frac{Spin(7)_{1} \times SU(2)_{-28}}{\mathcal{A}} \cong \frac{(E_{8})_{2} \times SU(2)_{-2} \times (E_{7})_{-2}}{\mathcal{A}'} \cong \frac{(E_{7})_{2} \times SU(8)_{-2}}{\mathcal{A}''} \nonumber \\[0.4cm]
    \cong & \frac{SU(2)_{2} \times SU(2)_{1} \times SU(2)_{-3}}{\mathbb{Z}_{2}} \cong \frac{(E_{7})_{1} \times (E_{7})_{1} \times (E_{7})_{-2}}{\mathbb{Z}_{2}} \cong (F_{4})_{1} \times Spin(9)_{-1} , \nonumber \\[0.4cm] 
    \cong & \frac{USp(6)_{1} \times USp(4)_{-1} \times SU(2)_{-1}}{\mathbb{Z}_{2}} \cong \frac{SU(3)_{2} \times SU(2)_{-2} \times U(1)_{-12}}{\mathbb{Z}_{6}},
\end{align}
for some appropriate algebra objects $\mathcal{A}$, $\mathcal{A}'$, $\mathcal{A}''$. The last case is notable since the gauging is by an abelian anyon, but it cannot be interpreted as a common center. Rather it is a quantum one-form symmetry present due to the specific Chern-Simons levels.

\section{Explicit Checks via Non-Abelian Anyon Condensation} \label{ExplicitChecks}

In this section, we present a detailed description of some computations involving non-abelian anyon condensation, specifically with the aim of providing a non-trivial check of the dualities claimed above. 

First, we summarize some consistency conditions that will allow us to pin down the resulting theory after non-abelian anyon condensation. We follow the heuristic rules of \cite{Bais:2008ni} for this purpose. Non-abelian anyon condensation was formalized in \cite{kong2014anyon}, but the rules of \cite{Bais:2008ni} are more useful for practical calculations. Following the logic of the three-step gauging procedure reviewed in Section \ref{sec:intro}, we aim to obtain the spectrum and topological spins of the anyons in the gauged theory without working out the full MTC data.  The rules of \cite{Bais:2008ni} enable this approach. In successive subsections, we consider many explicit examples of interest. In the following, we refer as \textit{parent theory} to the original theory before condensation/gauging, and as \textit{child theory} to the one obtained after condensation/gauging. For obvious reasons, we also refer to the parent and child as uncondensed and condensed theories, respectively. As in the case of standard gaugings of abelian anyons (one-form global symmetries) only bosons can condense, paralleling the fact that for abelian anyons the topological spins capture the `t Hooft anomalies \cite{Gaiotto:2014kfa, Gomis:2017ixy, Hsin:2018vcg}.\footnote{Relatedly, commutative algebras in a MTC decompose necessarily in terms of simple anyons with trivial topological spin. In particular then, special symmetric commutative Frobenius algebras decompose in terms of simple anyons with trivial topological spin.}  \\

When we perform anyon condensation, the simple anyons of the parent theory are generically split into many terms associated with excitations in the child theory:
\begin{equation}
    a \longrightarrow \sum_{i} n_{i}^{a}a_{i}, \quad n_{i}^{\ c} \in \mathbb{N}. \label{restriction}
\end{equation}
It is useful at this intermediate stage to distinguish between genuine line operators, and non-genuine line operators which necessarily arise at the end of topological surface operators \cite{Aharony:2013hda}. Below we often refer to genuine line operators as unconfined excitations, and non-genuine line operators as confined. Shortly we will see how to differentiate between confined excitations and unconfined ones on the right-hand side of a restriction.

The labels $a_{i}$ on the right-hand side of \eqref{restriction} do not necessarily correspond to genuine line operators in the child, and furthermore, many labels $a_{i}$ corresponding to different anyons $a$ in the parent theory do not necessarily correspond to different excitations in the child theory. Below we will write certain consistency conditions for non-abelian anyon condensation, and imposing such conditions may imply identifications between these many labels. In the following, we will refer to the decomposition \eqref{restriction} as \textit{the restriction of the anyon $a$}.\footnote{In this context, some references refer to the splitting $a \to \sum_{i} n_{i}^{\ a} a_{i}$ as ``branching''. We avoid this terminology to prevent confusion with the \textit{a priori} different concept of branching rules of affine Lie algebras.} 

We call the different elements $a_{i}$ on the right-hand side of the restriction \eqref{restriction} the components of the anyon $a$ in the parent theory. When an anyon $a$ restricts to a single label with unit multiplicity on the right-hand side (i.e., it does not ``split''), we abuse notation and call the corresponding component on the right-hand side $a$ as well. When the latter happens, the corresponding (non-necessarily genuine) line operator in some sense ``descends trivially'' from the parent theory. We adopt the previous nomenclature in the next subsections.

We assume that when we condense non-abelian anyons, the resulting theory has fusion rules fulfilling the following standard requirements: associativity, existence of a unique vacuum, and existence of unique conjugate representations with a unique way to annihilate to the vacuum.  Additionally, condensation is subject to the following rules \cite{Bais:2008ni}: 

1) A sector that condenses has a component that is indistinguishable from the vacuum sector in the condensed phase. Specifically:
\begin{equation}
    c \longrightarrow (c_{1} \equiv 0) + \sum_{i>1}n_{i}^{\ c}c_{i}, \quad n_{i}^{\ c} \in \mathbb{N},
\end{equation}
where we assume the vacuum component to have multiplicity one. 

2) We require fusion of the old and new labels to be compatible with the restriction, i.e., restricting to the resulting theory and fusion must commute:
\begin{equation}
    a \times b = \sum_{c} N_{ab}^{c} c \Longrightarrow \Big( \sum_{i}n_{i}^{\ a}a_{i} \Big) \times \Big( \sum_{j}n_{j}^{\ b}b_{j} \Big) = \sum_{c,k}N_{ab}^{c}n_{k}^{\ c}c_{k}.
\end{equation}
Additionally, if $a$ and $\bar{a}$ are conjugates to each other in the parent:
\begin{equation}
    a \longrightarrow \sum_{i} n_{i}^{\ a}a_{i} \Longrightarrow \bar{a} \longrightarrow \sum_{i}n_{i}^{\ a}\bar{a}_{i}.
\end{equation}
In order for these compatibility conditions to hold, it may be necessary to identify two different labels $a_{i}$ and $b_{j}$ on the right sides of the restrictions of two anyons $a$ and $b$ in the parent. 

As a consequence of these assumptions the quantum dimensions are preserved under restriction:
\begin{equation}
    a \longrightarrow \sum_{b} n_{b}^{\ a}b \Longrightarrow d_{a} = \sum_{b} n_{\ b}^{a}d_{b},
\end{equation} 
which will be a crucial tool below when doing explicit computations. 

3) Confined and deconfined excitations $c_{i}$ are distinguished by their lift to anyons $c$ in the parent theory. If the set of all the components that we identify to a certain $c_{i}$ lift to anyons in the parent theory that do not all have the same topological spin, the corresponding $c_{i}$ confines and thus it does not correspond to a simple object in the MTC data describing the condensed theory. Notice that a priori this condition is sufficient but not necessary to distinguish unconfined excitations from confined ones. Namely, there could be $c_{i}$ that lift to anyons all with the same topological spin and nevertheless in principle it could happen that $c_{i}$ confines. In this work we find that all examples are consistent when we assume that we can distinguish unconfined from confined excitations by the above criterion.

\subsection{Checking the $SU(2)_{1} \cong \big( (G_{2})_{1} \times SU(2)_{-3} \big)/\mathcal{Z}(\mathbf{Fib})$ Duality}  \label{IntroductionExample}

In this first check we analyze the straightforward example described in Section \ref{sec:introG2} and Section \ref{noninvertibleextensionsubsection} through direct, in-depth computation of non-abelian anyon condensation.

The spectra of $(G_{2})_{1}$ and $SU(2)_{3}$ are given in Tables \ref{g2lv1table} and \ref{su2lv3table} respectively. We call $\phi$ the only non-trivial anyon of $(G_{2})_{1}$, which obeys Fibonacci fusion rules. $SU(2)_{k}$ for general $k$ consists of $k+1$ lines labeled from $0$ to $k$, and with fusion rules given by
\begin{equation}
    \Lambda_{1} \times \Lambda_{2} = \sum_{\Lambda = |\Lambda_{1}-\Lambda_{2}|}^{\mathrm{min}(\Lambda_{1}+\Lambda_{2},2k-\Lambda_{1}-\Lambda_{2})}  \Lambda, \label{SU2kFusionRules}
\end{equation}
where the sum is restricted such that $\Lambda_{1} + \Lambda_{2} - \Lambda$ is even. We denote the lines in $(G_{2})_{1} \times SU(2)_{-3}$ in the obvious way: $(0,i)$ and $(\phi,i)$ with $i=0,1,2,3$.

\begin{table}[t]
\centering
\begin{tabular}[h]{|p{4cm}|p{4cm}|p{4cm}| }
\hline 
\multicolumn{3}{|c|}{$(G_{2})_{1}$} \\
\hline
Line label & Quantum Dimension & Conformal Weight \\
\hline
0 & $d_{0} = 1$ & $h_{0}=0$ \\
$\phi$ & $d_{\phi} = \frac{1+\sqrt{5}}{2}$ & $h_{\phi}=2/5$ \\
\hline
\end{tabular}
\caption{$(G_{2})_{1}$ data. In the table and in the text $\phi$ denotes the unique non-trivial line of $(G_{2})_{1}$ with Fibonacci fusion rule $\phi \times \phi = 0 + \phi$.}  \label{g2lv1table}
\end{table}

\begin{table}[!b]
\centering
\begin{tabular}[h]{|p{4cm}|p{4cm}|p{4cm}| }
\hline 
\multicolumn{3}{|c|}{$SU(2)_{3}$} \\
\hline
Line label & Quantum Dimension & Conformal Weight \\
\hline
0 & $d_{0} = 1$ & $h_{0}=0$ \\
1 & $d_{1} = \frac{1 + \sqrt{5}}{2}$ & $h_{1}=3/20$ \\
2 & $d_{2} = \frac{1 + \sqrt{5}}{2}$ & $h_{2}=2/5$ \\
3 & $d_{3} = 1$ & $h_{3}=3/4$ \\
\hline
\end{tabular}
\caption{$SU(2)_{3}$ data.}  \label{su2lv3table}
\end{table}

It is easy to check that there is only one non-trivial boson $(\phi,2)$ in the spectrum of $(G_{2})_{1} \times SU(2)_{-3}$, and it is non-abelian. Let us assume that such anyon can condense, which is the statement that
\begin{equation} \label{(phi,2)restriction}
    (\phi,2) \longrightarrow 0 + (\phi,2)_{2}, \hspace{.2in}d_{(\phi,2)_{2}} = \frac{1 + \sqrt{5}}{2}.
\end{equation}
 The simplest way to see that on top of $0$ there must be a single further label on the right-hand side of \eqref{(phi,2)restriction} is to check the conservation of the quantum dimension. Since $d_{0} = 1$, the remaining quantum dimension is $\frac{1+\sqrt{5}}{2}$, which is too small to allow a splitting. For future reference, another method to check that only a single line is allowed on top of $0$ in \eqref{(phi,2)restriction} is to inspect the fusion of $(\phi,2)$ with itself:
\begin{equation}
    (\phi,2) \times (\phi,2)= (0,0) + (0,2) + (\phi,0) + (\phi,2) \longrightarrow 0 + 0 + \ldots,
\end{equation}
where the arrow means that we have taken the restriction of the anyons on the right-hand side of the equality, and since $(\phi,2)$ condenses, a vacuum arises on the right-hand side. This means that we need $(\phi,2)$ to split into two components on the left-hand side. Since one of them is the vacuum 0 by assumption, the other one must be a single line with quantum dimension $\frac{1 + \sqrt{5}}{2}$. Furthermore, since the vacuum $0$ is self-conjugate, we deduce that $(\phi,2)_{2}$ must also be self-conjugate.

We also observe that all anyons of the form $(0,i)$ cannot split, since they do not have large enough quantum dimension. Similarly, the anyons $(\phi,0)$ and $(\phi,3)$ cannot split.

We can deduce the fate of $(\phi,2)_{2}$ upon condensation by studying its fusion with $(0,2)$:
\begin{equation}
    (0,2) \times (\phi,2) = (\phi,0) + (\phi,2) \longrightarrow 0 + \ldots
\end{equation}
This implies that $(0,2)$, which we already deduced does not split, is conjugate with one component of $(\phi,2)$. But $(0,2)$ is self-conjugate and it cannot be identified with the vacuum, since it has neither the appropriate quantum dimension, nor the spin to condense. Therefore, we must identify $(0,2) \cong (\phi,2)_{2}$. Now, note that $(\phi,2)_{2}$ lifts to $(\phi,2)$ while $(0,2)$ in the child theory lifts to $(0,2)$ in the parent theory, but $(0,2)$ and $(\phi,2)$ have different topological spins, so it follows that $(0,2) \cong (\phi,2)_{2}$ confine. A completely analogous argument shows that $(\phi,0)$ belongs in the restriction of $(\phi,2)$ and we have the identification $(\phi,0) \cong (\phi,2)_{2} \cong (0,2)$.

Since $(\phi,1)$ has sufficient quantum dimension to split we must check if it does. Computing its self-fusion:
\begin{equation}
    (\phi,1) \times (\phi,1) = (0,0) + (0,2) + (\phi,0) + (\phi,2) \longrightarrow 0 + 0 + \ldots,
\end{equation}
we deduce that it splits into two components $(\phi,1) \to (\phi,1)_{1} + (\phi,1)_{2}$.

Now that we know the splitting pattern we have to assign quantum dimensions. It is straightforward to check that
\begin{equation}
    (\phi,3) \times (\phi,1) = (0,2) + (\phi,2) \longrightarrow 0 + \ldots,
\end{equation}
implies that $(\phi,3)$ belongs in the restriction of $(\phi,1)$ since $(\phi,3)$ is self-conjugate, and since $d_{(\phi,3)} = (1 + \sqrt{5})/2$, by conservation of the quantum dimension it must be that the other component of $(\phi,1)$ has unit quantum dimension. That is, $d_{(\phi,1)_{1}} = (1 + \sqrt{5})/2$ and $d_{(\phi,1)_{2}} = 1$. 

From the fusions
\begin{align}
    &(0,1) \times (\phi,1) = (\phi,0) + (\phi,2) \longrightarrow 0 + \ldots, \\[0.3cm]
    &(0,3) \times (\phi,1) = (\phi,2) \longrightarrow 0 + \ldots,
\end{align}
we deduce by similar arguments as above that $(0,1),(0,3) \in (\phi,1)$. From the quantum dimensions $d_{(0,1)} = (1 + \sqrt{5})/2$ and $d_{(0,3)}=1$ we obtain that there is a single way to fit $(0,1)$ and $(0,3)$ in the restriction of $(\phi,1)$, implying the identifications $(\phi,3) \cong (\phi,1)_{1} \cong (0,1)$ and $(\phi,1)_{2} \cong (0,3)$. 

All in all, we obtain the condensation pattern:
\begin{equation}
    (0,0) \to (0,0), \quad (0,1) \to (0,1), \quad (0,2) \to (0,2), \quad (0,3) \to (0,3), \nonumber
\end{equation}
\begin{equation}
    (\phi,0) \to (0,2), \quad (\phi,1) \to (0,1) + (0,3), \quad (\phi,2) \to 0 + (0,2), \quad (\phi,3) \to (0,1), \nonumber
\end{equation}
from which it is straightforward to check that only $(0,0)$ and $(0,3)$ lift to anyons in the parent that have common topological spin and thus do not confine. They also have the correct fusion rules, spins and quantum dimensions to match those of $SU(2)_{1}$, as expected. 

\subsection{Checking the $SU(2)_{1} \times SU(2)_{2} \cong \big( M(5,4) \times SU(2)_{3} \big) / \mathcal{A}$ Duality} \label{cosetinversionformulachecksection}

In this subsection we check the coset inversion formulas \eqref{MTCEmbedding}-\eqref{IsolatingMTCcoset} observed on mathematical grounds in Section \ref{GeneralNonAbelianAnyonCondesationPicture}. Specifically we consider the example given in Eqn. \eqref{MinimalModelCheck}, and check that
\begin{equation}
    SU(2)_{k} \times SU(2)_{1} \cong \frac{M(k+3,k+2) \times SU(2)_{k+1}}{\mathcal{A}}, \quad k \geq 1,
\end{equation}
in the case $k=2$. When $k=1$ the gauging is abelian and the expression is easily checked by the three-step gauging rule \cite{Moore:1989yh, Hsin:2018vcg}. The case $k=2$ is more interesting, since the gauging is by a non-invertible one-form symmetry.

When $k=2$, $M(5,4)$ corresponds to the Tricritical Ising Model, whose spectrum can be found in Table \ref{tricritical}. Its non-trivial fusion rules are
\begin{align}
    & \varepsilon'' \times \varepsilon'' = 0, \quad \hspace{-0.5cm}&&\varepsilon'' \times \varepsilon' = \varepsilon, \quad \quad \hspace{0.25cm} \varepsilon'' \times \varepsilon = \varepsilon', \quad \varepsilon'' \times \sigma = \sigma, \quad \varepsilon'' \times \sigma' = \sigma', \\[0.3cm]
    &\varepsilon' \times \varepsilon' = 0 + \varepsilon', \quad
    \hspace{-0.5cm} &&\varepsilon' \times \varepsilon = \varepsilon + \varepsilon'', \quad
    \varepsilon' \times \sigma = \sigma + \sigma', \quad
    \varepsilon' \times \sigma' = \sigma,  \\[0.3cm]
    &\varepsilon \times \varepsilon = 0 + \varepsilon', \quad
    \hspace{-0.5cm} &&\varepsilon' \times \sigma = \sigma + \sigma', \quad
    \varepsilon \times \sigma' = \sigma,  \\[0.3cm]
    &\sigma \times \sigma = 0 + \varepsilon + \varepsilon' + \varepsilon'', \quad
    \hspace{-0.5cm} &&\sigma \times \sigma' = \varepsilon + \varepsilon', \quad
    \sigma' \times \sigma' = 0 + \varepsilon''.
\end{align}
Meanwhile, the spectrum of $SU(2)_{3}$ and that of the expected result $SU(2)_{1} \times SU(2)_{2}$ are shown in tables \ref{su2lv3table} and \ref{su2lv1timessu2lv2table} respectively. Their fusion rules are easily derived from Eqn. \eqref{SU2kFusionRules}. In the following we denote the anyons in the obvious way as in the product $M(5,4) \times SU(2)_{3}$.

\begin{table}[t]
\centering
\begin{tabular}[h]{|p{4cm}|p{4cm}|p{4cm}| }
\hline 
\multicolumn{3}{|c|}{$\mathrm{Tricritical \ Ising \ Model} \ M(5,4)$} \\
\hline
Line label & Quantum Dimension & Conformal Weight  \\
\hline
0 & $d_{0} = 1$ & $h_{0} = 0$ \\
$\varepsilon$ & $d_{\varepsilon} = \frac{1+\sqrt{5}}{2}$ & $h_{\varepsilon} =  1/10$ \\
$\varepsilon'$ & $d_{\varepsilon'} = \frac{1+\sqrt{5}}{2}$ & $h_{\varepsilon'} = 3/5$ \\
$\varepsilon''$ & $d_{\varepsilon''} = 1$ & $h_{\varepsilon''} = 3/2$ \\
$\sigma$ & $d_{\sigma} = \sqrt{2} \big( \frac{1+\sqrt{5}}{2} \big)$ & $h_{\sigma} = 3/80$ \\
$\sigma'$ & $d_{\sigma'} = \sqrt{2}$ & $h_{\sigma'} = 7/16$ \\
\hline
\end{tabular}
\caption{Data of the Tricritical Ising Model.}  \label{tricritical}
\end{table}

\begin{table}[!b]
\centering
\begin{tabular}[h]{|p{4cm}|p{4cm}|p{4cm}| }
\hline 
\multicolumn{3}{|c|}{$SU(2)_{2} \times SU(2)_{1}$} \\
\hline
Line label & Quantum Dimension & Conformal Weight  \\
\hline
(0,0) & $d_{(0,0)} = 1$ & $h_{(0,0)} = 0$ \\
(0,1) & $d_{(0,1)} = 1$ & $h_{(0,1)} = 1/4$ \\
(1,0) & $d_{(1,0)} = \sqrt{2}$ & $h_{(1,0)} = 3/16$ \\
(1,1) & $d_{(1,1)} = \sqrt{2}$ & $h_{(1,1)} = 7/16$ \\
(2,0) & $d_{(2,0)} = 1$ & $h_{(2,0)} = 1/2$ \\
(2,1) & $d_{(2,1)} = 1$ & $h_{(2,1)} = 3/4$ \\
\hline
\end{tabular}
\caption{$SU(2)_{2} \times SU(2)_{1}$ Data}  \label{su2lv1timessu2lv2table}
\end{table}

We start noticing that in the product $M(5,4) \times SU(2)_{3}$ there is a single boson $(\varepsilon',2)$, and it is non-abelian. Then, condensing $(\varepsilon',2)$ is the statement that we have the splitting
\begin{equation} \label{varepsilonprime2splitting}
    (\varepsilon',2) \to 0 + (\varepsilon',2)_{2},
\end{equation}
with $d_{(\varepsilon',2)_{2}} = (1+\sqrt{5})/2$ by conservation of the quantum dimension, so $(\varepsilon',2)$ cannot split further.

Now notice that the following pairs of anyons
\begin{align}
    &\{(0,0), (\varepsilon',2)\}, \quad \{(\varepsilon'',3), (\varepsilon,1)\}, \quad \{(\sigma', 3), (\sigma,1)\}, \nonumber \\[0.3cm] &\{(\sigma',0), (\sigma,2)\}, \quad \{(\varepsilon'',0), (\varepsilon,2)\}, \quad \{(0,3), (\varepsilon',1)\} \label{anyonlist}
\end{align}
share the same topological spins as the anyons in $SU(2)_{1} \times SU(2)_{2}$, respectively in the same order as shown in Table \ref{su2lv1timessu2lv2table}. Based on this observation, consider the second pair $\{(\varepsilon'',3), (\varepsilon,1)\}$ (since the first pair is basically the condensing boson), and study the fusion rules in this pair:
\begin{equation}
    (\varepsilon,1) \times (\varepsilon,1) = (0,0) + (0,2) + (\varepsilon',0) + (\varepsilon',2) \longrightarrow 0 + 0 + \ldots
\end{equation}
\begin{equation}
    (\varepsilon'',3) \times (\varepsilon,1) = (\varepsilon',2) \longrightarrow 0.
\end{equation}
The first fusion rule tells us that $(\varepsilon,1)$ splits in two, while the second one implies that $(\varepsilon'',3) \in (\varepsilon,1)$. So, we have the splitting
\begin{equation}
    (\varepsilon,1) \to (\varepsilon'',3) + (\varepsilon,1)_{2}.
\end{equation}
Using the same arguments, same conclusions follow for each of the pairs mentioned above; namely, the second anyon in a pair splits in two, and one of its components corresponds to the first anyon in the same pair.

Checking for confinement given this structure is now easy. Take for example the anyons $(\varepsilon',0)$ and $(0,2)$ and consider their fusion with $(\varepsilon',2)$:
\begin{equation}
    (\varepsilon',0) \times (\varepsilon',2) = (0,2) + (\varepsilon',2) \longrightarrow 0 + \ldots,
\end{equation}
\begin{equation}
    (0,2) \times (\varepsilon',2) = (\varepsilon',0) + (\varepsilon',2) \longrightarrow 0 + \ldots.
\end{equation}
This means that $(\varepsilon',0), (0,2) \in  (\varepsilon',2)$, and it is easy to see that there is only one way to match quantum dimensions with the restriction \eqref{varepsilonprime2splitting}. We must have the identification
\begin{equation}
    (0,2) \cong (\varepsilon',0) \cong (\varepsilon',2)_{2}.
\end{equation}
Clearly, all these excitations lift to anyons of different topological spin in the parent theory, so we deduce their confinement in the child theory.

A similar argument runs for the remaining excitations. That is, for any anyon not listed in \eqref{anyonlist} we can find an anyon that appears second in a pair of \eqref{anyonlist} such that from their fusion we can deduce that the former anyon belongs in the restriction of the latter anyon. This implies the confinement of all excitations, except for those that appear in the first entry of the pairs in \eqref{anyonlist}, which exactly matches the spectra of $SU(2)_{2} \times SU(2)_{1}$ shown in Table \ref{su2lv1timessu2lv2table}, as desired.

\subsection{Checking the $U(1)^{\mathrm{Orb}}_{6} \cong (SU(4)_{2} \times Spin(4)_{-4})/\mathcal{A}$ Duality} \label{orbifoldexample}

This is an important example of a Maverick coset which we will verify upon non-invertible anyon condensation gives a theory on the orbifold branch of $c=1$ theories; namely, the orbifold of the $U(1)_{6}$ free boson, which we denote $U(1)^{\mathrm{Orb}}_{6}$. This allows us to write the following Maverick duality of Chern-Simons-like theories:
\begin{equation}
    U(1)^{\mathrm{Orb}}_{6} \cong \frac{SU(4)_{2} \times Spin(4)_{-4}}{\mathcal{B}_{3}},
\end{equation}
for some appropriate gauging by an algebra object $\mathcal{B}_{3}$ that we write below. This example was studied in Section \ref{maverickdualities} above, as the first example in the second infinite family \eqref{secondmaverickfamily} of Maverick cosets.

\begin{table}[t]
\centering
\begin{tabular}[h]{|p{4cm}|p{4cm}|p{4cm}| }
\hline 
\multicolumn{3}{|c|}{$SU(4)_{2}$} \\
\hline
Line label & Quantum Dimension & Conformal Weight \\
\hline
$\mathbf{1}$ & $d_{\mathbf{1}}=1$ & $h_{\mathbf{1}}=0$ \\
$\overline{\mathbf{10}}$ & $d_{\overline{\mathbf{10}}}=1$ & $h_{\overline{\mathbf{10}}}=3/4$ \\
$\mathbf{20}'$ & $d_{\mathbf{20}'}=1$ & $h_{\mathbf{20}'}=1$ \\
$\mathbf{10}$ & $d_{\mathbf{10}}=1$ & $h_{\mathbf{10}}=3/4$ \\
$\overline{\mathbf{4}}$ & $d_{\overline{\mathbf{4}}}=\sqrt{3} $ & $h_{\overline{\mathbf{4}}}=5/16$ \\
$\mathbf{20}$ & $d_{\mathbf{20}}=\sqrt{3}$ & $h_{\mathbf{20}}=13/16$ \\
$\overline{\mathbf{20}}$ & $d_{\overline{\mathbf{20}}}=\sqrt{3}$ & $h_{\overline{\mathbf{20}}}=13/16$ \\
$\mathbf{4}$ & $d_{\mathbf{4}}=\sqrt{3}$ & $h_{\mathbf{4}}=5/16$ \\
$\mathbf{15}$ & $d_{\mathbf{15}}=2$ & $h_{\mathbf{15}}=2/3$ \\
$\mathbf{6}$ & $d_{\mathbf{6}}=2$ & $h_{\mathbf{6}}=5/12$ \\
\hline
\end{tabular}
\caption{$SU(4)_{2}$ data.}  \label{su4lv2table}
\end{table}

To check this example, we follow similar manipulations as in previous examples. The spectrum of $SU(4)_{2}$ is shown in Table \ref{su4lv2table}, and the subset of the fusion rules that we will use below can be described as follows. The line $\mathbf{10}$ generates a $\mathbf{Z}_{4}$ symmetry such that we have the orbits
\begin{align}
    &\mathbf{10}^{2} = \mathbf{20}', \quad \mathbf{10}^{3} = \overline{\mathbf{10}}, \quad \mathbf{10} \times \mathbf{15} = \mathbf{6}, \quad \mathbf{10} \times \mathbf{6} = \mathbf{15}, \\
    &\mathbf{10} \times \mathbf{4} = \overline{\mathbf{20}},  \quad \mathbf{10} \times \overline{\mathbf{20}} = \mathbf{20}, \quad \mathbf{10} \times \mathbf{20} = \overline{\mathbf{4}}, \quad \mathbf{10} \times \overline{\mathbf{4}} = \mathbf{4}.
\end{align}
The line $\mathbf{15}$ will be important and is such that
\begin{equation}
    \mathbf{15} \times \mathbf{15} = \mathbf{1} + \mathbf{15} + \mathbf{20}'.
\end{equation}
Finally, we will also need the fusion
\begin{equation}
    \mathbf{4} \times \overline{\mathbf{4}} = \mathbf{1} + \mathbf{15}.
\end{equation}
Other fusion rules may be found using the Kac software program \cite{Kac}. The spectrum of $SU(2)_{4}$ is shown in Table \ref{su2lv4table}, whose fusion rules can be read-off from Eqn. \eqref{SU2kFusionRules}. In this subsection we write $(\mathbf{R},i,j)$ where $\mathbf{R}$ labels the line in representation $\mathbf{R}$ of $SU(4)$, and the $i,j = 0, \ldots, 4$ labels the corresponding line in the $SU(2)_{-4}$ factor. The $\mathbf{10}$, $\mathbf{20'}$, and $\overline{\mathbf{10}}$ act as simple currents, so if we understand the fate of $(\mathbf{1},i,j)$, $(\mathbf{15},i,j)$, and $(\mathbf{4},i,j)$ upon gauging/condensation we can deduce the rest by acting with such simple currents. For future reference, the spectrum of $U(1)^{\mathrm{Orb}}_{6}$ is presented in Table \ref{Orb6table}.

\begin{table}[!b]
\centering
\begin{tabular}[h]{|p{4cm}|p{4cm}|p{4cm}| }
\hline 
\multicolumn{3}{|c|}{$SU(2)_{4}$} \\
\hline
Line label & Quantum Dimension & Conformal Weight \\
\hline
0 & $d_{0} = 1$ & $h_{0}=0$ \\
1 & $d_{1} = \sqrt{3}$ & $h_{1}=1/8$ \\
2 & $d_{2} = 2$ & $h_{2}=1/3$ \\
3 & $d_{3} = \sqrt{3}$ & $h_{3}=5/8$ \\
4 & $d_{4} = 1$ & $h_{4}=1$ \\
\hline
\end{tabular}
\caption{$SU(2)_{4}$ data.}  \label{su2lv4table}
\end{table}

In the product $SU(4)_{2} \times SU(2)_{-4} \times SU(2)_{-4}$ we have a set of eight abelian bosons
\begin{align}
    &(\mathbf{1},0,0), &&(\mathbf{1},0,4), &&&(\mathbf{1},4,0), &&&&(\mathbf{1},4,4), \nonumber \\[0.3cm]
    &(\mathbf{20}',0,0), &&(\mathbf{20}',0,4), &&&(\mathbf{20}',4,0), &&&&(\mathbf{20}',4,4), \nonumber
\end{align}
and a set of five non-abelian bosons
\begin{equation}
    (\mathbf{10},1,3), \quad (\mathbf{10},3,1), \quad (\mathbf{\bar{10}},1,3), \quad (\mathbf{\bar{10}},3,1), \quad (\mathbf{15},2,2).
\end{equation}
We first show that $(\mathbf{10},1,3)$ cannot condense. To show this let us assume it does and find an inconsistency. Consider the fusion
\begin{equation}
    (\mathbf{10},1,3) \times (\mathbf{\bar{10}},1,3) = (\mathbf{1},0,0) + (\mathbf{1},0,2) + (\mathbf{1},2,0) + (\mathbf{1},2,2), \label{1013fusion1013}
\end{equation}
which shows that $(\mathbf{10},1,3)$ and $(\mathbf{\bar{10}},1,3)$ are conjugates in the parent theory, so if one condenses and splits so does the other. If $(\mathbf{10},1,3)$ condenses it necessarily splits since $d_{(\mathbf{10},1,3)} = 3$, but this is inconsistent with \eqref{1013fusion1013} since there are no non-trivial bosons in the right-hand side. It follows that $(\mathbf{10},1,3)$ cannot condense. By a similar argument $(\mathbf{\bar{10}},1,3)$, $(\mathbf{10},3,1)$, and $(\mathbf{\bar{10}},3,1)$ do not condense.

\begin{table}[t]
\centering
\begin{tabular}[h]{|p{4cm}|p{4cm}|p{4cm}| }
\hline 
\multicolumn{3}{|c|}{$U(1)^{\mathrm{Orb}}_{6}$} \\
\hline
Line label & Quantum Dimension & Conformal Weight \\
\hline
0 & $d_{0}=1$ & $h_{0}=0$ \\
1 & $d_{1}=1$ & $h_{1}=3/4$ \\
2 & $d_{2}=1$ & $h_{2}=1$ \\
3 & $d_{3}=1$ & $h_{3}=3/4$ \\
4 & $d_{4}=2 $ & $h_{4}=1/12$ \\
5 & $d_{5}=2$ & $h_{5}=1/3$ \\
6 & $d_{6}=\sqrt{3}$ & $h_{6}=1/16$ \\
7 & $d_{7}=\sqrt{3}$ & $h_{7}=9/16$ \\
8 & $d_{8}=\sqrt{3}$ & $h_{8}=9/16$ \\
9 & $d_{9}=\sqrt{3}$ & $h_{9}=1/16$ \\
\hline
\end{tabular}
\caption{$U(1)^{\mathrm{Orb}}_{6}$ data.}  \label{Orb6table}
\end{table}

The only non-abelian boson that we can potentially condense is therefore $(\mathbf{15},2,2)$. But notice that we cannot condense all abelian bosons on top of this non-abelian one, since 
\begin{equation}
    (\mathbf{15},2,2) \times (\mathbf{15},2,2) = \mathrm{All \ abelian \ bosons} + (\mathbf{15},2,2) + \ldots,
\end{equation}
and the quantum dimension of $(\mathbf{15},2,2)$ is $d_{(\mathbf{15},2,2)} = 8$, meaning that it can split into at most eight labels. However, we have eight abelian bosons, so if all abelian bosons condense on top of $(\mathbf{15},2,2)$ we find nine vacua on the right-hand side of the previous fusion, leading to an inconsistency.

\begin{table}[!b]
\centering
\begin{tabular}[h]{|p{4cm}|p{4cm}|p{4cm}| }
\hline 
\multicolumn{3}{|c|}{$SO(4)_{4}$} \\
\hline
Line label & Quantum Dimension & Conformal Weight \\
\hline
(0,0) & $d_{(0,0)}=1$ & $h_{(0,0)}=0$ \\
(0,4) $\cong$ (4,0) & $d_{(0,4)}=1$ & $h_{(0,4)}=1$ \\
(2,0) $\cong$ (2,4) & $d_{(2,0)}=2$ & $h_{(2,0)}=1/3$ \\
(1,1) $\cong$ (3,3) & $d_{(1,1)}=3$ & $h_{(1,1)}=1/4$ \\
(1,3) $\cong$ (3,1) & $d_{(1,3)}=3$ & $h_{(1,3)}=3/4$ \\
(0,2) $\cong$ (4,2) & $d_{(0,2)}=2$ & $h_{(0,2))}=1/3$ \\
$(2,2)_{1}$ & $d_{(2,2)_{1}}=2$ & $h_{(2,2)_{1}}=2/3$ \\
$(2,2)_{2}$ & $d_{(2,2)_{2}}=2$ & $h_{(2,2)_{2}}=2/3$ \\
\hline
\end{tabular}
\caption{$SO(4)_{4}$ data.}  \label{so4lv4table}
\end{table}

For the sake of presentation let us first take as condensing bosons
\begin{equation} \label{condensingbosonsorbifold}
    \quad (\mathbf{1},4,4) \to 0, \quad (\mathbf{20}',0,4) \to 0, \quad (\mathbf{20}',4,0) \to 0, \quad (\mathbf{15},2,2) \to 0 + \ldots,
\end{equation}
or mathematically, the algebra object $\mathcal{B}_{3}$ given by the non-simple anyon
\begin{equation}
    \mathcal{B}_{3} = (\mathbf{1},0,0) + (\mathbf{1},4,4) + (\mathbf{20}',0,4) + (\mathbf{20}',4,0) + (\mathbf{15},2,2),
\end{equation}
and check that gauging gives rise to the spectrum of the $U(1)^{\mathrm{Orb}}_{6}$. After that we will see that other options are not consistent.

Notice that since $(\mathbf{1},4,4) \to 0$ condenses, the lines of the form $(\mathbf{1},a,b)$ arrange according to the gauging of $Spin(4)_{4}$ down to $SO(4)_{4}$. We will use this fact repeatedly later, so for the reader's convenience, we have summarized the spectrum of $SO(4)_{4}$ in terms of the lines of the parent in Table \ref{so4lv4table}. Entries that do not appear there confine, in accordance with the gauging of $Spin(4)_{4}$ to $SO(4)_{4}$. The theory $SO(4)_{4}$ can be understood easily from the three-step gauging rule \cite{Moore:1989yh, Hsin:2018vcg}, so we do not reproduce such details here. Notice that although the lists of anyons in Table \ref{so4lv4table} do not confine in $SO(4)_{4}$, their associated lines of the form $(\mathbf{1},a,b)$ in the current example may still confine because there could be additional identifications that imply their confinement.

To study how $(\mathbf{15},2,2)$ restricts we study the fusion with anyons of the form $(\mathbf{1},a,b)$ with $a$ and $b$ even:
\begin{align}
    (\mathbf{15},2,2) \times (\mathbf{1},0,2) &= (\mathbf{15},2,0) + (\mathbf{15},2,2) \label{1522fusion102}+ (\mathbf{15},2,4) \\[0.3cm]
    (\mathbf{15},2,2) \times (\mathbf{1},2,0) &= (\mathbf{15},0,2) + (\mathbf{15},2,2) + (\mathbf{15},4,2) \\[0.3cm]
    (\mathbf{15},2,2) \times (\mathbf{1},0,4) &= (\mathbf{15},2,2) \\[0.3cm]
    (\mathbf{15},2,2) \times (\mathbf{1},2,2) &= (\mathbf{15},0,0) + (\mathbf{15},0,2) + (\mathbf{15},0,4) \label{1522fusion122} \\[0.3cm]
    &+(\mathbf{15},2,0) + (\mathbf{15},2,2) + (\mathbf{15},2,4) \nonumber \\[0.3cm]
    &+(\mathbf{15},4,0) + (\mathbf{15},4,2) + (\mathbf{15},4,4). \nonumber
\end{align}

From the condensation of $Spin(4)_{4}$ to obtain $SO(4)_{4}$ we know that $(\mathbf{1},2,0)$, $(\mathbf{1},0,2)$ and $(\mathbf{1},2,2)_{1}$ are not identified with each other. For instance, if we assume $(\mathbf{1},0,2)$ and $(\mathbf{1},2,0)$ identify, we obtain that
\begin{equation}
    (\mathbf{1},0,2) \times (\mathbf{1},0,2) = (\mathbf{1},0,0) + (\mathbf{1},0,2) + (\mathbf{1},0,4),
\end{equation}
and
\begin{equation}
    (\mathbf{1},0,2) \times (\mathbf{1},2,0) = (\mathbf{1},2,2)
\end{equation}
would imply that $(\mathbf{1},2,2)$ condenses, which is not possible since $(\mathbf{1},2,2)$ does not have the correct topological spin to do so.

The set of fusions \eqref{1522fusion102}-\eqref{1522fusion122} upon restricting $(\mathbf{15},2,2)$ on the right-hand side imply that $(\mathbf{1},0,2)$, $(\mathbf{1},2,0)$, $(\mathbf{1},0,4)$, and one component of $(\mathbf{1},2,2)$ must belong to the restriction of $(\mathbf{15},2,2)$.\footnote{It is straightforward to check by computing self-fusions that $(\mathbf{1},0,2)$, $(\mathbf{1},2,0)$, $(\mathbf{1},0,4)$ do not split.} Using also the previous argument that $(\mathbf{1},0,2)$, $(\mathbf{1},2,0)$, and  $(\mathbf{1},2,2)$ cannot identify with each other, we obtain that the restriction of $(\mathbf{15},2,2)$ takes the form
\begin{equation} \label{1522restriction}
    (\mathbf{15},2,2) \longrightarrow 0 + (\mathbf{15},2,2)_{2} + (\mathbf{15},2,2)_{3} + (\mathbf{15},2,2)_{4} + (\mathbf{15},2,2)_{4},
\end{equation}
with
\begin{align}
    &(\mathbf{15},2,2)_{2} \cong (\mathbf{1},0,4) \cong (\mathbf{1},4,0), \quad d_{(\mathbf{1},0,4)} = 1, \\[0.3cm] &(\mathbf{15},2,2)_{3} \cong (\mathbf{1},0,2) \cong (\mathbf{1},4,2), \quad d_{(\mathbf{1},0,2)} = 2, \\[0.3cm] &(\mathbf{15},2,2)_{4} \cong (\mathbf{1},2,0) \cong (\mathbf{1},2,4), \quad d_{(\mathbf{1},2,0)} = 2, \\[0.3cm] &(\mathbf{15},2,2)_{5} \cong (\mathbf{1},2,2)_{1}, \hspace{2cm} d_{(\mathbf{1},2,2)_{1}} = 2,
\end{align}
with no further splittings since we have saturated the conservation of quantum dimension. In the restriction above we had to make a choice between $(\mathbf{1},2,2)_{1}$ or $(\mathbf{1},2,2)_{2}$ to appear in the restriction of $(\mathbf{15},2,2)$. Since in $SO(4)_{4}$ the lines $(2,2)_{1}$ and $(2,2)_{2}$ are symmetric between each other, the choice is actually immaterial, and by definition we have chosen $(\mathbf{1},2,2)_{1}$ to be the one appearing in the restriction of $(\mathbf{15},2,2)$.

In passing, let us note that now $(\mathbf{1},0,2)$, $(\mathbf{1},2,0)$, and $(\mathbf{1},2,2)_{1}$ have an additional lift to $(\mathbf{15},2,2)$, which implies that while $(\mathbf{1},0,2)$, $(\mathbf{1},2,0)$, and $(\mathbf{1},2,2)_{1}$ were unconfined in the $SO(4)_{4}$ theory, in the current example they confine. 

Next let us show that $(\mathbf{10},3,1)$ confines. This is easily seen by acting with our condensing abelian bosons:
\begin{align}
    (\mathbf{1},4,4) \times (\mathbf{10},3,1) &= (\mathbf{10},1,3) \label{144action1031} \\[0.3cm]
    (\mathbf{20}',0,4) \times (\mathbf{10},3,1) &= (\overline{\mathbf{10}},3,3) \\[0.3cm]
    (\mathbf{20}',4,0) \times (\mathbf{10},3,1) &= (\overline{\mathbf{10}},1,1). \label{2040action1031}
\end{align}
Using that the abelian bosons condense, we deduce the identifications $(\mathbf{10},3,1) \cong (\mathbf{10},1,3) \cong (\overline{\mathbf{10}},1,1) \cong (\overline{\mathbf{10}},3,3)$. From this is easy to see that $(\mathbf{10},3,1)$ and the rest of the labels on the list confine. A completely analogous argument 
allows us to deduce the identifications
\begin{align}
    (\mathbf{10},1,1) \cong (\mathbf{10},3,3) \cong (\overline{\mathbf{10}},1,3) \cong (\overline{\mathbf{10}},3,1),& \label{1011ids} \\[0.3cm]
    (\mathbf{1},1,1) \cong (\mathbf{1},3,3) \cong (\mathbf{20}',1,3) \cong (\mathbf{20}',3,1),& \\[0.3cm]
    (\mathbf{1},1,3) \cong (\mathbf{1},3,1) \cong (\mathbf{20}',1,1) \cong (\mathbf{20}',3,3),& \label{113ids}
\end{align}
from which in turn we can deduce the corresponding confinement of all the labels in the lists above. The remaining anyons of the form $(\mathbf{1},i,j)$ that we have not treated here all confine, as they already confined as $(i,j)$ in $SO(4)_{4}$.

In summary, the anyons of the form $(\mathbf{1},i,j)$ that do not confine are $(\mathbf{1},0,0)$, $(\mathbf{1},0,4)$, and $(\mathbf{1},2,2)_{2}$. It is straightforward to see that they match the anyons labelled by 0, 2 and 5 respectively in Table \ref{Orb6table} showing the spectrum of the $U(1)^{\mathrm{Orb}}_{6}$ theory, both in their conformal weight as in their quantum dimensions. As another check, the fusion rules for these lines in the orbifold theory are indeed the same as those of the corresponding lines in the $SO(4)_{4}$ theory. 

We now move to work out lines of the form $(\mathbf{15},a,b)$. Notice that the condensing abelian bosons relate $(\mathbf{15},a,b)$ and $(\mathbf{6},a,b)$, so studying the former fixes the latter. Start by analyzing $(\mathbf{15},0,0)$:
\begin{equation}
    (\mathbf{15},0,0) \times (\mathbf{15},0,0) = (\mathbf{1},0,0) + (\mathbf{15},0,0) + (\mathbf{20}',0,0).
\end{equation}
Thus $(\mathbf{15},0,0)$ does not split since $(\mathbf{20}',0,0)$ does not condense. From
\begin{equation}
    (\mathbf{15},0,0) \times (\mathbf{15},2,2) = (\mathbf{15},2,2) + \ldots
\end{equation}
we deduce that upon restricting $(\mathbf{15},2,2)$ on the right-hand side, since $(\mathbf{15},0,0)$ is self-conjugate, it belongs to the restriction of $(\mathbf{15},2,2)$. In particular, it must be identified with one of the quantum dimension two components of $(\mathbf{15},2,2)$. As such, it follows that $(\mathbf{15},0,0)$ confines.

To figure out what component we have to identify $(\mathbf{15},0,0)$ with precisely, consider
\begin{align}
    (\mathbf{15},2,2) = (\mathbf{15},0,0) \times (\mathbf{1},2,2) &\longrightarrow (\mathbf{15},0,0) \times (\mathbf{1},2,2)_{1} +  (\mathbf{15},0,0) \times (\mathbf{1},2,2)_{2} \nonumber \\[0.3cm] &\longrightarrow 0 \ldots,
\end{align}
where in the upper arrow we have restricted the right-hand side, while in the lower arrow we have restricted the left-hand side. Since $(\mathbf{1},2,2)_{1}$ and $(\mathbf{1},2,2)_{2}$ are self-conjugate, and $(\mathbf{1},2,2)_{1} \in (\mathbf{15},2,2)$ but $(\mathbf{1},2,2)_{2} \notin (\mathbf{15},2,2)$ it must be that
\begin{equation} \label{1500identification}
    (\mathbf{15},0,0) \cong (\mathbf{1},2,2)_{1}.
\end{equation}
In passing, acting with the condensing abelian bosons as in \eqref{144action1031}-\eqref{2040action1031} and \eqref{1011ids}-\eqref{113ids}, we obtain the additional identifications
\begin{equation} \label{idsof1500}
    (\mathbf{15},0,0) \cong (\mathbf{1},2,2)_{1} \cong (\mathbf{15},4,4) \cong (\mathbf{15},0,4) \cong (\mathbf{15},4,0).
\end{equation}

With the result \eqref{113ids} is straightforward to extract the content of the lines of the form $(\mathbf{15},a,b)$, since we can do
\begin{equation}
    (\mathbf{15},a,b) = (\mathbf{15},0,0) \times (\mathbf{1},a,b) \cong (\mathbf{1},2,2)_{1} \times (\mathbf{1},a,b),
\end{equation}
and then we can proceed to use the fusion rules of $SO(4)_{4}$ to derive the splitting of $(\mathbf{15},a,b)$ in terms of $(\mathbf{1},a,b)$'s which we already know. Of course, this means that $(\mathbf{15},a,b)$ will not give us any new lines in the spectrum of the child theory, since all of them can be identified in terms of lines that we have already considered. To check that none of the unconfined lines already obtained confine upon lifting them to $(\mathbf{15},a,b)$'s it is useful to know the explicit results
\begin{align}
    (\mathbf{15},0,2)& \longrightarrow (\mathbf{1},2,0) + (\mathbf{1},2,2)_{2}, \\[0.3cm]
    (\mathbf{15},2,0)& \longrightarrow (\mathbf{1},0,2) + (\mathbf{1},2,2)_{2}, \\[0.3cm]
    (\mathbf{15},4,2)& \longrightarrow (\mathbf{1},2,0) + (\mathbf{1},2,2)_{2}, \\[0.3cm]
    (\mathbf{15},2,4)& \longrightarrow (\mathbf{1},0,2) + (\mathbf{1},2,2)_{2}, \\[0.3cm]
    (\mathbf{1},2,2)&  \longrightarrow (\mathbf{1},2,2)_{1} + (\mathbf{1},2,2)_{2}, \\[0.3cm]
    (\mathbf{20}',2,2)& \longrightarrow (\mathbf{1},2,2)_{1} + (\mathbf{1},2,2)_{2}. 
\end{align}
Thus, the unconfined lines on the right-hand-side of $(\mathbf{15},0,2)$, $(\mathbf{15},2,0)$, $(\mathbf{15},2,4)$, $(\mathbf{15},4,2)$, $(\mathbf{1},2,2)$, and $(\mathbf{20}',2,2)$ are all the same; namely $(\mathbf{1},2,2)_{2}$, which is easy to see that it does not confine when considering the new splittings above, as all left-hand sides share the same topological spin. Similarly, from the fusion rules of $SU(2)_{4} \times SU(2)_{4}$ it is straightforward to check that $(\mathbf{1},0,4)$, $(\mathbf{1},4,0)$ are never contained in  $(\mathbf{1},2,2) \times (\mathbf{1},a,b)$, for any $a,b$ other than $a=b=2$, and thus also not contained in $(\mathbf{15},a,b) = (\mathbf{1},2,2)_{1} \times (\mathbf{1},a,b)$, so there are no new liftings that could confine $(\mathbf{1},0,4)$.

Before working-out the lines of the form $(\mathbf{4},a,b)$, let us apply the simple currents $(\overline{\mathbf{10}},0,0)$ to the unconfined spectra found thus far consisting of $(\mathbf{1},0,0)$, $(\mathbf{1},0,4)$ and $(\mathbf{1},2,2)_{1}$. We find
\begin{equation}
    (\overline{\mathbf{10}},0,0) \times (\mathbf{1},0,0) = (\overline{\mathbf{10}},0,0),
\end{equation}
\begin{equation}
    (\overline{\mathbf{10}},0,0) \times (\mathbf{1},0,4) = (\overline{\mathbf{10}},0,4),
\end{equation}
\begin{equation}
    (\overline{\mathbf{10}},0,0) \times (\mathbf{1},2,2)_{2} = (\overline{\mathbf{10}},2,2)_{2}.
\end{equation}
It is straightforward to see that further actions of the simple current will just permute the four lines on the right-hand sides above. Of course, we can apply $(\overline{\mathbf{10}},0,0)$ to all the rest of the lines we have found before, but this would either give confined sectors, or unconfined ones that are related to the lines on the right-hand side above by identifications already obtained at the level of $(\mathbf{1},a,b)$'s and $(\mathbf{15},a,b)$'s. Thus the action of the simple currents provide us with three new unconfined excitations in the child theory: $(\overline{\mathbf{10}},0,0)$, $(\overline{\mathbf{10}},0,4)$, and $(\overline{\mathbf{10}},2,2)_{2}$. We can easily check that they match, both in topological spin and quantum dimension to the lines labelled as 1, 3 and 4 in Table \ref{Orb6table} outlining the spectrum of $U(1)^{\mathrm{Orb}}_{6}$.

We move now to study the restriction of lines of the form $(\mathbf{4},a,b)$. First, notice that using our condensing abelian bosons as in \eqref{144action1031}-\eqref{2040action1031}, \eqref{1011ids}-\eqref{113ids}, or as in Eqn. \eqref{idsof1500} we can derive the identifications
\begin{equation}
    (\mathbf{4}, 1, 1) \cong (\mathbf{4},3,3) \cong (\mathbf{20},1,3) \cong (\mathbf{20},3,1),
\end{equation}
\begin{equation}
    (\mathbf{4}, 1, 3) \cong (\mathbf{4},3,1) \cong (\mathbf{20},1,1) \cong (\mathbf{20},3,3),
\end{equation}
\begin{equation}
    (\mathbf{\bar{4}},1,1) \cong (\mathbf{\bar{4}},3,3) \cong (\overline{\mathbf{20}},1,3) \cong (\overline{\mathbf{20}},3,1),
\end{equation}
\begin{equation}
    (\mathbf{\bar{4}},1,3) \cong (\mathbf{\bar{4}},3,1) \cong (\overline{\mathbf{20}},1,1) \cong (\overline{\mathbf{20}},3,3).
\end{equation}
The identifications imply that we can concentrate our attention to the leftmost anyons in each line. It is easy to check that these identification do not lead to the confinement of any of the anyons involved.

Consider now the fusion rule
\begin{align} \label{413fusionbar413}
    (\mathbf{4},1,3) \times (\mathbf{\bar{4}},1,3) & = (\mathbf{1},0,0) + (\mathbf{1},0,2) + (\mathbf{1},2,0) + (\mathbf{1},2,2) \\[0.3cm] 
    &+ (\mathbf{15},0,0) + (\mathbf{15},0,2) + (\mathbf{15},2,0) + (\mathbf{15},2,2), \nonumber
\end{align}
which upon restriction implies the splittings
\begin{equation}
    (\mathbf{4},1,3) \longrightarrow (\mathbf{4},1,3)_{1} + (\mathbf{4},1,3)_{2},
\end{equation}
\begin{equation}
    (\mathbf{\bar{4}},1,3) \longrightarrow \overline{(\mathbf{4},1,3)_{1}} + \overline{(\mathbf{4},1,3)_{2}},
\end{equation}
and a similar fusion implies the splittings $(\mathbf{4},1,1) \to(\mathbf{4},1,1)_{1} + (\mathbf{4},1,1)_{2}$, $(\mathbf{\bar{4}},1,1) \to \overline{(\mathbf{4},1,1)_{1}} + \overline{(\mathbf{4},1,1)_{2}}$. We can now consider the crossed fusion rule
\begin{equation}
    (\mathbf{4},1,3) \times (\mathbf{\bar{4}},1,1) = (\mathbf{15},2,2) + \ldots \longrightarrow 0+\ldots,
\end{equation}
which implies that one component of $(\mathbf{4},1,3)$ identifies with one component of $(\mathbf{\bar{4}},1,1)$. Let us define the subindex 2 in the previous splitting to be such that this holds. We find then
\begin{align}
    &(\mathbf{4},1,1) \longrightarrow (\mathbf{4},1,1)_{1} + (\mathbf{4},1,1)_{2}, \\[0.3cm]
    &(\mathbf{4},1,3) \longrightarrow (\mathbf{4},1,3)_{1} + (\mathbf{4},1,1)_{2}, \\[0.3cm]
    &(\mathbf{\bar{4}},1,1) \longrightarrow \overline{(\mathbf{4},1,1)_{1}} + \overline{(\mathbf{4},1,1)_{2}}, \\[0.3cm]
    &(\mathbf{\bar{4}},1,3) \longrightarrow \overline{(\mathbf{4},1,3)_{1}} + \overline{(\mathbf{4},1,1)_{2}} ,
\end{align}
from which it is easy to see that $(\mathbf{4},1,1)_{2}$ confines. Because of this, the lines with subindex 1 in the previous restrictions share the same quantum dimension.

Next we must determine the assignment of quantum dimensions. The easiest way to do this at this point is to use the formula \eqref{ChiralCentralChargeMTCFormula} relating the central charge with topological spins and quantum dimensions:
\begin{equation}
    e^{i\frac{\pi}{4}c} = \frac{1}{\sqrt{\sum_{i}d_{i}^{2}}}\sum_{i}d^{2}_{i}\theta_{i},
\end{equation}
and plugging in $c=1$ and the values for the unconfined excitations, whose topological spins and quantum dimensions we know (except of course for the quantum dimension that we want to determine). The equation can then be solved for the remaining unknown quantum dimension, which we assign to the unconfined excitations above. The result is that the assignments must be $d_{(\mathbf{4},1,1)_{1}} = \sqrt{3}$ and $d_{(\mathbf{4},1,1)_{2}} = 2\sqrt{3}$.

A slickly argument that does not involve using an additional formula nor using the central charge or other quantities as input, but that uses the same manipulations with fusions and quantum dimensions that we have used until now is given as follows. Split the quantum dimensions as:
\begin{equation}
    3\sqrt{3} = \frac{3\sqrt{3}}{a} + \frac{3\sqrt{3}}{a}(a-1), \quad a>1,
\end{equation}
and recall that quantum dimensions satisfy the fusion algebra as in Eqn. \eqref{FusionAlgebraofQuantumDims}, meaning that the assignments of quantum dimensions must be consistent with the fusion \eqref{413fusionbar413}.\footnote{In \eqref{413fusionbar413} we have written the fusion of $(\mathbf{4},1,3)$ and its conjugate, but an analog fusion holds for $(\mathbf{4},1,1)$ and its conjugate.} However, we have already determined the quantum dimensions on the right-hand side of \eqref{413fusionbar413}, and they are given by 1s and 2s. It is actually sufficient to know that the right-hand side of \eqref{413fusionbar413} gives an integer quantum dimension for any fusion product on the left-hand side.

Thus, we must have that $27/a^{2} \in \mathbb{N}_{>0}$, and $(a-1)27/a^{2} \in \mathbb{N}_{>0}$, where we may think of these two conditions as arising from the $(\mathbf{4},1,1)_{1} \times (\mathbf{4},1,1)_{1}$ and $(\mathbf{4},1,1)_{1} \times (\mathbf{4},1,1)_{2}$ fusion products, respectively. The first condition gives that $a=3\sqrt{3}/\sqrt{n}$ for some positive integer $n$, and on the second condition this gives $(3\sqrt{3n}-n) \in \mathbb{N}_{>0}$. It is now direct to see that the only possible solutions are $n = 3, 12$, which lead however to the same splitting of the quantum dimension:
\begin{equation}
    3\sqrt{3} = \sqrt{3} + 2\sqrt{3}.
\end{equation}
To decide which quantum dimension to assign to $(\mathbf{4},1,1)_{1}$, notice that the fusion product of two genuine line operators must be genuine line operators. This means that in \eqref{413fusionbar413}, the product of the quantum dimensions in the fusion $(\mathbf{4},1,1)_{1} \times (\mathbf{4},1,1)_{1}$ is bounded by the sum of the quantum dimensions of the unconfined excitations on the right-hand side. It is straightforward to check that the latter sums to nine, so the only consistent assignment is the one we found above: $d_{(\mathbf{4},1,1)_{1}} = \sqrt{3}$ and $d_{(\mathbf{4},1,1)_{2}} = 2\sqrt{3}$.

In summary, the unconfined excitations that we have found in the $(\mathbf{4},i,j)$ sector are $(\mathbf{4},1,1)_{1}$, $(\mathbf{4},1,3)_{1}$, and their conjugates. All of them have the same quantum dimension $d_{(\mathbf{4},1,1)_{1}} = \sqrt{3}$. We can check now that this matches with the lines 6,7,8,9 in the $U(1)^{\mathrm{Orb}}_{6}$ spectrum shown in Table \ref{Orb6table}.

Using the condensing abelian bosons as above is easy to show that the remaining anyons of the form $(\mathbf{4},i,j)$ confine. For example, $(\mathbf{4},0,0) \times (\mathbf{20}',0,4) = (\mathbf{20},0,4)$ and $(\mathbf{20}',0,4) \to 0$, but $(\mathbf{4},0,0)$ has topological spin $e^{2 \pi i 5/16}$ while $(\mathbf{20},0,4)$ has topological spin $e^{2 \pi i 13/16}$, implying that $(\mathbf{4},0,0)$ confines.

\subsubsection*{Inconsistency of Condensing Different Abelian Bosons}

We conclude this subsection indicating how making a putative different choice of condensing bosons (including the non-abelian one) other than \eqref{condensingbosonsorbifold} leads to an inconsistency.

We have already shown that condensing all abelian bosons leads to an inconsistency, so we need to take a subset of them closed under fusion. First suppose we were to take $(\mathbf{1},4,4)$, $(\mathbf{20}',0,0)$, $(\mathbf{20}',4,4)$ to condense on top of the non-abelian boson $(\mathbf{15},2,2)$. Since $(\mathbf{20}',0,0)$ condenses:
\begin{equation} \label{1500fusion1500}
    (\mathbf{15},0,0) \times (\mathbf{15},0,0) = (\mathbf{1},0,0) + (\mathbf{20}',0,0) + (\mathbf{15},0,0) \longrightarrow 0 + 0 + \ldots,
\end{equation}
so $(\mathbf{15},0,0)$ must split into two components each with unit quantum dimension. However, because $(\mathbf{1},4,4) \to 0$, $(\mathbf{1},2,2) \to (\mathbf{1},2,2)_{1} + (\mathbf{1},2,2)_{2}$ each with quantum dimension 2, since this follows from the gauging of $Spin(4)_{4}$ to $SO(4)_{4}$. This is inconsistent since the left-hand side of
\begin{equation} \label{1500fusion122}
    (\mathbf{15},2,2) = (\mathbf{15},0,0) \times (\mathbf{1},2,2).
\end{equation}
condenses, but the right hand side decomposes into fusions of components of quantum dimensions one and two. Since conjugates need to have the same quantum dimension, there is no way to accommodate a pair of conjugates that fuse to a vacuum.

Now suppose we were to take $(\mathbf{1},4,4)$, $(\mathbf{1},0,4)$, $(\mathbf{1},4,0)$ as condensing abelian bosons. From \eqref{1500fusion1500} we see that $(\mathbf{15},0,0)$ does not split in this case. However now
\begin{equation}
    (\mathbf{1},2,2) \times (\mathbf{1},2,2) = (\mathbf{1},0,0) + (\mathbf{1},0,4) + (\mathbf{1},4,0) + (\mathbf{1},4,4) + \ldots \longrightarrow 0+0+0+0+\ldots,
\end{equation}
implies $(\mathbf{1},2,2)$ splits into four unit quantum dimension components. The same argument as before using \eqref{1500fusion122} implies an inconsistency.

The case when we try to condense $(\mathbf{1},0,4)$, $(\mathbf{20}',4,0)$ and $(\mathbf{20}',4,4)$ is a bit different. First, notice that if we take $(\mathbf{1},0,4)$ to condense then the right $SU(2)_{4}$ factor condenses to $SU(3)_{1}$, and $(\mathbf{1},i,2) \to (\mathbf{1},i,2_{1}) + (\mathbf{1},i,2_{2})$, for any $i=0 , \ldots , 4$. We now observe that the fusion products of $(\mathbf{15},2,2)$ with $(\mathbf{1},2,0)$, $(\mathbf{1},4,0)$, $(\mathbf{1},0,2)$, $(\mathbf{1},2,2)$ and $(\mathbf{1},4,2)$ all will have a $(\mathbf{15},2,2)$ on the right-hand side and thus will have a vacuum after restriction. In particular it follows that $(\mathbf{1},4,0), (\mathbf{1},4,2_{1}), (\mathbf{1},2,0) \in (\mathbf{15},2,2)$. \footnote{Since the fusion rules in $SU(3)_{1}$ are symmetric between $2_{1}$ and $2_{2}$ we could have chosen $(\mathbf{1},4,2_{2})$ here instead. It is easy to see that the same conclusions follow.} It is easy to check that $(\mathbf{1},2,0)$ splits, and notice that $(\mathbf{1},4,0)$ cannot possibly be identified with $(\mathbf{1},4,2_{1})$, as if this was the case we could fuse both sides with $(\mathbf{1},4,0)$ obtaining an identification of $(\mathbf{1},4,2_{1})$ with $(\mathbf{1},0,0)$ which is not possible since $(\mathbf{1},4,2_{1})$ does not have the topological spin to condense. It follows that the restriction of $(\mathbf{15},2,2)$ must be of the form
\begin{equation}
    (\mathbf{15},2,2) \longrightarrow 0 + (\mathbf{1},4,0) + (\mathbf{1},4,2_{1}) + (\mathbf{1}, 2, 0) + b,
\end{equation}
for some $b$ with quantum dimension $d_{b} = 3$. However we now take that $(\mathbf{1},2,2_{1})$ must also be in the restriction $(\mathbf{15},2,2)$, and the only candidate that matches the quantum dimension above is $(\mathbf{1}, 2, 0)$, which is however inconsistent since identifying $(\mathbf{1}, 2, 0)$ with $(\mathbf{1},2,2_{1})$ implies upon fusing with $(\mathbf{1}, 2, 0)$ that $(\mathbf{1},0,0) + (\mathbf{1},2,0) + (\mathbf{1},4,0)$ is identified with $(\mathbf{1},0,2_{1}) + (\mathbf{1},2,2_{1}) + (\mathbf{1},4,2_{1})$, but none of these last anyons can condense, thus finding an inconsistency.

Finally, had we taken a single abelian boson to condense on top of the non-abelian one, $(\mathbf{15},2,2)$ would have split into just three components. Either $(\mathbf{1},4,0)$ or $(\mathbf{1},0,4)$ would not condense and it would belong in the restriction of $(\mathbf{15},2,2)$. So, the third anyon in the restriction would have quantum dimension 6. It is then easily seen that either $(\mathbf{1},2,0)$ or $(\mathbf{1},0,2)$ do not split and belong to $(\mathbf{15},2,2)$, but there is no component in the decomposition that matches the quantum dimension, leading to an inconsistency.

\section{Revisiting Conformal Embeddings and Level-Rank Dualities} \label{conformalembeddings}

In this section we revisit the conformal embeddings of \cite{Hsin:2016blu, Aharony:2016jvv, Cordova:2017vab, Cordova:2018qvg} and their relation to gauging of non-invertible symmetries. We have already touched upon this matter in Section \ref{noninvertibleextensionsubsection}, and here we will generalize that discussion to other conformal embeddings. In particular, in the case of the exceptional conformal embeddings of \cite{Cordova:2018qvg}, we will see that running an analogous argument to that of the classical embeddings of \cite{Hsin:2016blu, Aharony:2016jvv, Cordova:2017vab} will already lead us to the consideration of non-abelian anyon condensation to make the dualities work.

\subsection{Revisiting Classical Conformal Embeddings}

Let us start motivating our discussion with the simple case of that of the unitary series of conformal embeddings
\begin{equation}
    SU(N)_{k} \times SU(k)_{N} \hookrightarrow SU(Nk)_{1}, \quad N,k \in \mathbb{N}_{\geq 2},
\end{equation}
whose standard 3D TQFT duality interpretation, as reviewed at the end of Section \ref{cosetinterfacebulkboundaryreview}, is \cite{Hsin:2016blu}:
\begin{equation} \label{standardunitaryduality}
    SU(k)_{N} \cong \frac{SU(Nk)_{1} \times SU(N)_{-k}}{\mathbb{Z}_{N}}.
\end{equation}

To relate conformal embeddings to the gauging of non-invertible symmetries, we point out that the following alternative results also applies:
\begin{equation} \label{unitarynoninvertbile}
    \boxed{SU(Nk)_{1} \cong \frac{SU(N)_{k} \times SU(k)_{N}}{\mathcal{A}_{N,k}}} 
\end{equation}
where $\mathcal{A}_{N,k}$ is some gaugable algebra object (for brevity, on the rest of this section when we write a letter in calligraphic font we always mean some appropriate algebra object that we can use to gauge, and from now on we will not remark on what such objects stand for). The previous statement has been discussed in many mathematical references \cite{Frohlich:2003hm, Davydov:2011pp,Huang:2005gs, Huang:2014ixa, kong2014anyon, davydov2013witt}, but in our context it is most quickly understood from the coset inversion formula \eqref{MTCEmbedding} with $\mathcal{M} = SU(Nk)_{1}$, $\mathcal{C} = SU(k)_{N}$ and $\mathcal{M}' = SU(N)_{k}$. In this sense, \eqref{standardunitaryduality} is nothing but the standard form of cosets, Eqn. \eqref{IsolatingMTCcoset} of the coset inversion formulas, while \eqref{unitarynoninvertbile} is ``the parent statement'' Eqn. \eqref{MTCEmbedding} translated to a form more akin to physics, and for the particular example at hand. Clearly, the unitary groups play no role in the previous argument, and as such the same would hold for any other conformal embeddings. We explore this in the next sections below.

As we will see shortly, in the form \eqref{unitarynoninvertbile} the algebra object that we need to gauge is generically that of a non-invertible symmetry, although in particular cases it may simplify to some abelian gauging. For instance, it is easy to see that if we apply this form of the conformal embedding of unitary groups for $N=k=2$ we obtain
\begin{equation} \label{422}
    SU(4)_{1} \cong \frac{SU(2)_{2} \times SU(2)_{2}}{\mathbb{Z}_{2}},
\end{equation}
where the $\mathbb{Z}_{2}$ algebra is given by $\mathcal{A} = (0,0) + (2,2)$, where $(i,j)$ stands for the spin $i/2$ and $j/2$ representations of each $SU(2)_{2}$ factor. Eqn. \eqref{422} can be verified by a simple use of the three-step gauging procedure \cite{Moore:1989yh}. 

The simplest instance where we need to consider gauging by a non-invertible one-form symmetry to make \eqref{unitarynoninvertbile} valid occurs at $N=3$ and $k=2$, which we verify by a direct non-abelian anyon condensation computation on the following example. Further conformal embeddings will be studied in Section \ref{continuingclassicalembeddings} and Section \ref{FurtherConformalEmbeddings} below once we finish illustrating this example.

\subsubsection{Checking the $SU(6)_{1} \cong (SU(3)_{2} \times SU(2)_{3})/\mathcal{A}$ Duality} \label{unitaryexample}

In this subsection we check by direct computation that $SU(6)_{1}$ can be found as a non-abelian anyon condensation of $SU(3)_{2} \times SU(2)_{3}$. This is the simplest example in the infinite tower of conformal embeddings
\begin{equation}
    SU(N)_{k} \times SU(k)_{N} \hookrightarrow SU(Nk)_{1}, \quad N,k \in \mathbb{N}_{\geq 2},
\end{equation}
where we can think of $SU(Nk)_{1}$ as a non-abelian anyon condensation of $SU(N)_{k} \times SU(k)_{N}$. Such conformal embeddings are particularly interesting because of their intimate relation to level-rank dualities \cite{Hsin:2016blu}.

We follow the same procedure as in Section \ref{ExplicitChecks}, so the reader may find useful to read that section first before going through this example. The spectrum of $SU(2)_{3}$ is given in Table \ref{su2lv3table}, while the spectrum of $SU(3)_{2}$ is given in Table \ref{su3lv2table}. In the following, we write $(\mathbf{R},i)$ for a line in $SU(3)_{2} \times SU(2)_{3}$, where $\mathbf{R}$ labels a line in representation $\mathbf{R}$ of $SU(3)$ and $i=0,1,2,3$ labels the corresponding line in $SU(2)_{3}$. We will need the fusion rules for $SU(2)_{k}$ which can be found above in Eqn. \eqref{SU2kFusionRules}, and the fusion rules for $SU(3)_{2}$:
\begin{equation*}
    \mathbf{3} \times \mathbf{3} = \bar{\mathbf{3}} + \mathbf{6}, \quad \mathbf{3} \times \bar{\mathbf{3}} = \mathbf{1} + \mathbf{8}, \quad \mathbf{3} \times \mathbf{8} = \mathbf{3} + \bar{\mathbf{6}}, \quad \mathbf{3} \times \mathbf{6} = \bar{\mathbf{3}}, \quad \mathbf{3} \times \bar{\mathbf{6}} = \mathbf{8}
\end{equation*}
\begin{equation*}
    \bar{\mathbf{3}} \times \bar{\mathbf{3}} = \mathbf{3} + \bar{\mathbf{6}}, \quad \bar{\mathbf{3}} \times \mathbf{8} = \bar{\mathbf{3}} + \mathbf{6}, \quad \bar{\mathbf{3}} \times \mathbf{6} = \mathbf{3} , \quad \bar{\mathbf{3}} \times \bar{\mathbf{6}} = \mathbf{8}, 
\end{equation*}
\begin{equation*}
    \mathbf{8} \times \mathbf{8} = \mathbf{1} + \mathbf{8}, \quad \mathbf{8} \times \mathbf{6} = \bar{\mathbf{3}}, \quad \mathbf{8} \times \bar{\mathbf{6}} = \mathbf{3},
\end{equation*}
\begin{equation*}
    \mathbf{6} \times \mathbf{6} = \bar{\mathbf{6}} , \quad \mathbf{6} \times \bar{\mathbf{6}} = \mathbf{1}, \quad  \bar{\mathbf{6}} \times \bar{\mathbf{6}} = \mathbf{6}.
\end{equation*}

\begin{table}[t]
\centering
\begin{tabular}[h]{|p{4cm}|p{4cm}|p{4cm}| }
\hline 
\multicolumn{3}{|c|}{$SU(3)_{2}$} \\
\hline
Line label & Quantum Dimension & Conformal Weight \\
\hline
$\mathbf{1}$ & $d_{\mathbf{1}}=1$ & $h_{\mathbf{1}}=0$ \\
$\mathbf{3}$ & $d_{\mathbf{3}}=\frac{1+\sqrt{5}}{2}$ & $h_{\mathbf{3}}=4/15$ \\
$\mathbf{\bar{3}}$ & $d_{\mathbf{\bar{3}}}=\frac{1+\sqrt{5}}{2}$ & $h_{\mathbf{\bar{3}}}=4/15$ \\
$\mathbf{8}$ & $d_{\mathbf{8}}=\frac{1+\sqrt{5}}{2}$ & $h_{\mathbf{8}}=3/5$ \\
$\mathbf{6}$ & $d_{\mathbf{6}}=1 $ & $h_{\mathbf{6}}=2/3$ \\
$\mathbf{\bar{6}}$ & $d_{\mathbf{\bar{6}}}=1$ & $h_{\mathbf{\bar{6}}}=2/3$ \\
\hline
\end{tabular}
\caption{$SU(3)_{2}$ data.}  \label{su3lv2table}
\end{table}

Notice that in the spectrum of $SU(3)_{2}$ the lines $\mathbf{6}$ and $\mathbf{\bar{6}}$ act as $\mathbb{Z}_{3}$ simple currents, meaning that if we know the fate of $(\mathbf{1},i)$ and $(\mathbf{8},i)$ upon gauging/condensation, we can deduce that of $(\mathbf{6},i)$, $(\mathbf{3},i)$, and their conjugates by acting with $(\mathbf{6},0)$ and $(\mathbf{\bar{6}},0)$. Thus, we concentrate on $(\mathbf{1},i)$ and $(\mathbf{8},i)$.

To begin, observe that there is only one non-trivial boson in the product theory $SU(3)_{2} \times SU(2)_{3}$; namely, the boson $(\mathbf{8},2)$, with quantum dimension $d_{(\mathbf{8},2)} = \frac{3+\sqrt{5}}{2}$. Let us assume this boson condenses, which has to be the case since it would be the only way to obtain $SU(6)_{1}$ from a condensation of $SU(3)_{2} \times SU(2)_{3}$.

Assuming that $(\mathbf{8},2)$ condenses is the statement that $(\mathbf{8},2)$ restricts as 
\begin{equation} \label{(8,2)restriction}
    (\mathbf{8},2) \to 0 + (\mathbf{8},2)_{2} \, ,
\end{equation}
where by conservation of the quantum dimension, $d_{(\mathbf{8},2)_{2}} = \frac{1+\sqrt{5}}{2}$. This is too small to allow a further splitting, so $(\mathbf{8},2)$ must restrict to just two components.

Let us now notice that the line $(\mathbf{1},2)$ cannot split as it does not have large enough quantum dimension to do so. With this observation, consider
\begin{equation}
    (\mathbf{8},2) \times (\mathbf{1},2) = (\mathbf{8},0) + (\mathbf{8},2) \longrightarrow 0 + \ldots,
\end{equation}
which shows that $(\mathbf{1},2) \in (\mathbf{8},2)$ since $(\mathbf{1},2)$ is self-conjugate. Matching quantum dimensions the only possibility is to have the identification $(\mathbf{1},2) \cong (\mathbf{8},2)_{2}$, which in turn implies the confinement of these components since they lift to anyons in the parent theory with different topological spins.

To study the result of $(\mathbf{1},1)$ and $(\mathbf{1},3)$ after condensation, notice that these lines cannot split since they do not have large enough quantum dimension, and consider the fusion rules
\begin{equation}
    (\mathbf{8},2) \times (\mathbf{1},1) = (\mathbf{8},1) + (\mathbf{8},3), \label{fusionrule1}
\end{equation}
and
\begin{align}
    (\mathbf{8},2) \times (\mathbf{1},3) &= (\mathbf{8},1) \\[0.3cm] &= (\mathbf{1},3) + (\mathbf{1},2) \times (\mathbf{1},3) =(\mathbf{1},3) + (\mathbf{1},1), \label{fusionrule2}
\end{align}
where in the last expression the first equality comes from doing the standard fusion on the parent theory and the second line comes from first restricting $(\mathbf{8},2) \to 0 + (\mathbf{1},2)$ on the left-hand side, and then performing the fusion of these components with $(\mathbf{1},3)$. Comparing both expressions, we obtain the restriction of $(\mathbf{8},1)$ into an abelian anyon $(\mathbf{1},3)$ and a non-abelian anyon $(\mathbf{1},1)$:
\begin{equation}
    (\mathbf{8},1) \longrightarrow (\mathbf{1},1) + (\mathbf{1},3).
\end{equation}
It is easy to check now that the topological spin of $(\mathbf{1},1)$ is not equal to that of $(\mathbf{8},1)$ in the parent theory, and thus it follows that $(\mathbf{1},1)$ confines. Meanwhile, the topological spins of $(\mathbf{8},1)$ and $(\mathbf{1},3)$ match.

Using this information in the first fusion rule \eqref{fusionrule1} we obtain
\begin{align}
    (\mathbf{8},2) \times (\mathbf{1},1) &= (\mathbf{1},1) + (\mathbf{1},3) + (\mathbf{8},3) \\[0.3cm] &= (\mathbf{1},1) + (\mathbf{1},1) + (\mathbf{1},3),
\end{align}
where in the first equality we have used that $(\mathbf{8},1) \to (\mathbf{1},1) + (\mathbf{1},3)$ on the right side of \eqref{fusionrule1}, and in the second equality we have instead first restricted $(\mathbf{8},2) \to 0 + (\mathbf{1},2)$. We must therefore identify $(\mathbf{1},1) \cong (\mathbf{8},3)$.

It remains to study $(\mathbf{8},0)$, which does not split since it does not have large enough quantum dimension. Now, inspect the fusion
\begin{align}
    (\mathbf{8},2) \times (\mathbf{8},0) &= (\mathbf{1},2) + (\mathbf{8},2) = (\mathbf{1},2) + 0 + (\mathbf{1},2) \\[0.3cm] &= (\mathbf{8},0) + (\mathbf{8},2) = (\mathbf{8},0) + 0 + (\mathbf{1},2),
\end{align}
where in the first line we have performed the fusion on the parent theory and then used the restriction $(\mathbf{8},2) \to 0 + (\mathbf{1},2)$, while in the second line we have first restricted on the left-hand side and later computed the corresponding fusions, after which we used the $(\mathbf{8},2)$ restriction again. It follows that we must identify $(\mathbf{8},0) \cong (\mathbf{1},2)$.

\begin{table}[t]
\centering
\begin{tabular}[h]{|p{4cm}|p{4cm}|p{4cm}| }
\hline 
\multicolumn{3}{|c|}{$SU(6)_{1}$} \\
\hline
Line label & Quantum Dimension & Conformal Weight  \\
\hline
0 & $d_{1} = 1$ & $h_{1}=0$ \\
1 & $d_{2} = 1$ & $h_{2}=5/12$ \\
2 & $d_{3} = 1$ & $h_{3}=2/3$ \\
3 & $d_{4} = 1$ & $h_{4}=3/4$ \\
4 & $d_{5} = 1$ & $h_{5}=2/3$ \\
5 & $d_{6} = 1$ & $h_{6}=5/12$ \\
\hline
\end{tabular}
\caption{$SU(6)_{1}$ data. The label $n$ on the left column labels the fully antisymmetric representation with a single column with $n$ boxes in the Young tableux.}  \label{su6lv1table}
\end{table}

Clearly, the only unconfined lines in the $(\mathbf{1},i)$ and $(\mathbf{8},i)$ sectors are $(\mathbf{1},0)$ and $(\mathbf{1},3)$ (up to identifications). Following our comments above, we can now consider the action of the simple currents of $SU(3)_{2}$ in the form $(\mathbf{6},0)$ and $(\mathbf{\bar{6}},0)$ over the confined and unconfined excitations already found, which generates the rest of the sectors not considered up to this point. The spectrum of unconfined excitations is seen to match that of the expected $SU(6)_{1}$ theory, whose spectrum is summarized in Table \ref{su6lv1table}. Furthermore, since in this example the lines in the child theory descend trivially from those of the parent, we can easily check the $\mathbb{Z}_{6}$ fusion rules expected of $SU(6)_{1}$ by computing them directly in the parent:
\begin{align}
    (\mathbf{6},3)^{2} = (\bar{\mathbf{6}},0), \quad (\mathbf{6},3)^{3} = (\mathbf{1},3), \quad (\mathbf{6},3)^{6} = (\mathbf{1},0).
\end{align}

The condensation computation thus points that indeed\footnote{To make this precise we would have to check the modular $S$-matrix, $F$-symbols and $R$-symbols, but the previous consistency conditions do not provide us with these, and we need external sources to point us to what the result should be.}
\begin{equation}
    \boxed{SU(6)_{1} \cong \frac{SU(3)_{2} \times SU(2)_{3}}{\mathcal{A}_{3,2}}}
\end{equation}
with the algebra element $\mathcal{A}_{3,2} = (\mathbf{1},0) + (\mathbf{8},2)$.

\subsubsection{Continuing Classical Embeddings} \label{continuingclassicalembeddings}

In the same way that we obtained the main duality for unitary groups via non-abelian anyon condensation; namely, Eqn. \eqref{unitarynoninvertbile} in the last section, we can find additional dualities based on other groups as long as they participate in some conformal embedding. The list of conformal embeddings can be found in \cite{davydov2013witt}, which we use in the following. 

Let us start this discussion by recalling the conformal embeddings based on Spin groups studied in \cite{Aharony:2016jvv,Cordova:2017vab} which are the next natural examples to consider:
\begin{align}
    & SO(N)_{K} \times SO(K)_{N} \hookrightarrow Spin(NK)_{1}, \quad N \  \mathrm{even}, \ K \ \mathrm{even}, \\[0.3cm]
    & Spin(N)_{K} \times SO(K)_{N} \hookrightarrow Spin(NK)_{1}, \quad N \  \mathrm{even}, \ K \ \mathrm{odd}, \\[0.3cm]
    & Spin(N)_{K} \times Spin(K)_{N} \hookrightarrow Spin(NK)_{1}, \quad N \  \mathrm{odd}, \ K \ \mathrm{odd}.
\end{align}
These conformal embeddings yield the following standard 3D TQFT duality interpretation, obtained by the common center argument in accordance with \cite{Aharony:2016jvv,Cordova:2017vab}:
\begin{align}
    Spin(N)_{K} \cong \frac{Spin(NK)_{1} \times Spin(K)_{-N}}{\mathbb{Z}_{2}}, \quad N \  \mathrm{odd}, \ K \ \mathrm{odd} \label{stdcosetduality1} \\[0.3cm]
    Spin(N)_{K} \cong Spin(NK)_{1} \times SO(K)_{-N}, \quad N \  \mathrm{even}, \ K \ \mathrm{odd} \label{stdcosetduality2} \\[0.3cm]
    SO(N)_{K} \cong \frac{Spin(NK)_{1} \times Spin(K)_{-N}}{B}, \quad N \  \mathrm{odd}, \ K \ \mathrm{even} \label{stdcosetduality3} \\[0.3cm]
    SO(N)_{K} \cong \frac{Spin(NK)_{1} \times SO(K)_{-N}}{\mathbb{Z}_{2}}, \quad N \  \mathrm{even}, \ K \ \mathrm{even}, \label{stdcosetduality4}
\end{align}
where $B$ above is such that $B = \mathbb{Z}_{2} \times \mathbb{Z}_{2}$ for $k=0 \mod 4$ and $B = \mathbb{Z}_{4}$ for $k=2 \mod 4$. These dualities (derived in \cite{Aharony:2016jvv}) may be interpreted as solving for the cosets as in Eqn. \eqref{IsolatingMTCcoset}. If we invert these expressions into the initial ones \eqref{MTCEmbedding}, which exhibit the embeddings of the two factors into the bigger algebra as we did in the unitary case, we add the dualities
\begin{align}
    Spin(NK)_{1} = \frac{SO(N)_{k} \times SO(k)_{N}}{\mathcal{A}_{N,k}}, \quad N \  \mathrm{even}, \ K \ \mathrm{even}, \label{spinduality1} \\[0.3cm]
    Spin(NK)_{1} = \frac{Spin(N)_{k} \times SO(k)_{N}}{\mathcal{A}_{N,k}}, \quad N \  \mathrm{even}, \ K \ \mathrm{odd}, \label{spinduality2} \\[0.3cm]
    Spin(NK)_{1} = \frac{Spin(N)_{k} \times Spin(k)_{N}}{\mathcal{A}_{N,k}}, \quad N \  \mathrm{odd}, \ K \ \mathrm{odd}, \label{spinduality3}
\end{align}
which generically involve gauging by a non-invertible symmetry. We verify this in appendix \ref{su26su26intospin91} in the $N=k=3$ case, which is the simplest example of the above dualities that involve non-invertible anyon condensation. 

A similar story holds for the dualities and embeddings associated with symplectic groups \cite{Aharony:2016jvv}:
\begin{equation}
    USp(2N)_{k} \times USp(2k)_{N} \hookrightarrow Spin(4Nk)_{1}, 
\end{equation}
which have the 3D TQFT duality interpretation 
\begin{equation} \label{symplecticduality}
    USp(2N)_{k} \cong \frac{Spin(4Nk)_{1} \times USp(2k)_{-N}}{\mathbb{Z}_{2}} \longleftrightarrow Spin(4NK)_{1} \cong \frac{USp(2N)_{k} \times USp(2k)_{N}}{\mathcal{A}_{N,k}},
\end{equation}
where the left duality can be obtained by the common-center argument, as in \cite{Aharony:2016jvv}. The right duality is obtained following the arguments outlined above both in the unitary and Spin cases, and it can be readily verified in the case $N=2$, $k=1$ where both sides are related by abelian anyon condensation. The gauging on the right duality is generically by a non-invertible symmetry, however, as the next simplest case $N=k=2$ already shows. In the product $USp(4)_{2} \times USp(4)_{2}$ there are both abelian and non-abelian bosons, but it can be checked that no abelian gauging is sufficient to give $Spin(16)_{1}$ back.

In the previous examples the standard form of the cosets \eqref{standardunitaryduality}, \eqref{stdcosetduality1}-\eqref{stdcosetduality4} and the left duality on \eqref{symplecticduality} studied in \cite{Hsin:2016blu, Aharony:2016jvv, Cordova:2017vab} all involve gauging by an abelian symmetry. However, it is well-known that there is yet an additional infinite series of conformal embeddings with a tensoring of two affine lie algebras embedding into a bigger algebra \cite{davydov2013witt}. Namely:
\begin{equation} \label{additionalembedding}
    SO(N)_{4} \times SU(2)_{N} \hookrightarrow USp(2N)_{1}.
\end{equation}
This is an interesting example, as suppose we tried to follow the same logic as in the aforementioned ``standard coset dualities'' and try to isolate $SU(2)_{N}$ in terms of $USp(2N)_{1}$ and $SO(N)_{-4}$, and running the common center procedure. For $N=2$ there are no novelties and we find $SU(2)_{2} \cong USp(4)_{1} \times U(1)_{-4} / \mathbb{Z}_{2}$.\footnote{Here we have used that $SO(2)_{k} \cong U(1)_{k}$.} However, already at $N=3$ we observe that there is no abelian boson in the spectrum of $USp(6)_{1} \times SO(3)_{-4}$, so the common center procedure manifestly does not work for this value of $N$. 

The cure to the puzzle just mentioned is clear based on what we have studied so far in this work, and it is to extend the common center procedure and allow for gauging by non-invertible symmetries. Indeed, $USp(6)_{1} \times SO(3)_{-4}$ has two non-abelian bosons in its spectrum, and we can condense them and find $SU(2)_{3}$ correspondingly. We study the details of this procedure in appendix \ref{SU23fromusP61SO3m4}. In general then, we have the duality
\begin{equation}
    SU(2)_{N} \cong \frac{USp(2N)_{1} \times SO(N)_{-4}}{\mathcal{A}_{N}},
\end{equation}
where as the prior example shows, we must allow for the gauging of a non-invertible one-form symmetry on the right-hand side.

Of course, as we did previously for the other conformal embeddings, we can still apply the coset inversion theorem and write the duality implied by \eqref{MTCEmbedding}. This gives:
\begin{equation}
    USp(2N)_{1} \cong \frac{SU(2)_{N} \times SO(N)_{4}}{\mathcal{A}_{N}},
\end{equation}
and as above, for $N=2$ the gauging is by an abelian symmetry, but at $N=3$ we already need non-abelian anyon condensation to make the duality valid.

\subsection{Further Conformal Embeddings} \label{FurtherConformalEmbeddings}

Now we move to discuss the rest of the conformal embeddings, with a focus on those that involve non-invertible symmetries. A list of the conformal embeddings can be found in \cite{davydov2013witt}. We begin considering conformal embeddings with a product of two affine lie algebras in the denominator, as in the previous subsection, but we study those associated to exceptional Lie algebras. These embeddings have been previously studied in \cite{Cordova:2018qvg}, and here we extend and understand the associated dualities in terms of non-invertible symmetries. The subset of these embeddings for which non-abelian anyon condensation is necessary to understand the dualities are
\begin{align}
    &SU(2)_{1} \times SU(2)_{3} \hookrightarrow (G_{2})_{1}, \quad
    &&SU(2)_{1} \times USp(6)_{1} \hookrightarrow (F_{4})_{1},
    &&&SU(3)_{1} \times SU(3)_{2} \hookrightarrow (F_{4})_{1}, \nonumber \\[0.3cm]
    &SO(3)_{4} \times (G_{2})_{1} \hookrightarrow (F_{4})_{1}, \quad
    &&SU(2)_{3} \times (F_{4})_{1} \hookrightarrow (E_{7})_{1}, 
    &&&USp(6)_{1} \times (G_{2})_{1} \hookrightarrow (E_{7})_{1}, \nonumber \\[0.3cm]
    &SU(2)_{7} \times (G_{2})_{2} \hookrightarrow (E_{7})_{1}, \quad
    &&SU(3)_{2} \times (G_{2})_{1} \hookrightarrow (E_{6})_{1}, \label{exceptionalconformalembeddings}
\end{align} 

Here we recognize the example $SU(2)_{1} \times SU(2)_{3} \hookrightarrow (G_{2})_{1}$ we elaborated upon in Section \ref{sec:introG2} and in Section \ref{noninvertibleextensionsubsection}. Basically the same argument follows through in all of these embeddings. Let us quickly recall this argument in an alternative example. For instance, if we take the third example in the first line and try to isolate $SU(3)_{2}$ we obtain $SU(3)_{2} \cong (F_{4})_{1} \times SU(3)_{-1}$, with no further gauging on the right-hand side and everything is in order. If we instead try to isolate $SU(3)_{1}$, as explained previously there is no abelian common center to gauge by, but we still have to take into account non-abelian anyon condensation, and we obtain
\begin{equation}
    SU(3)_{1} \cong \frac{(F_{4})_{1} \times SU(3)_{-2}}{\mathcal{Z}(\mathbf{Fib})},
\end{equation}
where by $\mathcal{Z}(\mathbf{Fib})$ we mean an algebra object that can be traced back to a Lagrangian algebra responsible for gauging away $(G_{2})_{1} \times (G_{2})_{-1}$ or $(F_{4})_{1} \times (F_{4})_{-1}$ down to the trivial theory as in Section \ref{sec:introG2}, for which reason we interpret it as gauging by a Drinfeld center of a Fibonacci fusion category. An analogous example of this same form was verified in detail by a direct non-abelian anyon condensation computation in Section \ref{IntroductionExample}.

All of the conformal embeddings in \eqref{exceptionalconformalembeddings} yield dualities that can be obtained by the same arguments, and with one exception, in all of them we gauge by some Fibonacci Drinfeld center. The only exception is the first example in the third line, where
\begin{equation}
    (E_{7})_{1} \cong \frac{SU(2)_{7} \times (G_{2})_{2}}{\mathcal{Z}\big((G_{2})_{2}\big)},
\end{equation}
which is however obtained by the same arguments. 

We discuss now those conformal embeddings with a simple group, or single affine Lie algebra in the denominator of the conformal embedding, such as in the $SU(3)_{1}/SU(2)_{4}$ example. We clearly cannot use the same arguments that were used to obtain the ``standard coset form'' of the dualities, since now we do not have two factors in the denominator. That is, we will not have an analog of expressions such as \eqref{standardunitaryduality} or their counterparts for algebras other than unitary. However, we still have an analog of the dualities in the form outlined in this work, Eqn. \eqref{unitarynoninvertbile}. Indeed, this is tantamount to using \eqref{MTCEmbedding} with $\mathcal{C}$ describing the trivial theory. In the example $SU(3)_{1}/SU(2)_{4}$ this is nothing but the statement that we can gauge $SU(2)_{4}$ by some algebra to obtain $SU(3)_{1}$. Indeed, it is well-known that 
\begin{equation}
    SU(3)_{1} \cong SU(2)_{4}/\mathbb{Z}_{2}.
\end{equation}
Clearly, the general story for an arbitrary conformal embedding will be essentially the same, but where we allow to gauge by a non-invertible symmetry as per \eqref{MTCEmbedding}.

With the previous remarks in mind let us summarize a few results. Exploring the conformal embeddings, we find the following infinite families of dualities
\begin{align}
     & Spin(N^{2}-1)_{1} \cong \frac{SU(N)_{N}}{\mathcal{A}_{N}} \label{firstinfinite} \\[0.4cm]
     & SU \big( N( N \pm 1)/2 \big)_{1} \cong   \frac{SU(N)_{N \pm 2}}{\mathcal{A}_{N}} \label{secondinfinite} \\[0.4cm]
     & Spin\big( N(N-1)/2 \big)_{1}  \cong  \frac{Spin(N)_{N-2}}{\mathcal{A}_{N}} \label{thirdinfinite} \\[0.4cm]
     & Spin\big( (N^{2}+N-2)/2 \big)_{1} \cong  \frac{Spin(N)_{N+2}}{\mathcal{A}_{N}}, \label{ourthinfinite}
\end{align}
Notice that the first of these corresponds to the conformal embedding used in \cite{Komargodski:2020mxz} in the context of 2D CFT to propose the IR phases of 2D Adjoint QCD. Here we use the conformal embedding to establish a duality of 3D TQFTs. There are various checks that can be performed in the series above. The case $N=4$ in \eqref{ourthinfinite} gives $Spin(9)_{1} \cong Spin(4)_{6}$, which we verify by a explicit computation in appendix \ref{su26su26intospin91} and is actually equivalent to the duality \eqref{spinduality3} for $N=3$.

We finish mentioning the isolated cases (i.e., no infinite family) of the conformal embeddings where there is a single affine Lie algebra in the denominator. The complete list of these cases can be obtained by reading \cite{davydov2013witt}, and yields the following list of dualities: 
\begin{align}
    & SU(16)_{1} \cong Spin(10)_{4} / \mathcal{A}, \quad 
    SU(27)_{1} \cong (E_{6})_{6}/ \mathcal{A}, \quad 
    Spin(70)_{1} \cong SU(8)_{10} / \mathcal{A}, \\[0.4cm]
    & USp(4)_{1} \cong SU(2)_{10} / \mathcal{A}, \quad USp(20)_{1} \cong SU(6)_{6}/ \mathcal{A}, \quad (E_{6})_{1} \cong SU(3)_{9} / \mathcal{A},  \\[0.4cm]
    &(E_{7})_{1} \cong SU(3)_{21}/\mathcal{A}, \quad (G_{2})_{1} \cong SU(2)_{28} / \mathcal{A}, \quad Spin(16)_{1} \cong Spin(9)_{2} / \mathcal{A},  \\[0.4cm]
    &Spin(128)_{1} \cong Spin(16)_{16} / \mathcal{A}, \quad Spin(42)_{1} \cong USp(8)_{7}/\mathcal{A}, \quad USp(32)_{1} \cong Spin(12)_{8}/\mathcal{A},  \\[0.4cm]
    &Spin(78)_{1} \cong (E_{6})_{12}/\mathcal{A}, \quad Spin(133)_{1} \cong (E_{7})_{18}/\mathcal{A} \quad Spin(248)_{1} \cong (E_{8})_{30}/\mathcal{A},  \\[0.4cm]
    & USp(14)_{1} \cong USp(6)_{5} / \mathcal{A}, \quad USp(56)_{1} \cong (E_{7})_{12} / \mathcal{A}, \quad (E_{8})_{1} \cong USp(4)_{12} / \mathcal{A}, \\[0.4cm]
    & Spin(26)_{1} \cong (F_{4})_{3} / \mathcal{A}, \quad Spin(52)_{1} \cong (F_{4})_{9}  / \mathcal{A}, \quad Spin(14)_{1} \cong (G_{2})_{4} / \mathcal{A}, \\[0.4cm]
    & (E_{6})_{1} \cong (G_{2})_{3} / \mathcal{A}.
\end{align}
The first example on the second line, $USp(4)_{1} \cong SU(2)_{10}/\mathcal{A}$, is a known example that has been studied in the literature mainly with the aim of testing and understanding the formalism of non-abelian anyon condensation (see for example \cite{Bais:2008ni, Hung:2015hfa, Eliens:2013epa, Yu:2021zmu,neupert2016boson}). The last example in this list $(E_{6})_{1} \cong (G_{2})_{3} / \mathcal{A}$ can be verified very easily by a non-abelian anyon condensation computation, which we outline in appendix \ref{E61fromG23}.

\subsubsection{Three-State Potts Model Maverick Coset from Level-Rank Duality} \label{TSPMfromLevelRank}

In this subsection we show the simplest duality implied by the Maverick cosets, Eqn. \eqref{TSPMDuality}, but instead of doing the calculation directly, we will first use our knowledge of the conformal embeddings to simplify the calculation. To understand the three-state Potts Model, we make use of the first of the embeddings in \eqref{exceptionalconformalembeddings}, and write
\begin{equation}
    SU(2)_{3} \cong (G_{2})_{1} \times SU(2)_{-1}.
\end{equation}
It is also straightforward to check the duality $U(1)_{6} \cong SU(3)_{1} \times SU(2)_{-1}$, so that we have
\begin{equation} \label{productduality}
    SU(2)_{3} \times U(1)_{-6} \cong SU(3)_{-1} \times (G_{2})_{1} \times SU(2)_{-1} \times SU(2)_{1}.
\end{equation}
Now we can use the first embedding on the second line of \eqref{exceptionalconformalembeddings}, and use our knowledge of non-abelian anyon condensation to isolate $(G_{2})_{1}$, obtaining:
\begin{equation} \label{DualitywithSO3lv4}
    (G_{2})_{1} \cong \frac{(F_{4})_{1} \times SO(3)_{-4}}{\mathcal{Z}(\mathbf{Fib})}.
\end{equation}
This duality may be obtained from the arguments outlined previously, but it is simple and sufficiently interesting that we verify it by a direct non-abelian anyon condensation on the next subsection.

We also use the third embedding on the first line of \eqref{exceptionalconformalembeddings} to claim
\begin{equation}
    SU(3)_{2} \cong (F_{4})_{1} \times SU(3)_{-1}.
\end{equation}

These expressions allows us to introduce the first $SU(3)_{-1}$ factor on the right-hand side of \eqref{productduality} into the quotient given by \eqref{DualitywithSO3lv4}, with the $\mathcal{Z}(\mathbf{Fib})$ denominator acting trivially on such $SU(3)_{-1}$ factor, and write
\begin{equation}
    SU(2)_{3} \times U(1)_{-6} \cong \frac{SU(3)_{2} \times SO(3)_{-4}}{\mathcal{Z}(\mathbf{Fib})} \times SU(2)_{-1} \times SU(2)_{1}.
\end{equation}
Finally, gauging the $\mathbb{Z}_{2}$ Drinfeld center $SU(2)_{1} \times SU(2)_{-1}$ we obtain
\begin{equation}
    \frac{SU(2)_{3} \times U(1)_{-6}}{\mathbb{Z}_{2}} \cong \frac{SU(3)_{2} \times SO(3)_{-4}}{\mathcal{Z}(\mathbf{Fib})}, \cong \frac{SU(3)_{2} \times SU(2)_{-8}}{\mathcal{A}},
\end{equation}
which is indeed what we have obtained above using Maverick coset considerations instead of several conformal embeddings/exceptional level-rank duality manipulations. On the last line we have used that by definition $SO(3)_{4} \cong SU(2)_{8}/\mathbb{Z}_{2}$.

\section*{Acknowledgements}

We thank Jimmy Huang, and Carolyn Zhang for helpful conversations. CC and DGS acknowledge support from the Simons Collaboration on Global Categorical Symmetries, the US Department of Energy Grant 5-29073, and the Sloan Foundation. DGS is also supported by a Bloomenthal Fellowship in the Enrico Fermi Institute at the University of Chicago.

\appendix

\section{Review of (2+1)D TQFTs as Modular Tensor Categories} \label{mathsappendix}

In this appendix we summarize a few definitions and results about (unitary) (2+1)D TQFTs in terms of modular tensor categories (MTC). We will limit ourselves to those results that are useful to follow the main text. For a more in-depth discussion of the definitions and results used here we refer to \cite{Frohlich:2003hm} which is particularly clear in our context. Other general references are \cite{Kitaev:2005hzj}, or Section 5 in \cite{Benini:2018reh}. 

\subsection{Modular Tensor Categories}

Axiomatically, a unitary bosonic (2+1)D TQFT is described by a unitary modular tensor category $\mathcal{C}$. Recall that a category in general is composed by a set of objects and a set of morphisms between them. We denote the objects of the category $\mathrm{Obj}(\mathcal{C})$ and the morphisms between objects $a$ and $b$ in $\mathrm{Obj}(\mathcal{C})$ as $\mathrm{Hom}(a,b)$. Throughout, we assume that in the categories we are concerned with, $\mathrm{Obj}(\mathcal{C})$ is a set, the category admits direct sums of objects, and morphism-sets are vector spaces over a field that in physics we take to be the complex numbers. We also assume: \\

Semi-simplicity: Any object can be written as a direct sum of finitely many simple objects, where simple objects $s$ are such that the self-junction space $\mathrm{Hom}(s,s)$ is one-dimensional. \\

Finiteness: The number of simple objects in the category is finite. \\

We denote the set of simple objects in a category as $\mathcal{I}$.

A UMTC is a special type of monoidal, or tensor category, which in turn is essentially a category equipped with a notion of tensor product in between its objects. That is, if $a,b$ are objects in $\mathcal{C}$, then there exists an object $a \otimes b$ in $\mathcal{C}$. The full definition of a monoidal, or tensor category is technically more involved, requiring many consistency conditions which precise details however we will not need. See \cite{ReferenceKey} for more details. 

To define a modular tensor category, we do find useful to first define a Ribbon category. A Ribbon category is a tensor category that meets the following extra requirements. First, to every object $a \in \mathrm{Obj(\mathcal{C})}$ there is an object $a^{\vee} \in \mathrm{Obj(\mathcal{C})}$, the (right) dual of $a$, such that there exists morphisms $b_{a} \in \mathrm{Hom}(\mathbf{1},a \otimes a^{\vee})$ and $d_{a} \in \mathrm{Hom}(a^{\vee} \otimes a, \mathbf{1})$. We also require the existence of certain morphisms called \textit{braiding} and \textit{twist}:
\begin{equation}
    \mathrm{Braiding:} \quad c_{a,b} \in \mathrm{Hom}(a \otimes b, b \otimes a),
\end{equation}
\begin{equation}
    \mathrm{Twist:} \quad \theta_{a} \in \mathrm{Hom}(a \otimes a).
\end{equation}
All these morphisms are subject to various consistency conditions that can be found, e.g., in \cite{Frohlich:2003hm}. They allow a useful pictorial notation, whereby objects are denoted by lines with a label $a$ and morphisms are read as a diagram from bottom to top:
\begin{equation*}
\vcenter{\hbox{\includegraphics[scale = 0.08]{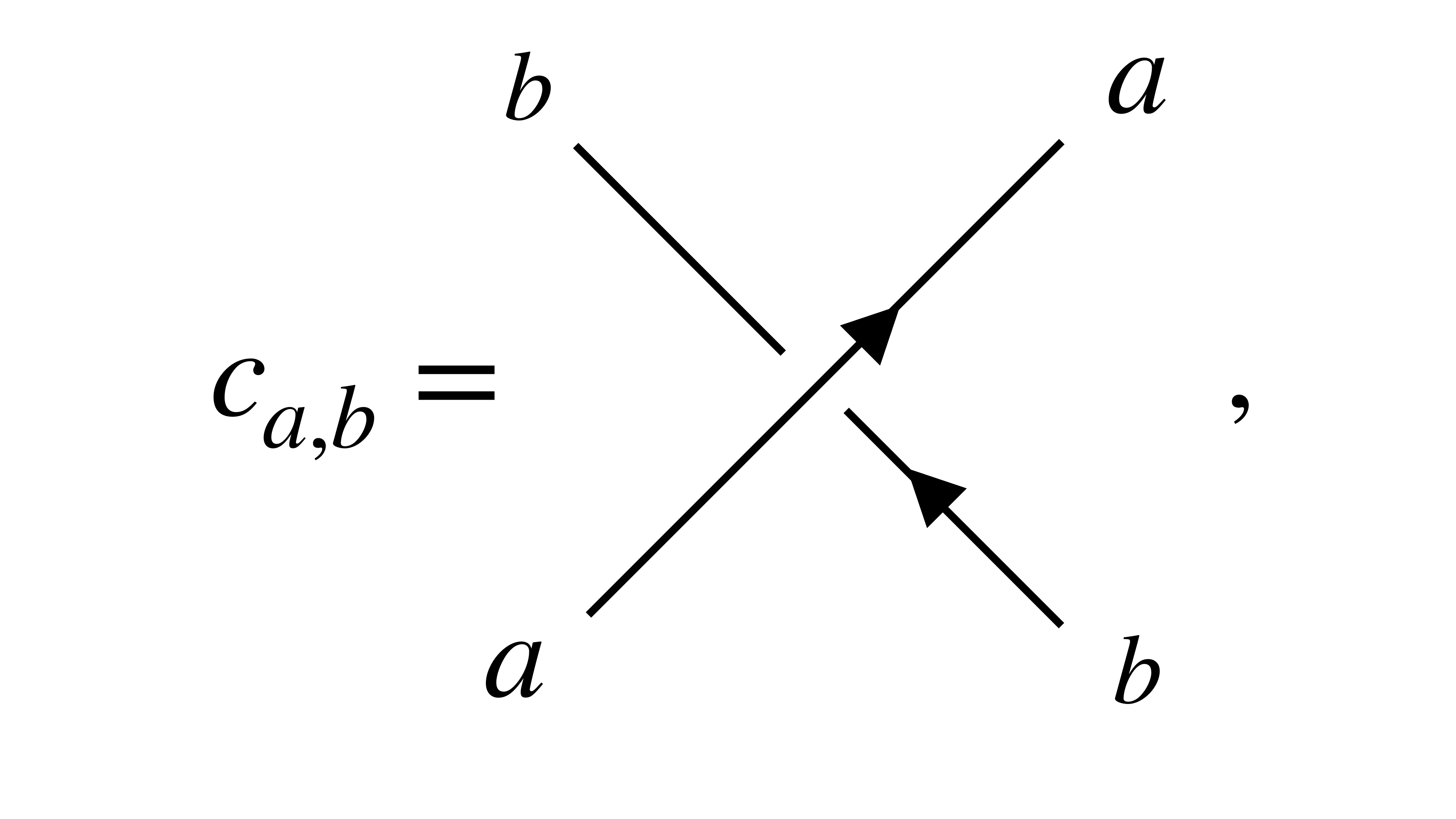}}} \quad \vcenter{\hbox{\includegraphics[scale = 0.08]{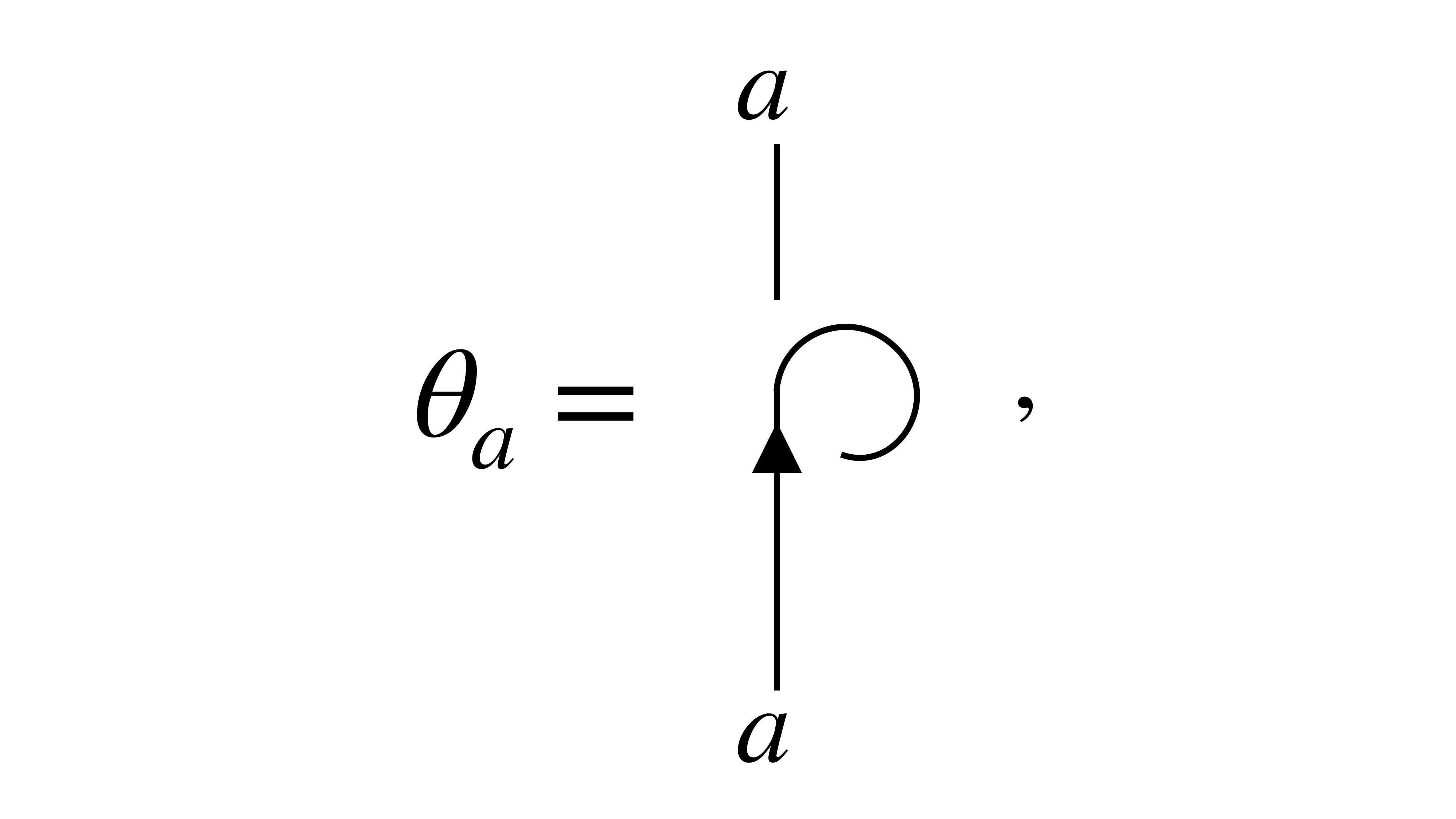}}}
\end{equation*}
\begin{equation}
    \vcenter{\hbox{\includegraphics[scale = 0.07]{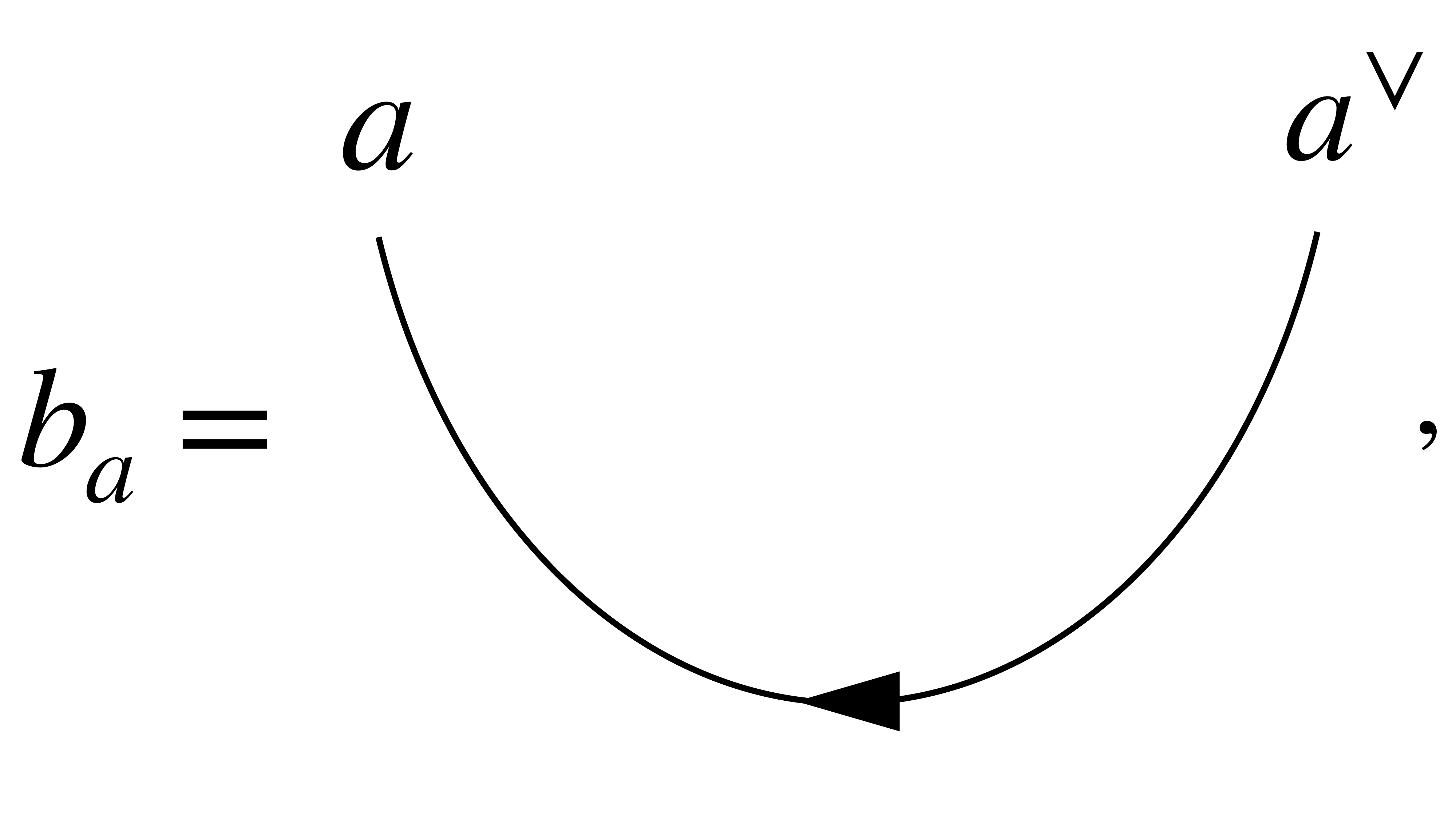}}} \hspace{2cm}
\vcenter{\hbox{\includegraphics[scale = 0.07]{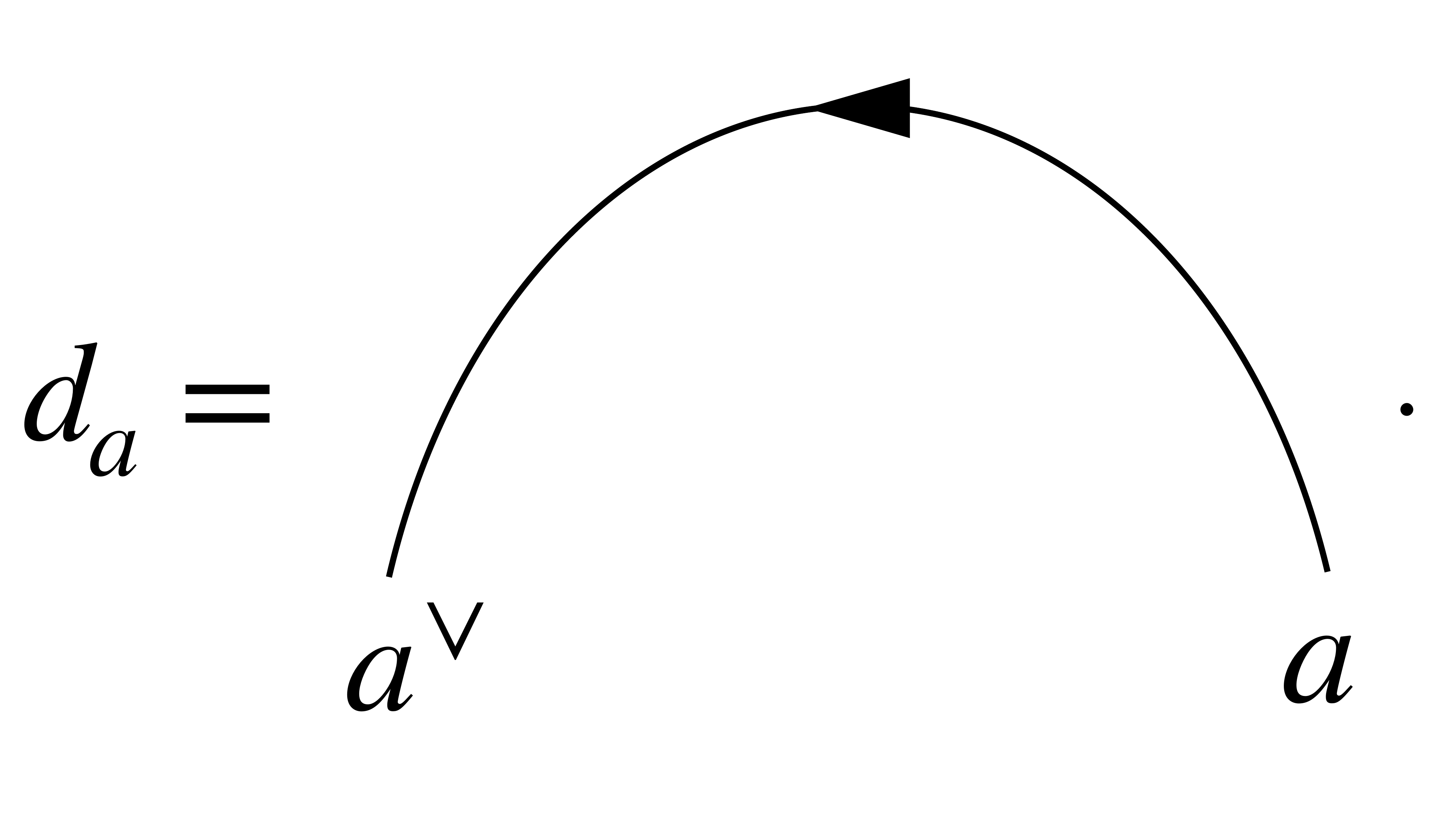}}}
\end{equation}

A MTC is then a ribbon category, subject to the assumptions stated above, and such that the matrix
\begin{equation}
    s_{a,b} \coloneqq \mathrm{Tr}(c_{a,b} \circ c_{b,a}) = \hspace{-0.1cm} \vcenter{\hbox{\includegraphics[scale = 0.065]{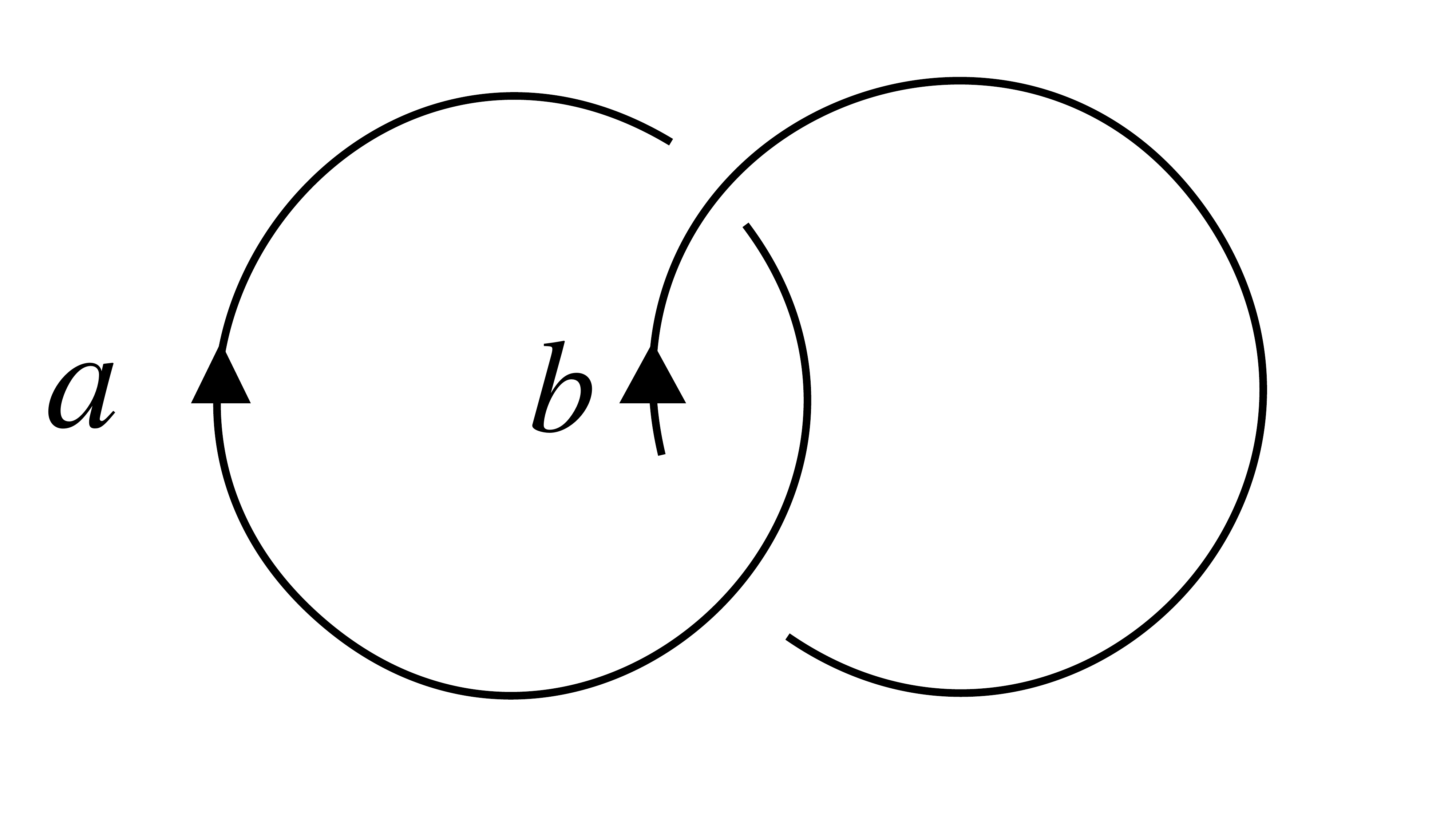}}} 
\end{equation}
with entries on the simple objects is non-degenerate.\footnote{Here we should be more precise with the definition of the trace. Intuitively, it is clear that we have to take the loops of the anyons involved. See \cite{Frohlich:2003hm} for a more precise version of this statement.} Imposing that this matrix is non-degenerate, one obtains the definition of a MTC. In physics one does not quite use $s$, but rather $S = S_{0,0} \, s$, with $S$ unitary and $S_{0,0} = 1/\mathcal{D}$, where $\mathcal{D}$ is the total quantum dimension (see below). The latter object can be recognized as the standard modular $S$-matrix encountered in physics satisfying $S_{ab} = S_{ba} = S_{\bar{a} b}^{*}$.

As a special type of monoidal (or fusion) category, a UMTC further consists of an associator or $F$-symbols dictating the associativity of line junctions, constrained by the so-called pentagon equations. Unitarity corresponds to the statement (among others) that a basis of the $F$-symbols exists such that the corresponding $F$-matrices are unitary. (See \cite{galindo2014braided} for a more precise discussion of unitarity in the context of ribbon categories.) We will not need all these details for our applications, so we refer the reader to the references above and the original works \cite{Moore:1988uz, Moore:1988qv} for details.

Let us now unpack the previous definitions in a more familiar context. Physically, the objects of the UMTC $\mathcal{C}$ correspond to the anyons, or line operators of the TQFT. The morphisms of $\mathcal{C}$ are associated to junctions of the lines. The tensor product is merely denoted by $\times$ and it corresponds to the well-known commutative fusion algebra
\begin{equation}
    a \times b = \sum_{c} N_{ab}^{c} \, c,
\end{equation}
which is also associative
\begin{equation}
    (a \times b) \times c = a \times (b \times c).
\end{equation}
The quantities $N_{ab}^{c}$ are non-negative integers called the fusion coefficients, and they satisfy $N_{ab}^{c} = N_{ba}^{c}$ as well as 
\begin{equation}
    \sum_{e}N_{ab}^{e} N^{d}_{ec} = \sum_{f} N^{d}_{af} N^{f}_{bc}
\end{equation}
by associativity. We require the existence of a unique transparent line operator, or unique identity object $0$:
\begin{equation}
    0 \times a = a \times 0 = a,
\end{equation}
and the existence of a unique conjugate sector $\bar{a}$ for any $a$ in $\mathcal{C}$, defined such that
\begin{equation} \label{conjugatedef}
    a \times \bar{a} = \bar{a} \times a = 0 + \sum_{c \neq 0} N_{a \bar{a}}^{c} \, c.
\end{equation}
These conjugate sectors, well-known in the physics context, are nothing but the dual objects defined above mathematically.

For our purposes, we define the quantum dimension of a simple anyon $a$ as the expectation value of the corresponding unknot:
\begin{equation}
d_{a} = \hspace{-0.1cm} \vcenter{\hbox{\includegraphics[scale = 0.035]{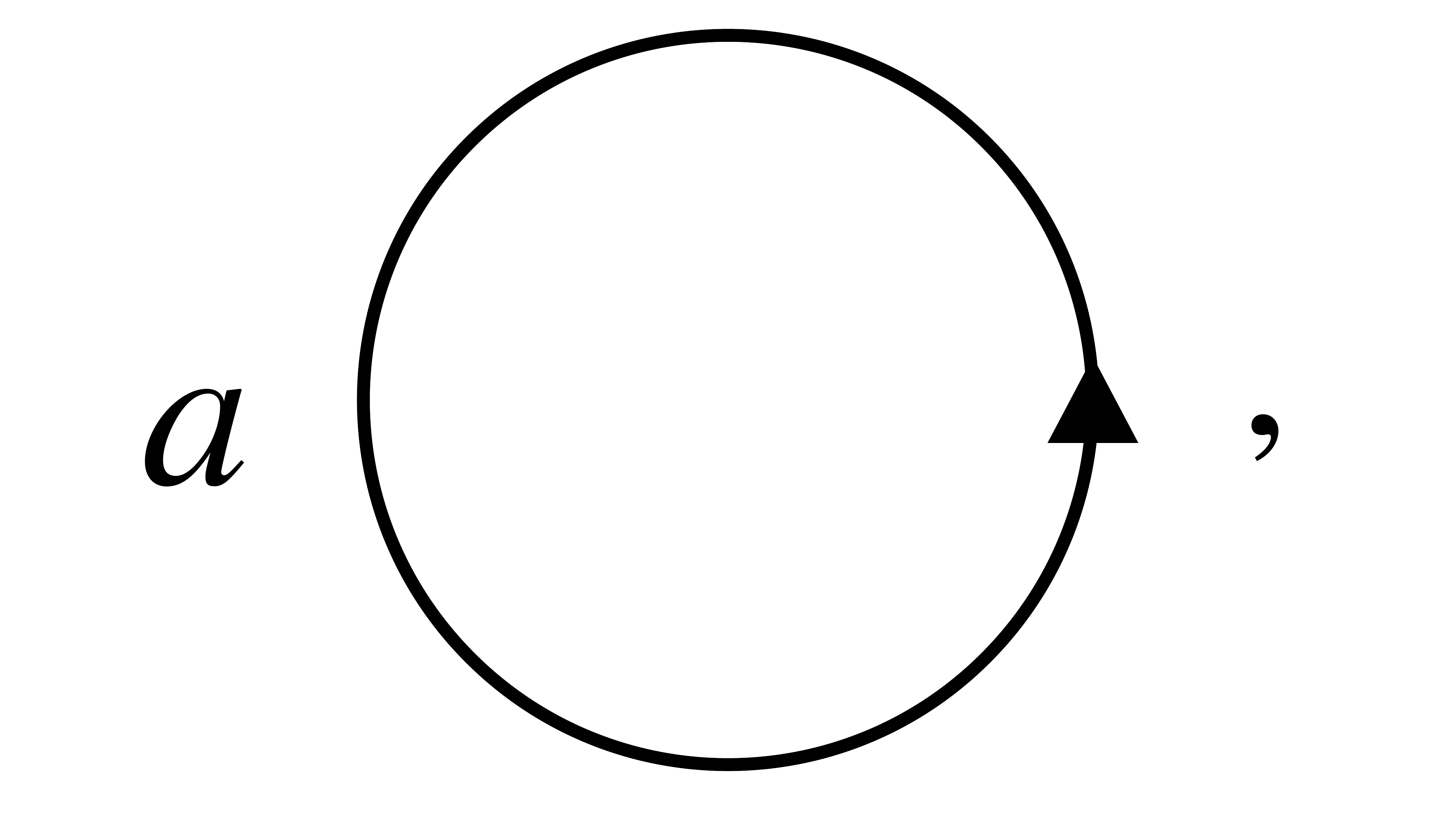}}}
\end{equation}
where an anyon and its conjugate satisfy $d_{a} = d_{\bar{a}}$. Unitarity requires that quantum dimensions are positive $d_{a} > 0$.

We say that a line operator, or an anyon $a$ is \textit{abelian} if $d_{a} = 1$. This is actually equivalent to the statement that fusion of any such anyon $a$ with an arbitrary anyon $b$ always gives back a single anyon with unit multiplicity. That is, for $a$ abelian $N_{ab}^{c} = \delta^{c}_{c'}$ for some simple $c'$ in $\mathcal{C}$. In particular, since the number of simple objects is finite, abelian anyons always generate some finite abelian group. Otherwise, we say the anyon is \textit{non-abelian}.

An important consequence of the associativity of the fusion algebra is that the matrices $\mathcal{N}_{a}$ defined by the entries $(\mathcal{N}_{a})_{b}^{\ c} =N_{ab}^{c}$ are mutually diagonalizable. Moreover, an eigenvector of the matrices $\mathcal{N}_{a}$ is the vector whose entries are the quantum dimensions, with eigenvalue $d_{a}$. This is equivalent to the statement that quantum dimensions follow the fusion algebra. That is:
\begin{equation} \label{FusionAlgebraofQuantumDims}
    d_{a} d_{b} = \sum_{c} N_{ab}^{c} \, d_{c} \, .
\end{equation}
It is also useful to define the \textit{total quantum dimension} of the UMTC $\mathcal{C}$:
\begin{equation} \label{totalquantumdimension}
    \mathcal{D}_{\mathcal{C}} = \sqrt{\sum_{a \in \mathcal{I}} d_{a}^{2}} \, .
\end{equation}

In a unitary MTC we can choose a Verlinde basis of anyons \cite{Moore:1988qv} in which the twist morphism is diagonalized on the basis of simple anyons. More precisely, the twist morphism is proportional by a phase to the identity morphism
\begin{equation}
\vcenter{\hbox{\includegraphics[scale = 0.1]{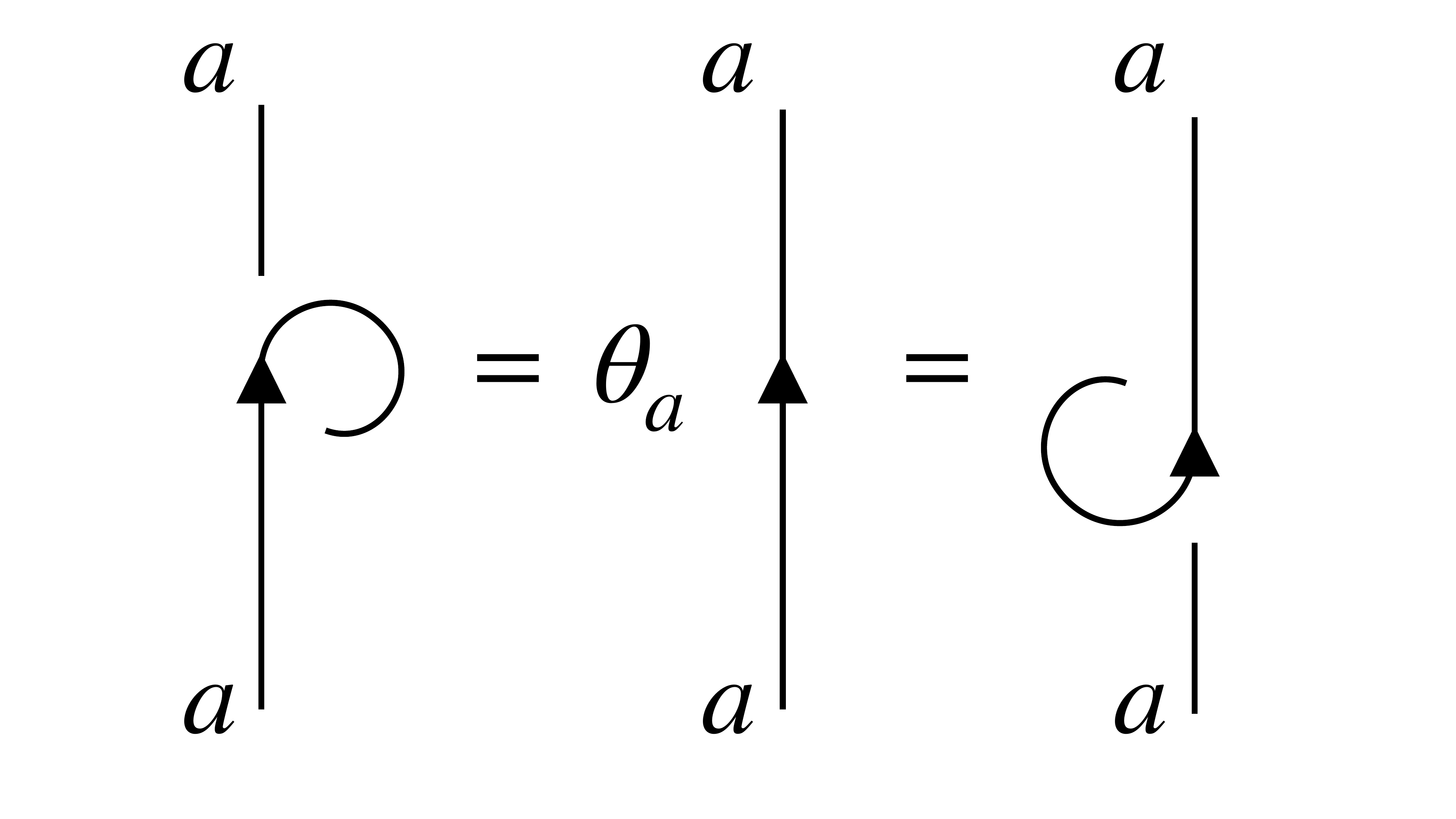}}}
\end{equation}
and we have abused notation and called the proportionality phase with the same symbol as the twist morphism $\theta_{a}$. This quantity is known as the \textit{topological spin of the anyon $a$}:
\begin{equation}
    \theta_{a} = e^{2 \pi i h_{a}},
\end{equation}
and we call $h_{a}$ to the spin of the line $a$, or abuse terminology from 2D CFT and call it the conformal weight of the line. In the text we use both nomenclatures interchangeably. The spin of a line operator is always a rational number \cite{VAFA1988421}. To avoid potential confusion, we stress the difference between spin and topological spin. In the main text we always keep the difference sharp, referring always to the quantity $\theta_{a}$ as topological spin, and the quantity in the exponent as spin, never interchangeably.

We have the following action of the anyons looping around each other:
\begin{equation} \label{braiding}
\vcenter{\hbox{\includegraphics[scale = 0.1]{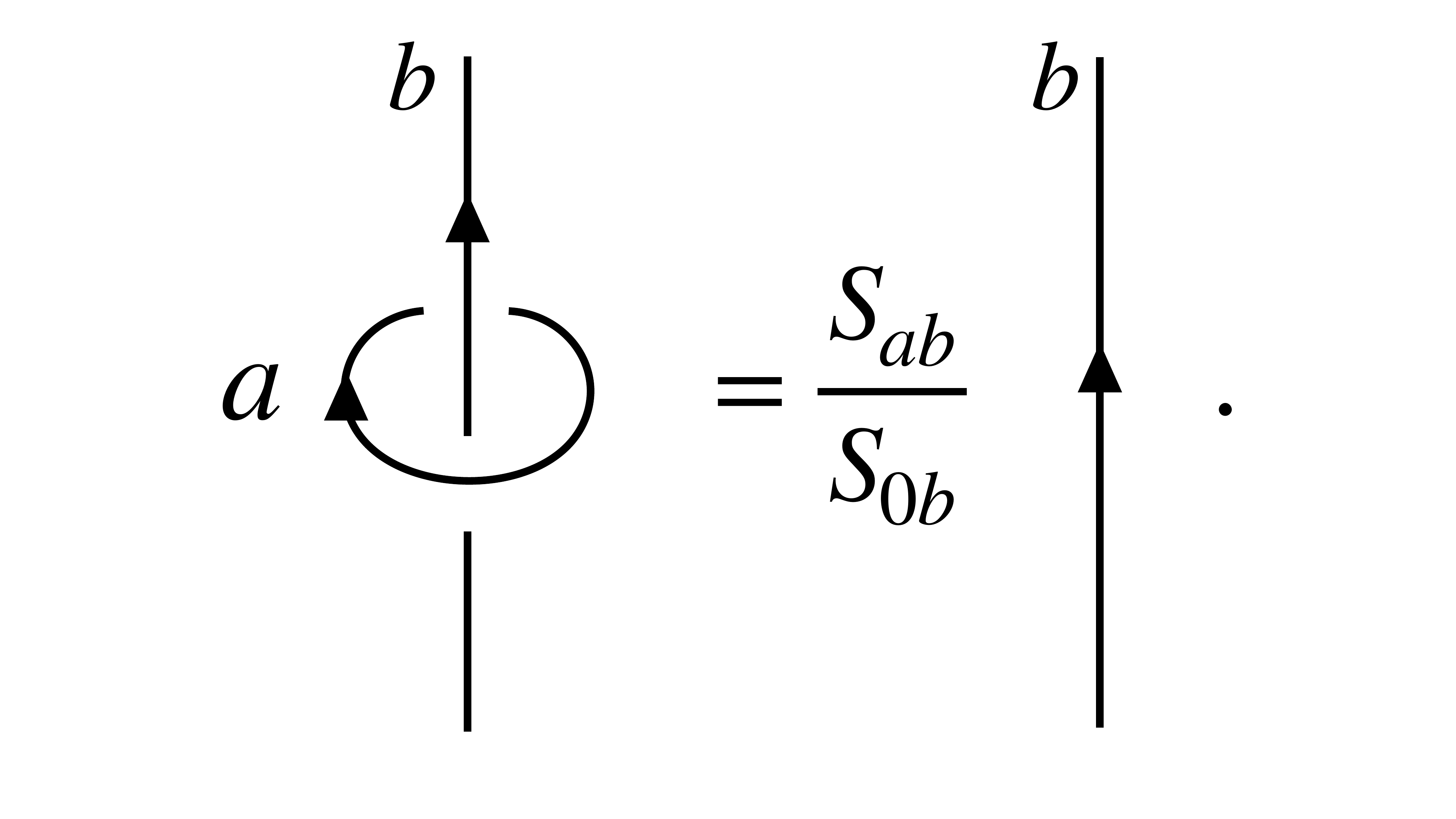}}}
\end{equation}
Taking $b$ to be the trivial anyon we find an expression for the quantum dimensions in terms of the modular $S$-matrix: $d_{a} = S_{a0}/S_{00}$.

The (chiral) central charge of the (2+1)D TQFT $c_{-}$ is defined by the topological spins and the quantum dimensions of the anyons by
\begin{equation} \label{ChiralCentralChargeMTCFormula}
    e^{i \frac{\pi}{4}c_{-}} = \frac{1}{\mathcal{D}} \sum_{a \in \mathcal{I}} d_{a}^{2} \, \theta_{a}.
\end{equation}
It is a non-trivial result that the combination on the right-hand side is a phase. Note that the MTC data only fixes the chiral central charge $c_{-}$, and thus the corresponding central charge of the edge modes, modulo eight. On the other hand, a given boundary CFT with known central charge does fix the chiral central charge of its TQFT bulk.

\section{Mathematical Results on Gauging, Cosets and Dualities} \label{MathematicsSection}

In this appendix we summarize some mathematical nomenclature and results from \cite{Frohlich:2003hm} that were claimed in Section \ref{GeneralNonAbelianAnyonCondesationPicture} and allow a better understanding of it. We separate this appendix in two subsections. In the first one we mostly summarize definitions, while in the second one we discuss certain theorems on Local Algebra Modules pertaining to the subject of cosets.

\subsection{Frobenius Algebras and Local A-Modules} \label{FrobeniusAlgebras}

An (associative) algebra (with unit) $\mathcal{A}$ in a Ribbon category $\mathcal{C}$ is a triple $(\mathcal{A},m,\eta)$ consisting of an object $\mathcal{A}$ in $\mathcal{C}$, a multiplication morphism $m \in \mathrm{Hom}(\mathcal{A} \otimes \mathcal{A}, \mathcal{A})$ and a unit morphism $\eta \in \mathrm{Hom}(\mathbf{1},\mathcal{A})$ such that
\begin{equation}
    m \circ (m \otimes \mathrm{id}_{\mathcal{A}}) = m \circ (\mathrm{id}_{\mathcal{A}} \otimes m), \quad \mathrm{and} \quad m \circ (\eta \otimes \mathrm{id}_{\mathcal{A}}) = \mathrm{id}_{\mathcal{A}} = m \circ (\mathrm{id}_{\mathcal{A}} \otimes \eta).
\end{equation}
The multiplication and unit are often denoted pictorially as follows:
\begin{equation} \label{Multiplicationandunit}
\vcenter{\hbox{\includegraphics[scale = 0.08]{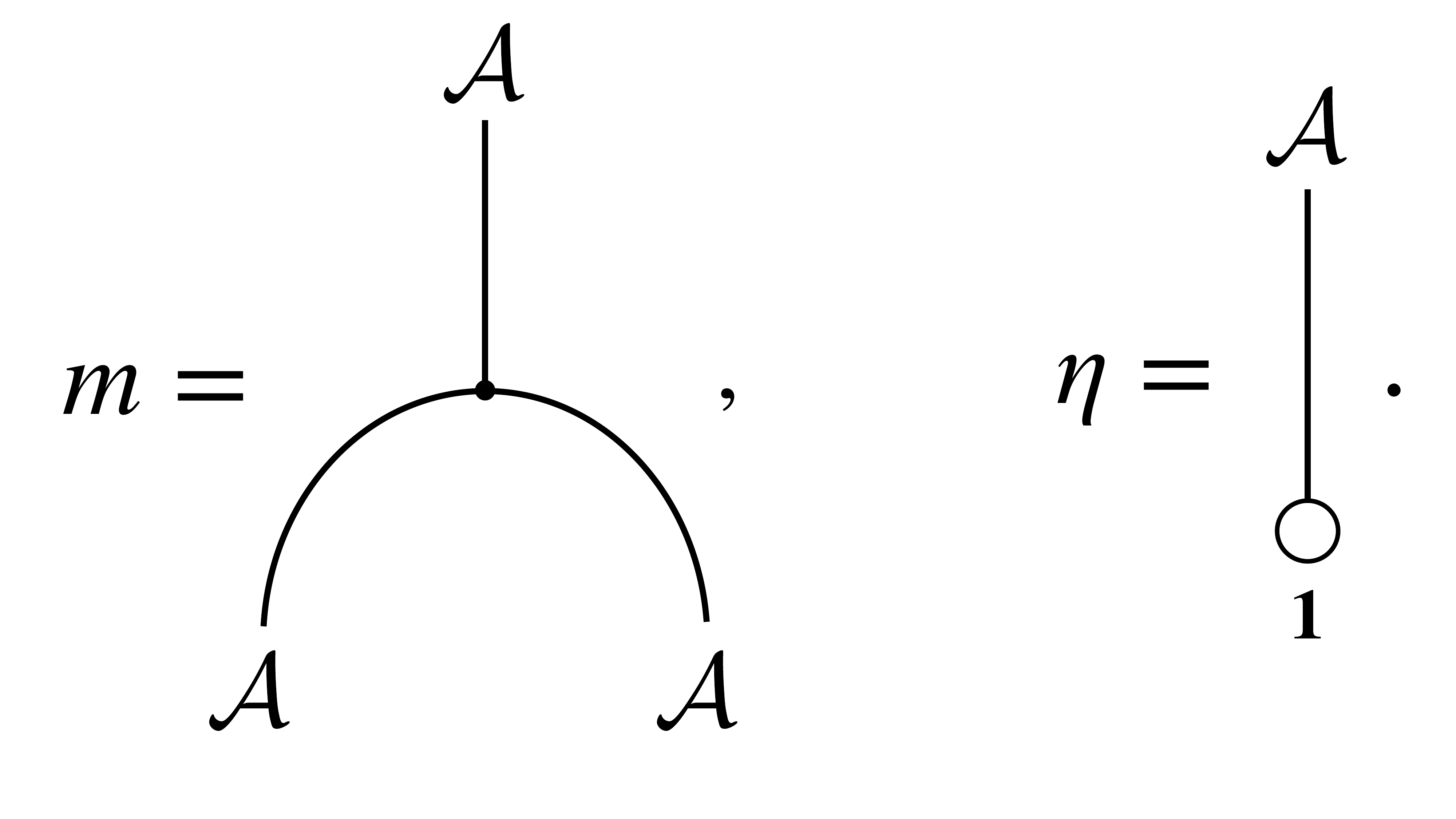}}} 
\end{equation}
Similarly, a coalgebra in $\mathcal{C}$ is a triple $(\mathcal{A},\Delta,\epsilon)$ consisting of an object $\mathcal{A}$, a comultiplication map $\Delta \in \mathrm{Hom}(\mathcal{A}, \mathcal{A} \otimes \mathcal{A})$ and a counit $\epsilon \in \mathrm{Hom}(\mathcal{A},\mathbf{1})$ possessing analogous coassociativity and counit properties as above, as well as similar pictorial notations. We say an algebra $\mathcal{A}$ in a braided tensor category is \textit{commutative} if $m \circ c_{\mathcal{A},\mathcal{A}} = m$, and similarly for cocommutativity. \\

A left module over an algebra $\mathcal{A} \in \mathrm{Obj}(\mathcal{C})$ is a pair $(M,\rho_{M})$ consisting of an object $M$ in $\mathcal{C}$ and a morphism $\rho_{M} \in \mathrm{Hom}(\mathcal{A} \otimes M,M)$ such that
\begin{equation}
    \rho_{M} \circ (m \otimes \mathrm{id}_{M}) = \rho_{M} \circ (\mathrm{id}_{M} \otimes \rho_{M}), \quad \mathrm{and} \quad \rho_{M} \circ (\eta \otimes \mathrm{id}_{M}) = \mathrm{id}_{M}.
\end{equation}
For brevity we will refer to left modules just as modules. The definition of right modules is analogous. Taking modules as objects, and the subspaces
\begin{equation}
    \{ f \in \mathrm{Hom}(N,M) | f \circ \rho_{N} = \rho_{M} \circ (\mathrm{id}_{\mathcal{A}} \otimes f)\}
\end{equation}
as morphisms from $(N,\rho_{N})$ to $(M,\rho_{M})$, one defines the category of (left) $\mathcal{A}$-modules (and similarly when considering right modules instead).

Relatedly, an $\mathcal{A}$-bimodule is a triple $(M,\rho^{l}_{M},\rho^{r}_{M})$ such that $(M,\rho^{l}_{M})$ is a left $\mathcal{A}$-module, $(M,\rho^{r}_{M})$ is a right $\mathcal{A}$-module, and the left and right actions of $\mathcal{A}$ commute. \\

We are interested in the following definitions:
\begin{itemize}
    \item An algebra endowed with a counit $\epsilon \in \mathrm{Hom}(\mathcal{A}, \mathbf{1})$ in a Ribbon category is called \textit{symmetric} if the two morphisms defined pictorially
     \begin{equation} \label{symmetric}
     \vcenter{\hbox{\includegraphics[scale = 0.125]{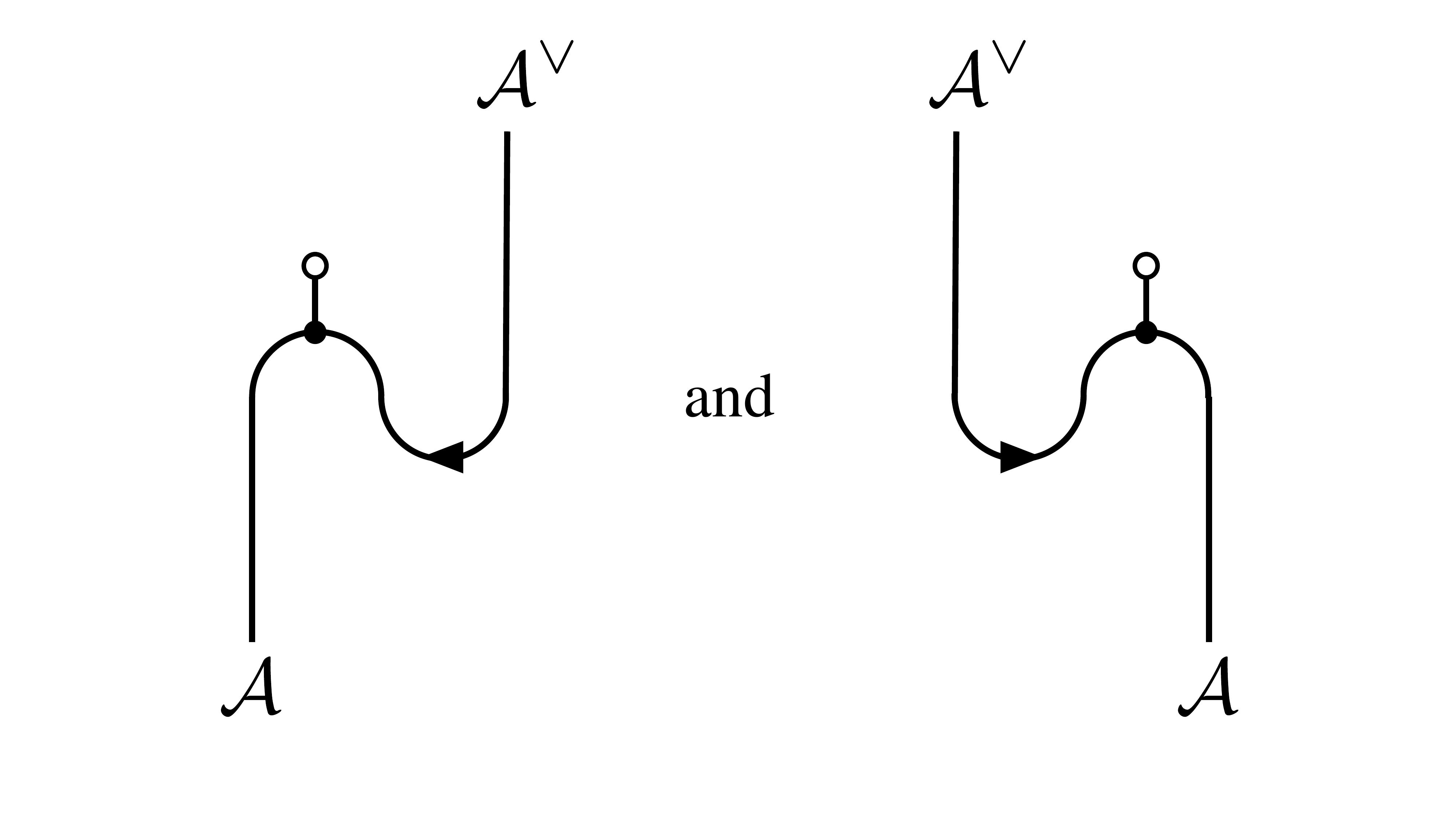}}} 
     \end{equation}
in $\mathrm{Hom}(\mathcal{A},\mathcal{A}^{\vee})$ are equal.

    \item An algebra in a tensor category is \textit{Frobenius} if we have a quintuple $(\mathcal{A},m,\eta,\Delta,\epsilon)$ such that $(\mathcal{A},m,\eta)$ is an algebra, $(\mathcal{A},\Delta,\epsilon)$ a coalgebra, satisfying the compatibility condition
    \begin{equation}
        (\mathrm{id}_{\mathcal{A}} \otimes m) \circ (\Delta \otimes \mathrm{id}_{\mathcal{A}}) = \Delta \circ m = (m \otimes \mathrm{id}_{\mathcal{A}}) \circ (\mathrm{id}_{\mathcal{A}} \otimes \Delta).
    \end{equation}

    \item We call a Frobenius algebra \textit{special} if $\epsilon \circ \eta = \beta_{\mathbf{1}} \, \mathrm{id}_{\mathbf{1}}$ and $m \circ \Delta = \beta_{\mathcal{A}} \, \mathrm{id}_{\mathcal{A}}$, for some non-zero numbers $\beta_{\mathbf{1}}$, $\beta_{\mathcal{A}}$.

    \item An algebra $\mathcal{A}$ is called \textit{simple} if all bimodule endomorphisms when we consider $\mathcal{A}$ as a bimodule over itself are proportional to the identity morphism.
\end{itemize}

\noindent The following results are useful to know:

\begin{itemize}
    \item A commutative symmetric Frobenius algebra $\mathcal{A}$ has trivial twist $\theta_{\mathcal{A}} = \mathrm{id}_{\mathcal{A}}$. Physically, this mathematical statement captures the fact that only bosons are non-anomalous, and thus, only them can participate in a commutative Frobenius algebra.

    \item Conversely, every commutative Frobenius algebra with trivial twist is symmetric.

    \item A commutative symmetric Frobenius algebra is also cocommutative.
\end{itemize}

A module $(M,\rho_{M})$ over a commutative symmetric special Frobenius algebra $\mathcal{A}$ in a Ribbon category is called \textit{local} if and only if
\begin{equation}
    \rho_{M} \circ P^{l/r}_{\mathcal{A}} (M) = \rho_{M},
\end{equation}
where $P^{l}_{\mathcal{A}}(a)$ is the following morphism defined pictorially by the combination of multiplications, comultiplications and braiding:
\begin{equation} \label{leftlocalmorphism}
\vcenter{\hbox{\includegraphics[scale = 0.12]{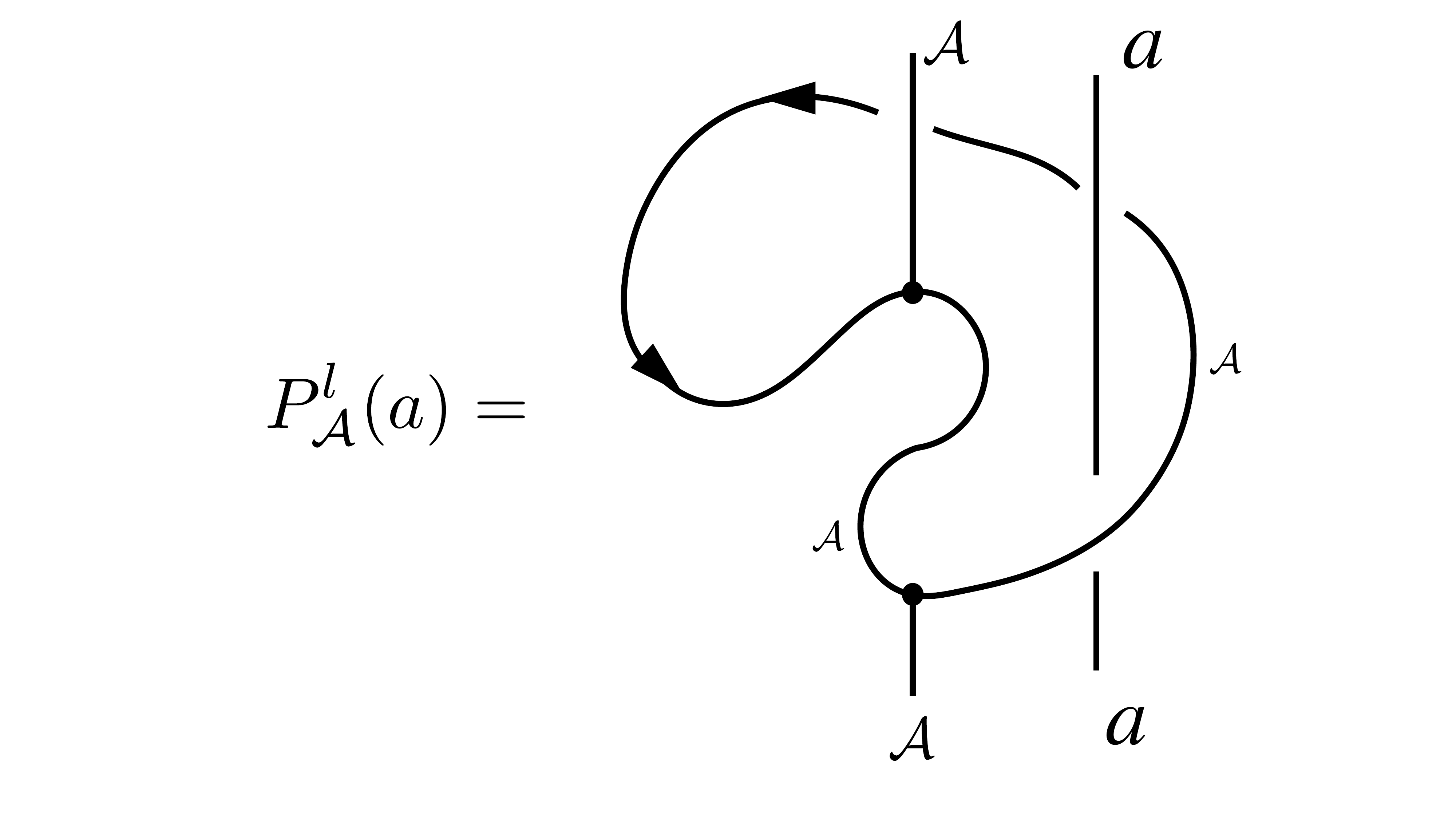}}} 
\end{equation}
and an analogous pictorial expression defines $P^{r}_{\mathcal{A}}(a)$ with all the braidings above reversed. \\

These are all the definitions we will need. The main result (or rather, a version of it) for which we need the previous definitions, and upon which the theory of non-invertible anyon condensation rests, is the following rather fundamental theorem of algebras over modular tensor categories: \textit{If we have a special symmetric commutative Frobenius algebra $\mathcal{A}$ in a modular tensor category $\mathcal{C}$, and if additionally $\mathcal{A}$ is a simple algebra, then the category $\mathcal{C}^{\mathrm{loc}}_{\mathcal{A}}$ of local $\mathcal{A}$-modules is also a modular tensor category.} The latter MTC is what we physically refer to as ``the gauged theory'' or ``the condensed theory.'' \\

Throughout this work we have mentioned multiple times the concept of a ``Lagrangian algebra.'' For our purposes a Lagrangian algebra is a special type of Frobenius algebra such that after condensation, we obtain the trivial theory as a result. If we condense just on half of spacetime then, we obtain a gapped boundary for the original theory before condensation. For a precise mathematical definition, see \cite{kong2014anyon}. An important statement is that any theory of the form $\mathcal{C} \boxtimes \bar{\mathcal{C}}$ has a Lagrangian algebra (the ``diagonal'' Lagrangian algebra) of the form
\begin{equation}
    \mathcal{A} = \sum_{c \in \mathcal{I}} \, (c, \tilde{c}) \, ,
\end{equation}
where the sum runs over all simple objects of $\mathcal{C}$ and $\tilde{c}$ stands for the image of $\mathcal{C}$ in $\bar{\mathcal{C}}$, where $\bar{\mathcal{C}}$ is the time-reversal of $\mathcal{C}$.

\subsection{Local Algebra Modules and Cosets}  \label{LocalAlgebraModulesandCosets}

Throughout the text we have given a rather physical picture of the ideas at play. Here we focus on a few more mathematical statements that are the basis for many claimed statements, mainly in Section \ref{noninvertibleextensionsubsection}. In Appendices \ref{mathsappendix} and \ref{FrobeniusAlgebras} we have already stated and unpacked with more precision many of the definitions used here.

In the following, we have to recall that the TQFT obtained by the gauging of a one-form symmetry in a 3D TQFT described by some MTC $\mathcal{C}$, is given by the category $\mathcal{C}^{\mathrm{loc}}_{\mathcal{A}}$ of local $\mathcal{A}$-modules for some special symmetric commutative Frobenius algebra object $\mathcal{A}$ over $\mathcal{C}$ (see Appendix \ref{FrobeniusAlgebras} for definitions). The category $\mathcal{C}^{\mathrm{loc}}_{\mathcal{A}}$ is again an MTC, so it appropriately describes a new 3D TQFT descending from the parent one described by $\mathcal{C}$. We stress that from the physics viewpoint the previous result does not necessarily restrict to the notion of gauging by some group-like symmetry, so the construction generically involves gauging by a non-invertible symmetry. An important constraint over $\mathcal{C}^{\mathrm{loc}}_{\mathcal{A}}$ is that its total quantum dimension must satisfy \cite{Frohlich:2003hm}:
\begin{equation} \label{quantumdimensionconstraint}
    \mathcal{D}_{\mathcal{C}^{\mathrm{loc}}_{\mathcal{A}}} = \frac{\mathcal{D}_{\mathcal{C}}}{\mathrm{dim}(\mathcal{A})^{2}},
\end{equation}
where $\mathrm{dim}(\mathcal{A})$ is defined as the sum of the quantum dimensions of each simple anyon in $\mathcal{A}$.

The primary focus of this section is to state and discuss the main result of \cite{Frohlich:2003hm} in our context. This result states that when a MTC $\mathcal{M}$ is written as the category of local $\mathcal{A}$-modules for some commutative symmetric special Frobenius algebra $\mathcal{A}$ in the direct product of two MTCs $\mathcal{C}$ and $\mathcal{M}'$:
\begin{equation} \label{originalcosetstatement}
    \mathcal{M} \cong (\mathcal{C} \boxtimes \mathcal{M}')^{\mathrm{loc}}_{\mathcal{A}},
\end{equation}
and if we require the algebra $\mathcal{A}$ in $\mathcal{C} \boxtimes \mathcal{M}'$ to be such that the only subobject of $\mathcal{A}$ of the form $a \times \mathbf{1}$ is $\mathbf{1} \times \mathbf{1}$, where $\mathbf{1}$ is the tensor unit, then there exists a commutative symmetric special Frobenius algebra $\mathcal{B}$ in the direct product $\mathcal{M} \boxtimes \overline{\mathcal{M}'}$, such that if we take the corresponding category of local modules, we obtain
\begin{equation} \label{cosetisolation}
    \mathcal{C} \cong (\mathcal{M} \boxtimes \overline{\mathcal{M}'})^{\mathrm{loc}}_{\mathcal{B}}.
\end{equation}
The algebra $\mathcal{B}$ can in principle be computed from the categories $\mathcal{M}$ and $\mathcal{M'}$, and the modularity of $\mathcal{C}$ can actually be derived from that of $\mathcal{M}$ and $\mathcal{M}'$. In practice, it is often found to be more useful to inspect for a Frobenius algebra object (a set of condensing bosons), and carry the non-abelian anyon condensation procedure.

The previous is often sufficient for most practical applications, but it is important to point out that if we do not require $\mathcal{A}$ to fulfill the condition below Eqn. \eqref{originalcosetstatement}, then a version of the previous theorem still holds. Namely, there exist certain algebras $\mathcal{T}_{1}$ and $\mathcal{T}_{2}$ such that if we take local modules:
\begin{equation} \label{actualcosettheorem}
    \mathcal{C}^{\mathrm{loc}}_{\mathcal{T}_{1}} \cong (\mathcal{M} \boxtimes \overline{\mathcal{M}'})^{\mathrm{loc}}_{\mathcal{T}_{2}}.
\end{equation}
Essentially, what the requirement over $\mathcal{A}$ below \eqref{originalcosetstatement} does is to trivialize the gauging on the left-hand side.

The previous mathematical results have a clear interpretation in physics. In the standard coset construction, Eqn. \eqref{originalcosetstatement} is the statement that an affine lie algebra embeds in another one, with the corresponding TQFTs described by the MTCs $\mathcal{M}'$ and $\mathcal{M}$ respectively. The coset theory, whose construction we reviewed in the CFT context in Section \ref{cosetinterfacebulkboundaryreview}, is then described in the corresponding TQFT setup by the MTC $\mathcal{C}$ whose content is in principle determined by the data of $\mathcal{M}$ and $\mathcal{M}'$ via \eqref{cosetisolation}. Notice however that the previous theorem in some sense generalizes the standard GKO coset construction since nowhere above it was needed that $\mathcal{M}$ or $\mathcal{M}'$ were MTCs associated with those of an affine Lie algebra. They may be arbitrary MTCs (see \cite{Bais:1987zk} for some work on generalizing the coset construction to higher spin currents in the context of 2D CFT). More importantly, a second point of generalization, stressed above, is that $\mathcal{B}$ in \eqref{cosetisolation} does not necessarily have to correspond to some abelian gauging. As seen in the previous section, one instance where this phenomenon takes place would be the Maverick cosets \cite{Fuchs:1995tq, Pedrini:1999iy, Dunbar:1992gh, Dunbar:1993hr,Bais:2008ni}.

One form of the statement that a duality exists follows from this mathematical perspective when the same MTC $\mathcal{C}$, not (necessarily) having a path-integral or gauge theory description, appears ``connecting'' two different pairs of MTCs $(\mathcal{M}_{1},\mathcal{M}'_{1})$ and $(\mathcal{M}_{2}, \mathcal{M}'_{2})$ having such a description. That is:
\begin{equation}
    \mathcal{M}_{1} \cong (\mathcal{C} \boxtimes \mathcal{M}'_{1})^{\mathrm{loc}}_{\mathcal{A}_{1}}, \quad \mathrm{and} \quad \mathcal{M}_{2} \cong (\mathcal{C} \boxtimes \mathcal{M}'_{2})^{\mathrm{loc}}_{\mathcal{A}_{2}},
\end{equation}
for some Frobenius algebras $\mathcal{A}_{1}$ and $\mathcal{A}_{2}$. By the theorem above we can isolate $\mathcal{C}$ in two different ways, and write
\begin{equation}
    (\mathcal{M}_{1} \boxtimes \overline{\mathcal{M}'_{1}})^{\mathrm{loc}}_{\mathcal{B}_{1}} \cong (\mathcal{M}_{2} \boxtimes \overline{\mathcal{M}'_{2}})^{\mathrm{loc}}_{\mathcal{B}_{2}}
\end{equation}
for some Frobenius algebras $\mathcal{B}_{1}$ and $\mathcal{B}_{2}$. For instance, these could correspond to two different descriptions of the parafermions. One of them could be a Maverick coset description, as in Section \ref{maverickdualities}, and the other could be the standard coset description $SU(2)_{k} \times U(1)_{-2k}/\mathbb{Z}_{2}$ \cite{DiFrancesco:1997nk}. This is essentially the path we followed for the first infinite family of Maverick dualities studied in Section \ref{maverickdualities}. It could also be that $\mathcal{C}$ itself admits a field theory description, in which case \eqref{cosetisolation} allows us to express such a description in terms of a different one based on $\mathcal{M}$ and $\mathcal{M}'$. The latter is what happens with the conformal embedding $SU(N)_{k} \times SU(k)_{N} \hookrightarrow SU(Nk)_{1}$ reviewed at the end of Section \ref{cosetinterfacebulkboundaryreview}. Here we can take a standard Chern-Simons description for $SU(k)_{N}$, but we can take an alternative one given by the right-hand side of \eqref{cosetisolation}. The latter is nothing but the $SU(Nk)_{1} \times SU(N)_{-k} / \mathbb{Z}_{N}$ expression of the same theory, which is essentially the content of the duality \eqref{standardunitaryembedding}.

The form \eqref{actualcosettheorem} of the theorem is sometimes useful. To illustrate this \cite{Frohlich:2003hm}, we can consider in the contexts of the conformal embeddings:
\begin{equation}
    \frac{(E_{8})_{1}}{SU(3)_{6} \times SU(2)_{16}},
\end{equation}
which implies that $(E_{8})_{1}$ can be written as in \eqref{originalcosetstatement} with $\mathcal{C} = SU(3)_{6}$ and $\mathcal{M}' = SU(2)_{16}$. In this case we need to use \eqref{actualcosettheorem}, and fortunately the gaugings are abelian so we can easily verify that
\begin{equation}
    SU(3)_{6}/\mathbb{Z}_{3} \cong (E_{8})_{1} \times SU(2)_{-16}/\mathbb{Z}_{2}.
\end{equation}
That is, both sides involve a non-trivial gauging, which is the main point of the theorem in the form \eqref{actualcosettheorem}.

\section{Further Examples on Non-Abelian Anyon Condensation}

In this appendix we present in detail various check example computations of non-invertible anyon condensation that pertain to the global subject of the paper but for the sake of organization we have summarized here instead. The reader may want to first look at the beginning of Section \ref{ExplicitChecks} where we outline the rules that we use below to verify such examples on non-invertible anyon condensation.

\subsection{$(G_{2})_{1} \times (G_{2})_{-1} \longrightarrow \mathbbm{1}$} \label{G21DrinfeldToVacuum}

This is a rather trivial example of condensation by a non-abelian anyon that condenses the theory to the trivial theory. The example is trivial in the sense that the theory is of the form $G_{k} \times G_{-k}$, which is known to condense to the vacuum \cite{muger2001subfactors,Davydov:2011pp,davydov2013witt}. Equivalently, a gapped boundary exists that separates the trivial theory from $G_{k} \times G_{-k}$, which hosts topological degrees of freedom described by the $G_{k}/G_{k}$ topological coset. For completeness, we write here the condensation computation explicitly, since this particular example is easy and it appears twice in this work, both in our example in Section \ref{sec:introG2} and in the derivation of the three-state Potts model Maverick duality from exceptional conformal embeddings.

The spectrum of $(G_{2})_{1}$ is shown in Table \ref{g2lv1table}. The lines in the double are denoted $(0,0)$, $(\phi,0)$, $(0,\phi)$, and $(\phi, \phi)$. Here we have abused notation and called $\phi$ to the non-trivial entry in both $(G_{2})_{1}$ and $(G_{2})_{-1}$, but where there is no ambiguity as we can tell them apart by their position in the ordered pair. 

Assume that the unique non-trivial boson $(\phi,\phi)$ condenses. That is:
\begin{equation}
    (\phi, \phi) \longrightarrow 0 + (\phi, \phi)_{2},
\end{equation}
or in mathematical terms, the algebra that we condense is
\begin{equation}
    \mathcal{A} = (0,0) + (\phi,\phi),
\end{equation}
which is nothing but the Lagrangian algebra given by the diagonal anyons that always exists in a theory of the form $\mathcal{C} \times \overline{\mathcal{C}}$, with $\mathcal{C} = (G_{2})_{1}$ here.

By conservation of the quantum dimension $d_{(\phi, \phi)_{2}} = \frac{1+\sqrt{5}}{2}$, implying that $(\phi, \phi)$ can only split into two lines. Further, since $(\phi,\phi)$ and 0 are self-conjugate, it follows that $(\phi, \phi)_{2}$ also is self-conjugate. It is easy to check that the other lines $(\phi,0)$ and $(0,\phi)$ cannot split and are also self-conjugate.

Since $(\phi, \phi)$ condenses, the fusion rules
\begin{equation}
    (\phi, \phi) \times (0, \phi) = (\phi, 0) + (\phi, \phi) \longrightarrow 0 + \ldots,
\end{equation}
\begin{equation}
    (\phi, \phi) \times (\phi, 0) = (0, \phi) + (\phi, \phi) \longrightarrow 0 + \ldots
\end{equation}
imply that $(\phi, \phi)$ contains in its restriction the conjugates of $(0, \phi)$ and $(\phi, 0)$, which are however self-conjugate. Since their quantum dimensions are $d_{(0, \phi)} =  d_{(\phi, 0)} = (1+\sqrt{5})/2$ they cannot be identified to the vacuum, and must be identified with $(\phi, \phi)_{2}$. In other words, we have the identification
\begin{equation}
    (0, \phi) \cong (\phi,0) \cong (\phi, \phi)_{2}.
\end{equation}
These labels therefore lift to lines in the parent theory with different topological spins, and so it follows that all such excitations confine. The only non-trivial line in the gauged theory is thus the vacuum, obtaining the expected result.

\subsection{Checking the $(G_{2})_{1} \cong \big( (F_{4})_{1} \times SO(3)_{-4} \big)/\mathcal{Z}(\mathbf{Fib})$ Duality} \label{ExampleforsimplestMaverick}

This calculation arises when we prove the duality suggested by the simplest Maverick coset \eqref{TSPMDuality} using exceptional conformal embeddings. See the previous subsection. The spectrum of $(F_{4})_{1}$ is that of $(G_{2})_{1}$, but with the only non-trivial line having spin $3/5$ instead of $2/5$. We call $\phi$ the only non-trivial anyon of $(F_{4})_{1}$, which obeys Fibonacci fusion rules. The spectrum of $SO(3)_{4}$ is shown in Table \ref{so3lv4table}, and its non-trivial fusion rules are
\begin{align} \label{so3lv4fusionrules}
    &2 \times 2 = 0+2+4_{1}+4_{2}, \quad 2 \times 4_{1} = 2 + 4_{2}, \quad 2 \times 4_{2} = 2 + 4_{1}, \nonumber \\[0.3cm]
    &4_{1} \times 4_{1} = 0 + 4_{1}, \quad 4_{1} \times 4_{2} = 2, \quad 4_{2} \times 4_{2} = 0 + 4_{2}.
\end{align}

\begin{table}[t]
\centering
\begin{tabular}[h]{|p{4cm}|p{4cm}|p{4cm}| }
\hline 
\multicolumn{3}{|c|}{$SO(3)_{4}$} \\
\hline
Line label & Quantum Dimension & Conformal Weight \\
\hline
0 & $d_{0}=1$ & $h_{0}=0$ \\
2 $\cong$ (4,0) & $d_{2} = (3+\sqrt{5})/2$ & $h_{2}=1/5$ \\
$4_{1}$ $\cong$ (2,4) & $d_{4_{1}} = (1 + \sqrt{5})/2$ & $h_{4_{1}}=3/5$ \\
$4_{2}$ $\cong$ (3,3) & $d_{4_{2}} = (1 + \sqrt{5})/2$ & $h_{4_{2}}=3/5$ \\
\hline
\end{tabular}
\caption{$SO(3)_{4}$ data.}  \label{so3lv4table}
\end{table}

We first need to determine the bosons that condense. It is easy to verify that the only bosons in the product $(F_{4})_{1} \times SO(3)_{-4}$ are $(\phi,4_{1})$ and $(\phi,4_{2})$, which are non-abelian. From the fusions
\begin{align}
    &(\phi,4_{1}) \times (\phi,4_{1}) = (0,0) + (0,4_{1}) + (\phi,0) + (\phi,4_{1}), \\[0.3cm]
    &(\phi,4_{1}) \times (\phi,4_{2}) = (0,2) + (\phi,2), \\[0.3cm]
    &(\phi,4_{2}) \times (\phi,4_{2}) = (0,0) + (0,4_{2}) + (\phi,0) + (\phi,4_{2}),
\end{align}
it is clear that $(\phi,4_{1})$ and $(\phi,4_{2})$ cannot simultaneously condense, so me must make a choice. However, lines $4_{1}$ and $4_{2}$ are symmetric in $SO(3)_{4}$, so the choice is actually immaterial. We choose $(\phi,4_{1})$ to condense:
\begin{equation}
    (\phi,4_{1}) \longrightarrow 0 + (\phi,4_{1})_{2}.
\end{equation}
The quantum dimension of $(\phi,4_{1})_{2}$ is $d_{(\phi,4_{1})_{2}} = (1+\sqrt{5})/2$, so there is no further splitting allowed by conservation of the quantum dimension. Since $(\phi,4_{2})$ does not condense, we can also see from the above fusion rules that it does not split and $d_{(\phi,4_{2})} = (3 + \sqrt{5})/2$.

The quantum dimensions do not allow $(0,4_{1})$ and $(0,4_{2})$ to split, but in principle they allow for $(0,2)$ to split. This is not the case, which can be verified from
\begin{equation}
    (0,2) \times (0,2) = (0,0) + (0,2) + (0,4_{1}) + (0,4_{2})
\end{equation}
since no non-trivial bosons appear on the right-hand side. Since none of these anyons split, the fusion
\begin{equation}
    (\phi,4_{1}) \times (0,4_{1}) = (\phi, 0) + (\phi, 4_{1}) \longrightarrow 0 +\ldots,
\end{equation}
implies the identification $(\phi,4_{1})_{2} \cong (0,4_{1})$, which in turn implies the confinement of such excitations.

We now consider the rest of the $(\phi,i)$ anyons that are not bosons. Start noticing that
\begin{equation}
    (\phi, 2) \times (\phi, 2) = (0,0) + (\phi, 4_{1}) + \ldots, \longrightarrow0+0+\ldots,
\end{equation}
implies that $(\phi, 2)$ splits: $(\phi, 2) \to (\phi, 2)_{1} + (\phi, 2)_{2}$. To assign quantum dimensions we can consider the fusion
\begin{equation}
    (\phi,2) \times (0,2) = (\phi,4_{1}) + \ldots \longrightarrow 0 + \ldots,
\end{equation}
so $(0,2)$ belongs to the restriction of $(\phi,2)$. Let $(\phi, 2)_{2}$ be by definition the component that identifies with $(0,2)$. The quantum dimensions must then be $d_{(\phi,2)_{1}} = (1 + \sqrt{5})/2$, and $d_{(\phi,2)_{2}} = (3 + \sqrt{5})/2$. With this knowledge it is now straightforward to take the fusions $(\phi,2) \times (0,4_{2})$ and $(\phi,2) \times (\phi,4_{2})$ and deduce the identifications $(\phi,2)_{1} \cong (0,4_{2})$ and $(\phi,2)_{2} \cong (\phi,4_{2})$ using the by-now usual arguments.

Finally, $(\phi,0)$ is self-conjugate and cannot split. Moreover, the fusion
\begin{equation}
   (\phi,0) \times (0,4_{1}) = (\phi,4_{1}) + \ldots \to 0+\ldots
\end{equation}
implies the identification $(\phi,0) \cong (0,4_{1})$, which in turn shows the confinement of such excitations. 

To summarize, we have the following condensation pattern:
\begin{align}
    &(0,0) \to 0, \quad (0,2) \to (0,2), \quad (0,4_{1}) \to(0,4_{1}), \quad (0,4_{2}) \to (0,4_{2}), \\[0.3cm] &(\phi, 0) \to (0,4_{1}), \quad (\phi,2) \to (0,4_{2}) + (0,2), \quad (\phi,4_{1}) \to 0 + (0,4_{1}), \quad (\phi,4_{2}) \to (0,2).
\end{align}

Studying the lifts implied by the previous restrictions to the anyons in the parent theory, we can check that the unconfined excitations are the vacuum and $(0,4_{2})$. These have indeed the correct quantum dimensions, fusion rules and spin to recognize the result as $(G_{2})_{1}$.

\subsection{$Spin(9)_{1} \cong \big( Spin(3)_{3} \times Spin(3)_{3} \big)/\mathcal{A}$} \label{su26su26intospin91}

In this subsection we consider an example in the infinite family of conformal embeddings corresponding to spin groups:
\begin{equation}
    \frac{Spin(Nk)_{1}}{Spin(N)_{k} \times Spin(k)_{N}}, \ \mathrm{for} \quad k,N \ \mathrm{odd}.
\end{equation}
This family of embeddings is intimately related to level-rank dualities for orthogonal groups (see \cite{Aharony:2016jvv,Cordova:2017vab, Cordova:2017kue}), and furthermore conformal embeddings with spin groups in the numerator have the very important property of describing the low-energy dynamics of 2D QCD \cite{Komargodski:2020mxz,Delmastro:2021otj,Delmastro:2022prj}.\footnote{More precisely, fermionic conformal embeddings $SO(N)_{1}/ H_{\tilde{k}}$ describe the low-energy dynamics of 2D QCD with gauge group $H$ and $\tilde{k}$ the index of the embedding, which in this case corresponds to the Dynkin index of the representation in which the fermions transform.}\footnote{Rigorously speaking this result is conjectural, but supported by highly non-trivial evidence on its support like the matching of anomalies.} For the specific case $N=k=\nu$ the previous conformal embeddings are also intrinsically related to anomalous theories for time-reversal symmetry that realize a value $\nu$ for such anomaly \cite{cheng2018microscopic, Cordova:2017kue} (once we add fermionic invertible factors).

We will consider the example $N=k=3$ as this gives rise to the simplest example where the numerator ($Spin(9)_{1}$) has interesting non-abelian fusion rules; namely, Ising fusion rules. In this case, there exists the exceptional isomorphism of chiral algebras $Spin(3)_{3} \cong SU(2)_{6}$. So the task consists of condensing some (non-abelian) boson(s) in $SU(2)_{6} \times SU(2)_{6}$, and matching the unconfined lines of the condensed theory with those of $Spin(9)_{1}$. The spectrum of $SU(2)_{6}$ is shown in Table \ref{su2lv6table}, and the fusion rules can be read from \eqref{SU2kFusionRules}. The lines in the product theory are denoted as $(i,j)$, where $i,j=0,\ldots,6$ is the label of a single factor.

There are four bosons in the product theory $SU(2)_{6} \times SU(2)_{6}$: $(0,0)$,  $(2,4)$, $(4,2)$, and $(6,6)$. The latter is the ``common center'' of the product, which is abelian. The remaining two non-trivial bosons are non-abelian. Merely condensing the common center in $Spin(4)_{6} \cong SU(2)_{6} \times SU(2)_{6}$ leads to the well-known answer $SO(4)_{6}$. Then, as expected, one needs to take at least one of the non-abelian bosons to condense in order to obtain $Spin(9)_{1}$.

\begin{table}[t]
\centering
\begin{tabular}[h]{|p{4cm}|p{4cm}|p{4cm}| }
\hline 
\multicolumn{3}{|c|}{$SU(2)_{6}$} \\
\hline
Line label & Quantum Dimension & Conformal Weight \\
\hline
0 & $d_{0} = 1$ & $h_{0}=0$ \\
1 & $d_{1} = \sqrt{2 + \sqrt{2}}$ & $h_{1}=3/32$ \\
2 & $d_{2} = 1 + \sqrt{2}$ & $h_{2}=1/4$ \\
3 & $d_{3} = \sqrt{2(2 + \sqrt{2})}$ & $h_{3}=15/32$ \\
4 & $d_{4} = 1 + \sqrt{2}$ & $h_{4}=3/4$ \\
5 & $d_{5} = \sqrt{2 + \sqrt{2}}$ & $h_{5}=35/32$ \\
6 & $d_{6} = 1$ & $h_{6}=3/2$ \\
\hline
\end{tabular}
\caption{$SU(2)_{6}$ data.}  \label{su2lv6table}
\end{table}

Assume that $(2,4)$ condenses; that is:
\begin{equation}
    (2,4) \longrightarrow 0 + \ldots
\end{equation}
It is easy to see that since
\begin{equation}
    (0,a) \times (0,a) = (0,0) + \sum_{i} (0,a_{i}),
\end{equation}
the lines $(0,a)$ and $(a,0)$ do not split and are self-conjugate. With this observation, consider the fusion
\begin{equation}
    (0,2) \times (2,4) = (2,2) + (2,4) + (2,6) \longrightarrow 0 + \ldots,
\end{equation}
which implies that $(0,2)$ belongs to the restriction of $(2,4)$. This means that $(2,4)$ restricts as
\begin{equation} \label{24splitting}
    (2,4) \longrightarrow 0 + (2,4)_{2} + (2,4)_{3},
\end{equation}
with $(2,4)_{2} \cong (0,2)$, and since $d_{(0,2)} = 1 + \sqrt{2}$ by conservation of the quantum dimension we must have that $d_{(2,4)_{3}} = 1 + \sqrt{2}$. Notice that a priori the quantum dimension is sufficiently large to allow $(2,4)$ to split into four lines, but this is ruled out by the fusion
\begin{equation}
    (2,4) \times (2,4) = (0,0) + (2,4) + (4,2) + \ldots,
\end{equation}
since there are not enough bosons on the right side that could potentially condense to accommodate four vacua. Furthermore, we also see from this fusion rule that $(2,4)$ can split as \eqref{24splitting} only if $(4,2)$ also condenses. Thus we have deduced that if $(2,4)$ condenses, $(4,2)$ also condenses.

By a similar argument, $(2,0)$ also belongs to the restriction of $(2,4)$. However, $(0,2)$ and $(2,0)$ cannot be identified with each other. To see this, consider the fusion with $(0,6)$ which has unit quantum dimension:
\begin{align}
    (0,6) \times (2,4) &= (2,2) = (0,2) \times (2,0) \overset{!}{=} (0,2) \times (0,2) = (0,0) + (0,2) + (0,4)  \\[0.3cm] &= (0,6) + \ldots,
\end{align}
where in the third equality in the first row we have (wrongly) assumed that $(0,2)$ and $(2,0)$ identify, and in the second row we have first restricted $(2,4)$ on the left-hand side and isolated the trivial fusion between the vacuum and $(0,6)$. Since both rows must agree, $(0,6)$ must identify with some line in the last equality of the first row, but only one line has unit quantum dimension; namely, the vacuum. This identification, however, is equivalent to the statement that $(0,6)$ condenses, which is not possible since it does not have the correct topological spin to do so. We thus reach the conclusion that $(0,2)$ and $(2,0)$ cannot be identified with each other.

Gathering this knowledge, we find the restrictions of $(2,4)$ and $(4,2)$ (the latter follows from the same arguments, once we know it condenses as we have shown above):
\begin{equation}
    (2,4) \longrightarrow 0 + (0,2) + (2,0),
\end{equation}
\begin{equation}
    (4,2) \longrightarrow 0 + (0,2) + (2,0).
\end{equation}
We now deduce a few more identifications. First:
\begin{equation}
    (2,4) \times (6,6) = (4,2) \Longrightarrow (6,6) + \ldots = 0 + \ldots,
\end{equation}
where we have used the restrictions of $(2,4)$ and $(4,2)$, and then isolated the fusion of $(6,6)$ with the vacuum on the left-hand side. Only the vacuum on the right-hand side has unit quantum dimension that can match that of $(6,6)$. Thus, we have deduced that $(6,6)$ condenses: $(6,6) \to 0$.  

In passing, notice that what we have just found is that the condensing algebra is
\begin{equation} \label{spin91algebraobject}
    \mathcal{A} = (0,0) + (2,4) + (4,2) + (6,6),
\end{equation}
and in these terms what we want to show is that
\begin{equation}
    Spin(9)_{1} \cong \frac{Spin(3)_{3} \times Spin(3)_{3}}{\mathcal{A}},
\end{equation}
for the algebra object $\mathcal{A}$ in \eqref{spin91algebraobject} above.

Now $(4,0) \times (0,2) = (4,2) \to 0 + \ldots$, but $(0,a)$ and $(a,0)$ are self-conjugate for any $a$, and thus we must have the identifications $(2,0) \cong (0,4)$, $(0,2) \cong (4,0)$. Similarly, $(2,6) \times (6,6) = (4,0)$, but $(6,6)$ restricts only to the vacuum, and thus we have the identification $(2,6) \cong (0,4)$. By similar arguments, we find, overall: $(2,6) \cong (4,0) \cong (0,2) \cong (6,4)$, and $(6,2) \cong (0,4) \cong (2,0) \cong (4,6)$. It is straightforward to check from these identifications that the corresponding labels confine on the child theory.

The simple currents $(0,6)$ and $(6,0)$ identify with each other since $(0,6) = (6,6) \times (6,0)$ and $(6,6)$ condenses. Using that $(4,4) = (6,6) \times (2,2)$ we obtain $(2,2) \cong (4,4)$, and using that $(2,2) = (0,6) \times (2,4)$ we obtain the restriction:
\begin{equation}
    (2,2) \longrightarrow (0,6) + (2,0) + (0,2).
\end{equation}
Notice that $(0,6)$ always lifts to anyons in the parent theory that share the same topological spin, and thus it does not confine.

We now study lines $(a,b)$ that have $a$ and $b$ odd integers. First, notice that since
\begin{equation}
    (1,1) \times (1,1) = (0,0) + (0,2) + (2,0) + (2,2), \label{11fusion11}
\end{equation}
(1,1) cannot split. A similar argument shows that $(1,5)$ also does not split. Since $(1,1) \times (6,6) = (5,5)$ and $(1,5) \times (6,6) = (5,1)$ and $(6,6)$ condenses, we find the identifications $(1,1) \cong (5,5)$ and $(1,5) \cong (5,1)$. Furthermore:
\begin{equation}
    (1,1) \times (1,5) = (2,4) + \ldots \longrightarrow 0 + \ldots,
\end{equation}
which means $(1,5)$ is conjugate to $(1,1)$ in the child theory, but $(1,1)$ is self-conjugate, which implies the further identifications $(1,1) \cong (1,5) \cong (5,1) \cong (5,5)$.

To make progress, study now the fusion rules
\begin{align}
    (3,3) \times (3,3) &= (0,0) + (2,4) + (4,2) + (6,6) + \ldots \longrightarrow 0+0+0+0+\ldots \label{33fusion33} \\[0.3cm] 
    (1,1) \times (3,3) &= (2,2) + (2,4) + (4,2) + (4,4) \longrightarrow 0+0+\ldots \label{11fusion33}
\end{align}
The first fusion rule allows two possibilities: either $(3,3)$ splits into four distinct labels or one label with multiplicity two. The first possibility is however not consistent with \eqref{11fusion33}, so it follows that $(3,3)$ restricts as $(3,3) \to 2(1,1)$ for consistency in between both \eqref{33fusion33} and \eqref{11fusion33}. This restriction also shows that $(1,1)$ confines, as it lifts to anyons in the parent theory with different topological spins.

Consider now the rest of the lines with two odd entries. In particular, let us consider the fusion rule
\begin{equation}
    (1,1) \times (1,3) = (0,2) + (0,4) + (2,2) + (2,4) \longrightarrow 0 + \ldots,
\end{equation}
which implies that $(1,1)$ belongs in the restriction of $(1,3)$ (a similar statement holds for $(3,1)$). In turn, by conservation of the quantum dimension, this implies that $(1,3)$ has a restriction of the form:
\begin{equation}
    (1,3) \longrightarrow (1,3)_{1} + (1,1),
\end{equation}
with $d_{(1,3)_{1}} = \sqrt{2}$. Clearly, there is no sufficient quantum dimension for further splitting.

Using the standard arguments: $(1,3) \times (6,6) = (5,3) \Rightarrow (1,3) \cong (5,3)$ and $(3,1) \times (6,6) = (3,5) \Rightarrow (3,1) \cong (3,5)$, meaning the restrictions we have to consider are
\begin{equation}
    (1,3) \longrightarrow (1,3)_{1} + (1,1),
\end{equation}
and
\begin{equation}
    (3,1) \longrightarrow (3,1)_{1} + (1,1).
\end{equation}

It is straightforward to check that the three fusion rules $(1,3) \times (1,3)$, $(1,3) \times (3,1)$, and $(3,1) \times (3,1)$ contain two vacua after restriction. Since we know $(1,1)$ is self-conjugate, this means that $(1,3)_{1}$ and $(3,1)_{1}$ must both be self-conjugate, but conjugate to each other as well. We have to then identify $(1,3)_{1} \cong (3,1)_{1}$. Clearly, the previous identifications do not lead to confinement of $(1,3)_{1}$.

Finally, the lines in $SU(2)_{6} \times SU(2)_{6}$ of the form $(2m,2n+1)$ with $m,n$ integer confine. For example, $(0,1) \times (6,6) = (6,5)$, and since $(6,6)$, condenses we have to identify $(0,1) \cong (6,5)$. However, these expressions lift to anyons of different topological spins and thus confine. The same argument holds for the rest of such lines.

All in all, the spectrum of unconfined lines that we have found is given by $(0,0)$, $(6,0)$, and $(1,3)_{1}$ with topological spins and quantum dimensions that match those of $Spin(9)_{1}$, in accordance with the conformal embedding $Spin(3)_{3} \times Spin(3)_{3} \hookrightarrow Spin(9)_{1}$. The previous can be readily verified by looking at the spectrum of $Spin(9)_{1}$, presented in Table \ref{spin9lv1table}.  \\

\begin{table}[t]
\centering
\begin{tabular}[h]{|p{4cm}|p{4cm}|p{4cm}| }
\hline 
\multicolumn{3}{|c|}{$Spin(9)_{1}$} \\
\hline
Line label & Quantum Dimension & Conformal Weight \\
\hline
0 & $d_{0} = 1$ & $h_{0}=0$ \\
1 & $d_{1} = 1$ & $h_{1}=1/2$ \\
2 & $d_{2} = \sqrt{2}$ & $h_{2}=9/16$ \\
\hline
\end{tabular}
\caption{$Spin(9)_{1}$ data.}  \label{spin9lv1table}
\end{table}

\subsubsection*{Fusion Rules}

In the current example the child theory has interesting fusion rules (Ising fusion rules), and the condensation procedure is also sufficiently straightforward to deduce them explicitly from the parent theory. We turn to do this next.

First, $(0,0)$ and $(6,0)$ descend trivially from the parent theory so they maintain $\mathbb{Z}_{2}$ fusion rules: 
\begin{equation}
    (0,6) \times (0,6) = (0,0). \label{isingfusion1}
\end{equation}
To deduce the fusion rule $(0,6) \times (1,3)_{1}$ we fuse $(0,6)$ with $(1,3)$:
\begin{align}
    (0,6) \times (1,3) &= (1,3) = (1,3)_{1} + (1,1) \\[0.3cm]
&= (0,6) \times (1,3)_{1} + (0,6) \times (1,1) = (0,6) \times (1,3)_{1} + (1,5) \cong (0,6) \times (1,3)_{1} + (1,1), \nonumber
\end{align}
where in the first row we have performed the fusion in the parent theory and restricted the result, while in the second row we have first restricted on the left-hand side, then performed the fusions from the lines that descend trivially from the parent, and then used the identification $(1,5) \cong (1,1)$. Since both rows must match, we have the fusion rule in the child theory:
\begin{equation}
    (0,6) \times (1,3)_{1} = (1,3)_{1}. \label{isingfusion2}
\end{equation}

Deducing $(1,3)_{1} \times (1,3)_{1}$ is slightly more complicated. We consider the fusion $(1,3) \times (1,3)$ in the parent theory:
\begin{align}
    (1,3) \times (1,3) &= (0,0) + (0,2) + (0,4) + (0,6) + (2,0) + (2,2) + (2,4) + (2,6)\label{line1} \\[0.3cm] &= (1,3)_{1}  \times (1,3)_{1} + (1,3)_{1} \times (1,1) + (1,1) \times (1,3)_{1}  + (1,1) \times (1,1), \label{line2}
\end{align}
where in the first row we have performed the fusion in the parent theory, while in the second row we have restricted the left-hand side first.

The matching of \eqref{line1} with \eqref{line2} is simplified if we use the explicit fusion in Eqn. \eqref{11fusion11}, finding:
\begin{equation}
    (0,4) + (0,6) + (2,4) + (2,6) = (1,3)_{1} \times (1,3)_{1} + (1,3)_{1} \times (1,1) + (1,1) \times (1,3)_{1}. \label{someeqn}
\end{equation}
We use now that if a genuine line operator fuses with another genuine line operator it must give genuine line operators, while if a genuine line operator fuses with a confining/non-genuine line operator, it must give confining/non-genuine line operators. Since $(1,1)$ confines and $(1,3)_{1}$ does not, the previous means that we can extract the fusion $(1,3)_{1} \times (1,3)_{1}$ by inspecting the unconfined excitations on the left hand side of \eqref{someeqn} after restriction. Doing this we obtain:
\begin{equation}
    (1,3)_{1} \times (1,3)_{1} = 0 + (0,6). \label{isingfusion3}
\end{equation}
As promised, the fusion rules \eqref{isingfusion1}, \eqref{isingfusion2}, and \eqref{isingfusion3} are indeed those of Ising.

\subsection{$\big( SU(4)_{1} \times SU(2)_{-10}\big)/\mathcal{A}$}
\label{Ising Maverick Duality}

\begin{table}[t]
\centering
\begin{tabular}[h]{|p{4cm}|p{4cm}|p{4cm}| }
\hline 
\multicolumn{3}{|c|}{$SU(4)_{1}$} \\
\hline
Line label & Quantum Dimension & Conformal Weight  \\
\hline
$\mathbf{1}$ & $d_{\mathbf{1}} = 1$ & $h_{\mathbf{1}}=0$ \\
$\mathbf{4}$ & $d_{\mathbf{4}} = 1$ & $h_{\mathbf{4}}=3/8$ \\
$\mathbf{6}$ & $d_{\mathbf{6}} = 1$ & $h_{\mathbf{6}}=1/2$ \\
$\mathbf{\bar{4}}$ & $d_{\mathbf{\bar{4}}} = 1$ & $h_{\mathbf{\bar{4}}}=3/8$ \\
\hline
\end{tabular}
\caption{$SU(4)_{1}$ data.}  \label{su4lv1table}
\end{table}

This is an amusing example that we touched upon in Section \ref{maverickdualities} of a Maverick duality expressing the Ising TQFT (or rather any of its many coset descriptions) in terms of $SU(4)_{1}$ and $SU(2)_{10}$ Chern-Simons gauge theories via non-abelian anyon condensation. The spectrum of $SU(4)_{1}$ can be found in Table \ref{su4lv1table} and that of $SU(2)_{10}$ can be found in Table \ref{su2lv10table}. $SU(4)_{1}$ follows $\mathbb{Z}_{4}$ fusion rules, while those of $SU(2)_{10}$ can be obtained from Eqn. \eqref{SU2kFusionRules}. The product theory has three non-trivial bosons: $(\mathbf{6},10), (\mathbf{1},6)$, and $(\mathbf{6},4)$. The first of these is abelian, while the other two are non-abelian. Condensing just the non-abelian boson $(\mathbf{1},6)$ is tantamount to condensing only the anyon $6$ in $SU(2)_{10}$, which leads to $SU(2)_{10}/\mathcal{A} \cong USp(4)_{1}$ where $\mathcal{A} = 0 + 6$. In the product theory, one would then obtain $SU(4)_{1} \times USp(4)_{-1}$. This points to the fact that one should instead condense all available bosons to obtain the Ising TQFT:
\begin{align}
    &(\mathbf{6},10) \longrightarrow 0, \\[0.3cm]
    &(\mathbf{1},6) \longrightarrow 0 + (\mathbf{1},6)_{2}, \\[0.3cm]
    &(\mathbf{6},4) \longrightarrow 0 + (\mathbf{6},4)_{2}, 
\end{align}
where the non-abelian anyons cannot be split further, as it is not allowed by the fusion rules. For example:
\begin{equation}
    (\mathbf{1},6) \times (\mathbf{1},6) = (\mathbf{1},0) + (\mathbf{1},6) + \ldots \longrightarrow 0 + 0 + \ldots,
\end{equation}
implies that $(\mathbf{1},6)$ can have at most a twofold split. It is also easy to see from $(\mathbf{6},10) \times (\mathbf{1},6) = (\mathbf{6},4)$ that $(\mathbf{1},6)$ and $(\mathbf{6},4)$ must share the same restriction: $(\mathbf{1},6)_{2} \cong (\mathbf{6},4)_{2}$. This finishes the discussion as far as the bosons go.

\begin{table}[!b]
\centering
\begin{tabular}[h]{|p{4cm}|p{4cm}|p{4cm}| }
\hline 
\multicolumn{3}{|c|}{$SU(2)_{10}$} \\
\hline
Line label & Quantum Dimension & Conformal Weight \\
\hline
0 & $d_{0} = 1$ & $h_{0}=0$ \\
1 & $d_{1} = \sqrt{2+\sqrt{3}}$ & $h_{1}=1/16$ \\
2 & $d_{2} = 1 + \sqrt{3}$ & $h_{2}=1/6$ \\
3 & $d_{3} = \sqrt{2} + \sqrt{2+\sqrt{3}}$ & $h_{3}=5/16$ \\
4 & $d_{4} = 2+\sqrt{3}$ & $h_{4}=1/2$ \\
5 & $d_{5} = 2\sqrt{2+\sqrt{3}}$ & $h_{5}=35/48$ \\
6 & $d_{6} = 2+\sqrt{3}$ & $h_{6}=1$ \\
7 & $d_{7} = \sqrt{2} + \sqrt{2+\sqrt{3}}$ & $h_{7}=21/16$ \\
8 & $d_{8} = 1 + \sqrt{3}$ & $h_{8}=5/3$ \\
9 & $d_{9} = \sqrt{2+\sqrt{3}}$ & $h_{9}=33/16$ \\
10 & $d_{10} = 1$ & $h_{10}=5/2$ \\
\hline
\end{tabular}
\caption{$SU(2)_{10}$ data.}  \label{su2lv10table}
\end{table}

We focus now on anyons that are fermions, i.e., that have topological spin $\theta = -1$. These are $(\mathbf{6},0)$, $(\mathbf{1},10)$, $(\mathbf{1},4)$, and $(\mathbf{6},6)$. Using similar fusion rule arguments as above we can deduce that the anyons in the parent $(\mathbf{1},4)$ and $(\mathbf{6},6)$ share the same restriction, and moreover:
\begin{equation}
    (\mathbf{6},0) \times (\mathbf{1},4) = (\mathbf{6},4) \longrightarrow 0 + \ldots,
\end{equation}
implies $(\mathbf{1},4) \to (\mathbf{6},0) + (\mathbf{1},4)_{2}$. We can show that the second component $(\mathbf{1},4)_{2}$ confines by studying the fusion with the boson $(\mathbf{1},6)$:
\begin{equation}
    (\mathbf{1},4) \times (\mathbf{1},6) = (\mathbf{1},6) + \ldots \longrightarrow 0 + \ldots
\end{equation}
so $(\mathbf{1},4)$ and $(\mathbf{1},6)$ must have one of their components identified. But $(\mathbf{6},0) \in (\mathbf{1},4)$ cannot condense (i.e., we cannot identify it to the vacuum), so the only consistent identification is
\begin{equation}
    (\mathbf{1},4)_{2} \cong (\mathbf{1},6)_{2},
\end{equation}
from which it is easy to study their lift to anyons in the parent and check that they confine. The remaining fermion is $(\mathbf{1},10)$, which identifies with $ (\mathbf{6},0)$ since $(\mathbf{6},10) \times (\mathbf{1},10) = (\mathbf{6},0)$ and $(\mathbf{6},10) \to 0$, so $(\mathbf{1},10) \cong (\mathbf{6},0)$. 

At this point the list of unconfined anyons are the trivial one, and one fermion $(\mathbf{6},0)$. Let us focus now on the spectrum of anyons in the parent with topological spin $\theta = e^{2 \pi i / 16}$. These correspond to $(\mathbf{4},3)$, $(\mathbf{\bar{4}},3)$, $(\mathbf{4},7)$, and $(\mathbf{\bar{4}},7)$. The quantum dimensions of all these anyons in the parent is $d_{(\mathbf{4},3)} = \sqrt{2} + \sqrt{2 + \sqrt{3}}$, and fusions of the form
\begin{equation}
    (\mathbf{4},3) \times (\mathbf{\bar{4}},3) = (\mathbf{1},0) + (\mathbf{1},2) + (\mathbf{1},4) + (\mathbf{1},6) \longrightarrow 0 + 0 + \ldots,
\end{equation}
allows to conclude that this set of anyons all split in two. Furthermore, because $(\mathbf{6},10) \to 0$, the fusions $(\mathbf{6},10) \times (\mathbf{4},3) = (\mathbf{\bar{4}},7)$ and $(\mathbf{6},10) \times (\mathbf{\bar{4}},3) = (\mathbf{4},7)$ imply that the following pairs of anyons in the parent share their restriction:
\begin{equation}
    (\mathbf{4},3) \cong (\mathbf{\bar{4}},7), \quad \mathrm{and} \quad (\mathbf{4},7) \cong (\mathbf{\bar{4}},3)
\end{equation}

To deduce the splitting of the quantum dimensions we will consider the fusion with the anyon $(\mathbf{\bar{4}},9)$. This anyon is such that
\begin{equation}
    (\mathbf{4},9) \times (\mathbf{\bar{4}},9) = (\mathbf{1},0) + (\mathbf{1},2) \longrightarrow 0 + \ldots,
\end{equation}
so it does not split, and such that
\begin{equation}
    (\mathbf{4},9) \times (\mathbf{4},9) = (\mathbf{6},0+2) = (\mathbf{6},0) + (\mathbf{6},2),
\end{equation}
so it is not self-conjugate. That is, $(\mathbf{\bar{4}},9)$ corresponds to a different excitation in the child theory. Then:
\begin{equation}
    (\mathbf{\bar{4}},9) \times (\mathbf{4},3) = (\mathbf{1},6) + (\mathbf{1},8) \longrightarrow 0 + \ldots,
\end{equation}
implies that $(\mathbf{4},9) \in (\mathbf{4},3)$, so the splitting of the quantum dimensions in $(\mathbf{4},3)$ corresponds to $d_{(\mathbf{4},9)} = \sqrt{2+\sqrt{3}}$, and another component with quantum dimension $\sqrt{2}$. From the previous we also conclude that $(\mathbf{4},9)$ and $(\mathbf{\bar{4}},9)$ confine.
Finally, from the fusions
\begin{equation}
    (\mathbf{4},3) \times (\mathbf{4},7) = (\mathbf{6},4) + (\mathbf{6},6) + (\mathbf{6},8) + (\mathbf{6},10) \longrightarrow 0 + 0 + \ldots,
\end{equation}
\begin{equation}
    (\mathbf{4},3) \times (\mathbf{4},3) = (\mathbf{6},0) + (\mathbf{6},2) + (\mathbf{6},4) + (\mathbf{6},6) \longrightarrow 0 + \ldots,
\end{equation}
we conclude that the restrictions must be
\begin{equation}
    (\mathbf{4},3) = (\mathbf{4},3)_{1} + (\mathbf{4},9),
\end{equation}
\begin{equation}
    (\mathbf{4},7) = (\mathbf{4},3)_{1} + (\mathbf{\bar{4}},9),
\end{equation}
with $(\mathbf{4},9)$ and $(\mathbf{\bar{4}},9)$ confining. $(\mathbf{4},3)_{1}$ is then the anyon that descends to the spin field with topological spin $\theta = e^{2 \pi i /16}$ in the Ising model.

It is now straightforward to argue for the confinement of the remaining excitations. For example, $(\mathbf{6},10) \times (\mathbf{1},3) = (\mathbf{6},7) \Longrightarrow (\mathbf{1},3) \cong (\mathbf{6},7)$. Studying the lift of their spin to the parent, it can readily be checked for the confinement of $(\mathbf{6},7)$ and $(\mathbf{1},3)$. A similar argument holds for the rest of the anyons, with the exception of $(\mathbf{4},5)$. To argue for the confinement of this anyon, notice
\begin{equation}
    (\mathbf{4},5) \times (\mathbf{4},5) = (\mathbf{6},4) +  (\mathbf{6},10) \longrightarrow 0 + 0 + \ldots, 
\end{equation}
so $(\mathbf{4},5)$ splits in two, and
\begin{align}
    & (\mathbf{4},9) \times (\mathbf{4},5) = (\mathbf{6},4) + (\mathbf{6},6) \longrightarrow 0 + \ldots, \\[0.3cm]
    & (\mathbf{\bar{4}},9) \times (\mathbf{4},5) = (\mathbf{1},4) + (\mathbf{1},6) \longrightarrow 0 + \ldots, 
\end{align}
imply that $(\mathbf{4},9), (\mathbf{\bar{4}},9) \in (\mathbf{4},5)$. As argued above, $(\mathbf{4},9)$ and $(\mathbf{\bar{4}},9)$ cannot identify with each other, so it must be that $(\mathbf{4},5) \longrightarrow (\mathbf{4},9) + (\mathbf{\bar{4}},9)$, and thus all components of $(\mathbf{4},5)$ confine.

Then, the unconfined excitations are $0$, $(\mathbf{6},0)$ and $(\mathbf{4},3)_{1}$, which as expected, gives the data of the Ising TQFT.

\subsection{$(E_{6})_{1} \cong (G_{2})_{3}/\mathcal{A}$} \label{E61fromG23}

\begin{table}[t]
\centering
\begin{tabular}[h]{|p{4cm}|p{4cm}|p{4cm}| }
\hline 
\multicolumn{3}{|c|}{$(G_{2})_{3}$} \\
\hline
Line label & Quantum Dimension & Conformal Weight  \\
\hline
$\mathbf{1}$ & $d_{\mathbf{1}} = 1$ & $h_{\mathbf{1}}=0$ \\
$\mathbf{7}$ & $d_{\mathbf{7}} = \frac{3+\sqrt{21}}{2}$ & $h_{\mathbf{7}}=2/7$ \\
$\mathbf{27}$ & $d_{\mathbf{27}} = \frac{7+\sqrt{21}}{2}$ & $h_{\mathbf{27}}=2/3$ \\
$\mathbf{77}$ & $d_{\mathbf{77}} = \frac{3+\sqrt{21}}{2}$ & $h_{\mathbf{77}}=8/7$ \\
$\mathbf{14}$ & $d_{\mathbf{14}} = \frac{3+\sqrt{21}}{2}$ & $h_{\mathbf{14}}=4/7$ \\
$\mathbf{64}$ & $d_{\mathbf{64}} = \frac{5+\sqrt{21}}{2}$ & $h_{\mathbf{64}}=1$ \\
\hline
\end{tabular}
\caption{$(G_{2})_{3}$ data.}  \label{G2lv3table}
\end{table}

\begin{table}[!b]
\centering
\begin{tabular}[h]{|p{4cm}|p{4cm}|p{4cm}| }
\hline 
\multicolumn{3}{|c|}{$(E_{6})_{1}$} \\
\hline
Line label & Quantum Dimension & Conformal Weight  \\
\hline
$\mathbf{1}$ & $d_{\mathbf{1}} = 1$ & $h_{\mathbf{1}}=0$ \\
$\mathbf{27}$ & $d_{\mathbf{27}} = 1$ & $h_{\mathbf{27}}=2/3$ \\
$\overline{\mathbf{27}}$ & $d_{\overline{\mathbf{27}}} = 1$ & $h_{\overline{\mathbf{27}}}=2/3$ \\
\hline
\end{tabular}
\caption{$(E_{6})_{1}$ data.}  \label{E6lv1table}
\end{table}

This is an interesting example of a single affine lie algebra embedding into another single one mentioned in Section \ref{FurtherConformalEmbeddings}. The spectrum of $(G_{2})_{3}$ can be found in Table \ref{G2lv3table} while the spectrum of $(E_{6})_{1}$ which is the expect result is presented in Table \ref{E6lv1table}. The $(E_{6})_{1}$ Chern-Simons theory has $\mathbb{Z}_{3}$ fusion rules. Those of $(G_{2})_{3}$ are fairly unwieldy to write down, so instead we just use them as we need them through the calculation.

Our starting theory $(G_{2})_{3}$ has only one non-trivial boson; namely $\mathbf{64}$, and it is non-abelian. We assume it condenses, and the fusion rule
\begin{equation}
    \mathbf{64} \times \mathbf{64} = \mathbf{1} + \mathbf{7} + \mathbf{27} + \mathbf{77} + \mathbf{14} + \mathbf{64} \longrightarrow 0 + 0 + \ldots
\end{equation}
needs $\mathbf{64}$ to split in two for consistency with the right-hand side. Thus, $\mathbf{64} \to 0 + \mathbf{64}_{2}$, with $d_{\mathbf{64}_{2}} = (3+\sqrt{21})/2$. We can now use the fusion rules
\begin{align}
    \mathbf{7}  &\times \mathbf{64} = \mathbf{27} + \mathbf{77} + \mathbf{14} + \mathbf{64} \longrightarrow 0 + \ldots \\[0.3cm]
    \mathbf{14} &\times \mathbf{64} = \mathbf{7} + \mathbf{27} + \mathbf{77} + \mathbf{64} \longrightarrow 0 + \ldots \\[0.3cm]
    \mathbf{77} &\times \mathbf{64} = \mathbf{7} + \mathbf{27} + \mathbf{14} + \mathbf{64} \longrightarrow 0 + \ldots
\end{align}
where we conclude that $\mathbf{7}$, $\mathbf{14}$, and $\mathbf{77}$ belong in the restriction of $\mathbf{64}$. Clearly, the only possibility is to have the identifications $\mathbf{7} \cong \mathbf{14} \cong \mathbf{77} \cong \mathbf{64}_{2}$ from which we also deduce the confinement of these excitations.

It only remains to study $\mathbf{27}$. Examining the self-fusion
\begin{equation}
    \mathbf{27} \times \mathbf{27} = \mathbf{1} + 2 (\mathbf{64}) + \ldots \longrightarrow 0 + 0 + 0 + \ldots,
\end{equation}
we see that $\mathbf{27}$ splits in three: $\mathbf{27} \to \mathbf{27}_{1} + \mathbf{27}_{2} + \mathbf{27}_{3}$. To assign quantum dimensions it is sufficient to study the fusion
\begin{equation}
    \mathbf{27} \times \mathbf{64} = \mathbf{64} + \ldots \longrightarrow 0 + \ldots,
\end{equation}
from which we deduce that one component of $\mathbf{27}$ must belong in the restriction of $\mathbf{64}$. Obviously, $\mathbf{27}$ cannot condense so one of its components, say $\mathbf{27}_{3}$, must identify with $\mathbf{64}_{2}$: $\mathbf{27}_{3} \cong \mathbf{64}_{2}$, so $\mathbf{27}_{3}$ confines, and we must have $d_{\mathbf{27}_{1}} = d_{\mathbf{27}_{2}} = 1$ and $d_{\mathbf{27}_{3}} = (3 + \sqrt{21})/2$.

The fusion rules of the remaining non-confined excitations 0, $\mathbf{27}_{1}$ and $\mathbf{27}_{2}$ can be deduced from the associativity of the fusion and the fact that all these components are abelian. Then, indeed they have the correct spins, fusion rules, and quantum dimensions to recognize the expected result $(E_{6})_{1}$. So we have indeed found that
\begin{equation}
    (G_{2})_{3}/\mathcal{A} = (E_{6})_{1},
\end{equation}
where $\mathcal{A} = \mathbf{1} + \mathbf{64}$.

\subsection{$SU(2)_{3} \cong \big( USp(6)_{1} \times SO(3)_{-4} \big)/\mathcal{A}$} \label{SU23fromusP61SO3m4}

\begin{table}[t]
\centering
\begin{tabular}[h]{|p{4cm}|p{4cm}|p{4cm}| }
\hline 
\multicolumn{3}{|c|}{$USp(6)_{1}$} \\
\hline
Line label & Quantum Dimension & Conformal Weight  \\
\hline
$\mathbf{1}$ & $d_{\mathbf{1}} \ \ = 1$ & $h_{\mathbf{1}} \ \ = 0$ \\
$\mathbf{14}'$ & $d_{\mathbf{14}'} = 1$ & $h_{\mathbf{14}'}=3/4$ \\
$\mathbf{14}$ & $d_{\mathbf{14}} \, = \frac{1 + \sqrt{5}}{2}$ & $h_{\mathbf{14}} = \, 3/5$ \\
$\mathbf{6}$ & $d_{\mathbf{6}} \ \ = \frac{1 + \sqrt{5}}{2}$ & $h_{\mathbf{6}} \ \ = 7/20$ \\
\hline
\end{tabular}
\caption{$USp(6)_{1}$ data.}  \label{USp6lv1table}
\end{table}

In this subsection we consider an example in the infinite family of conformal embeddings studied in Section \ref{continuingclassicalembeddings}:
\begin{equation}
     SO(N)_{4} \times SU(2)_{N} \hookrightarrow USp(2N)_{1},
\end{equation}
where we attempt to express $SU(2)_{N}$ in terms of $SO(N)_{4}$ and $USp(2N)_{1}$. We consider $N=3$ which is the simplest case in which non-abelian anyon condensation must be considered.

The spectrum of $USp(6)_{1}$ and $SO(3)_{4}$ can be found in Tables \ref{USp6lv1table} and \ref{so3lv4table} respectively. The fusion rules of $USp(6)_{1}$ are given as follows:
\begin{align}
    &\mathbf{14}' \times \mathbf{14}' = \mathbf{1}, \quad \mathbf{14}' \times \mathbf{14} = \mathbf{6}, \quad \mathbf{14}' \times \mathbf{6} = \mathbf{14}, \nonumber \\[0.3cm]
    &\mathbf{14} \times \mathbf{14} = \mathbf{1} + \mathbf{14}, \quad \mathbf{14} \times \mathbf{6} = \mathbf{14}' + \mathbf{6}, \quad \mathbf{6} \times \mathbf{6} = \mathbf{1} + \mathbf{14}.
\end{align}
The fusion rules of $SO(3)_{4}$ may be found in Eqn. \eqref{so3lv4fusionrules}.

It is easy to see that the product $USp(6)_{1} \times SO(3)_{-4}$ has two bosons, both of which are non-abelian: $(\mathbf{14},4_{1})$ and $(\mathbf{14},4_{2})$. Running a similar argument as in the beginning of Appendix \ref{ExampleforsimplestMaverick} we can see that only one of them can condense, and since $4_{1}$ and $4_{2}$ are symmetric between each other the choice is immaterial. We choose $(\mathbf{14},4_{1})$ to condense, and thus $(\mathbf{14},4_{2})$ does not split and has quantum dimension $d_{(\mathbf{14},4_{2})} = (3+\sqrt{5})/2$. To study how $(\mathbf{14},4_{1})$ splits consider the fusion and restriction $(\mathbf{14},4_{1}) \times (\mathbf{1},4_{1}) =  (\mathbf{14}, 0) + (\mathbf{14},4_{1}) \to 0$. Then, $(\mathbf{1},4_{1})$ belongs to the restriction of $(\mathbf{14},4_{1})$ and therefore confines.

We can study the fate of the boson $(\mathbf{14},4_{2})$ by computing the fusion with $(\mathbf{14},2)$:
\begin{align}
    (\mathbf{14},2) \times (\mathbf{14},2) &= (\mathbf{1},0) + (\mathbf{14},4_{1}) + \ldots \longrightarrow 0 + 0 + \ldots, \\[0.3cm]
    (\mathbf{14},4_{2}) \times (\mathbf{14},2) &= (\mathbf{1}, 2) + (\mathbf{1}, 4_{1}) + (\mathbf{14}, 2) + (\mathbf{14}, 4_{1}) \longrightarrow 0 + \ldots
\end{align}
The first fusion says that $(\mathbf{14},2)$ splits into two components $(\mathbf{14},2) \to (\mathbf{14},2)_{1} +  (\mathbf{14},2)_{2}$, and the second fusion implies the identification $(\mathbf{14},4_{2}) \cong (\mathbf{14},2)_{1}$\footnote{Here we have defined $(\mathbf{14},2)_{1}$ to be the component of $(\mathbf{14},2)$ that identifies.} and corresponding confinement of these excitations. Correspondingly, $d_{(\mathbf{14},2)_{1}} = (3 + \sqrt{5})/2$ and $d_{(\mathbf{14},2)_{2}} = (1 + \sqrt{5})/2$. We may now identify the remaining component of $(\mathbf{14},2)$ with $(\mathbf{1},4_{2})$, in accordance with the fusion
\begin{equation}
    (\mathbf{1}, 4_{2}) \times (\mathbf{14}, 2) = (\mathbf{14}, 2) +  (\mathbf{14}, 4_{1}) \longrightarrow 0 + \ldots
\end{equation}
and matching of quantum dimensions. So, indeed $(\mathbf{14},2)_{2} \cong (\mathbf{1},4_{2})$.

There are two additional identifications that we can deduce, namely
\begin{equation}
     (\mathbf{14},4_{2}) \times (\mathbf{1},2) = (\mathbf{14}, 2) + (\mathbf{14},4_{1}) \longrightarrow 0 + \ldots
\end{equation}
implies $(\mathbf{1},2) \cong (\mathbf{14},4_{2})$, and so $(\mathbf{1},2)$ confines. Also
\begin{equation}
    (\mathbf{14},0) \times (\mathbf{1},4_{1}) = (\mathbf{14},4_{1}) \longrightarrow 0 + \ldots
\end{equation}
implies $(\mathbf{14},0) \cong (\mathbf{1},4_{1})$, and since $(\mathbf{1},4_{1})$ confines, so does $(\mathbf{14},0)$.

We move-on now to consider anyons of the form $(\mathbf{14}',a)$ and $(\mathbf{6},a)$, with $a$ a label in $SO(3)_{-4}$. Studying self-fusions as usual we deduce that $(\mathbf{6},2)$ and $(\mathbf{6},4_{1})$ split into two components, while $(\mathbf{6},4_{2})$ does not split. The fusion
\begin{equation}
    (\mathbf{6},4_{2}) \times (\mathbf{6},2) = (\mathbf{14},4_{1}) + \ldots \longrightarrow 0 + \ldots,
\end{equation}
implies that $(\mathbf{6},2)$ restricts as $(\mathbf{6},2) \to (\mathbf{6},4_{2}) + (\mathbf{6},2)_{2}$, and as such $(\mathbf{6},4_{2})$ confines. The remaining component is identified with $(\mathbf{14}',4_{2})$ because of the fusion
\begin{equation}
    (\mathbf{14}',4_{2}) \times (\mathbf{6},2) = (\mathbf{14},2) + (\mathbf{14},4_{1}) \longrightarrow 0 + \ldots
\end{equation}
and the matching of the quantum dimensions, so we obtain the full restriction $(\mathbf{6},2) \to (\mathbf{14}',4_{2}) + (\mathbf{6},4_{2})$.

Similarly, we have the fusions $(\mathbf{14}',0) \times (\mathbf{6}, 4_{1}) = (\mathbf{14},4_{1}) \to 0 + \ldots$ and $(\mathbf{14}', 4_{1}) \times (\mathbf{6},4_{1}) = (\mathbf{14},0) + (\mathbf{14},4_{1}) \to 0 + \ldots$. From this we find the restriction $(\mathbf{6},4_{1}) \to (\mathbf{14}',0) +(\mathbf{14}',4_{1})$, and in turn we deduce the confinement of $(\mathbf{14}',4_{1})$. 

Finally, from the fusions $(\mathbf{14}', 2) \times (\mathbf{6}, 4_{2})$ and $(\mathbf{6}, 0) \times (\mathbf{14}', 4_{1})$ we may deduce the identifications $(\mathbf{14}',2) \cong (\mathbf{6},4_{2})$ and $(\mathbf{6}, 0) \cong (\mathbf{14}', 4_{1})$ and the corresponding confinement of such excitations.

All in all, considering identifications we get the unconfined excitations $(\mathbf{14}', 4_{2})$, $(\mathbf{1},4_{2})$ and $(\mathbf{14}',0)$ on top of the vacuum. This correctly reproduces the spectrum of the expect result $SU(2)_{3}$ (presented in Table \ref{su2lv3table}) according to the conformal embedding \eqref{additionalembedding}.

\bibliographystyle{JHEPmod}
\bibliography{references}

\end{document}